\documentclass[tradiabstract]{aa}
\usepackage[dvipsnames]{xcolor}
\usepackage[varg]{txfonts}
\usepackage{epstopdf}
\usepackage{multicol}
\usepackage{natbib}
\usepackage{rotating}
\usepackage{graphicx}
\usepackage{rotating}
\usepackage{multirow}
\usepackage{url}
\usepackage{amssymb}

\usepackage{comment}

\bibpunct{(}{)}{;}{a}{}{,} 

\usepackage[flushleft]{threeparttable}

\usepackage{caption}

\definecolor{green}{rgb}{0,0.5,0}

\newcommand*\mean[1]{\overline{#1}}

\newcommand{\zabs}{\ensuremath{z_{\rm abs}}}

\newcommand{\HH}{\mbox{H$\rm _2$}}

\newcommand{\lya}{\mbox{${\rm Ly}\alpha$}}
\newcommand{\Ha}{\mbox{${\rm H}\alpha$}}
\newcommand{\Hb}{\mbox{${\rm H}\beta$}}

\newcommand{\CI}{\ion{C}{i}}

\newcommand{\CrII}{\ion{Cr}{ii}}

\newcommand{\FeII}{\ion{Fe}{ii}}
\newcommand{\HI}{\ion{H}{i}}
\newcommand{\MgI}{\ion{Mg}{i}}
\newcommand{\MgII}{\ion{Mg}{ii}}
\newcommand{\MnII}{\ion{Mn}{ii}}

\newcommand{\NiII}{\ion{Ni}{ii}}

\newcommand{\SiII}{\ion{Si}{ii}}

\newcommand{\ZnII}{\ion{Zn}{ii}}

\newcommand{\OIII}{\ion{O}{iii}}

\newcommand{\kms}{\ensuremath{{\rm km\,s^{-1}}}}
\newcommand{\cmsq}{\ensuremath{{\rm cm}^{-2}}}

\begin{document}

\title{
Chemical enrichment and host galaxies of extremely-strong intervening DLAs towards quasars \thanks{Based on observations performed with Very Large Telescope of the European Southern Observatory under Prog.~ID 095.A-0224(A) using the X-shooter spectrograph.}}
\subtitle{Probing same galactic environments as DLAs associated to $\gamma$-ray burst afterglows ?}
\titlerunning{Spectroscopic observations of extremely-strong damped Lyman-$\alpha$ absorbers.}

\author{A.~Ranjan\inst{1}
   \and P.~Noterdaeme\inst{1}
   \and J.-K.~Krogager\inst{1}
   \and P.~Petitjean\inst{1}
   \and R.~Srianand\inst{2}
   \and S.~A.~Balashev\inst{3}
   \and N.~Gupta\inst{2}
   \and C.~Ledoux\inst{4}
}

\institute{
Institut d'Astrophysique de Paris, UMR\,7095, CNRS-SU, 98bis boulevard Arago, 75014 Paris, France
  \and
  Inter-University Centre for Astronomy and Astrophysics, Post Bag 4, Ganeshkhind, 411 007, Pune, India
\and Ioffe Institute, Polytekhnicheskaya 26, 194021 Saint Petersburg, Russia
\and European Southern Observatory, Alonso de C\'ordova 3107, Vitacura, Casilla 19001, Santiago, Chile
}

\date{\today}

\abstract{

We present the results from VLT/X-shooter spectroscopic observations of 11 extremely strong intervening damped Lyman-$\alpha$ absorbers (ESDLAs) initially selected as high $N(\HI)$ (i.e. $\ge 5 \times 10^{21}$~\cmsq\ ) candidates from the Sloan Digital Sky Survey (SDSS).
We confirm the high \HI\ column densities which we measure to be in the range $\log N(\HI)=21.6-22.4$. Molecular hydrogen is detected with high column densities ($N(\HH) \ge 10^{18}$~\cmsq) in five out of eleven systems, three of which are reported here for the first time, and we obtain conservative upper limits on $N(\HH)$ for the remaining six systems. We also measure the column density of various metal species (\ZnII, \FeII, \SiII, \CrII, \CI), quantify the absorption-line kinematics ($\Delta v_{90}$), and estimate the extinction of the background quasar light ($A_V$) due to dust in the absorbing gas.

We compare the chemical properties of this sample of ESDLAs, supplemented with literature measurements, to that of DLAs located at the redshift of long-duration $\gamma$-ray bursts (GRB-DLAs).

We confirm that the two populations are almost indistinguishable in terms of chemical enrichment and gas kinematics. In addition, we find no marked differences in the incidence of H$_2$. All this suggests that ESDLAs and GRB-DLAs probe similar galactic environments. 

We search for the {galaxy} counterparts of ESDLAs and find associated emission lines in three out of eleven systems, two of which are reported here for the first time (at $\zabs$~=~2.304 and 2.323 towards the quasars SDSS\,J002503.03+114547.80 and SDSS\,J114347.21+142021.60, respectively). The measured separations between the {quasar} sightlines and {the emission associated with the ESDLA galaxy} (for a total of 5 sightlines) are all very small $(\rho < 3$~kpc). Since our observations are complete up to $\rho \sim 7$~kpc, we argue that the emission counterparts of the remaining systems are more likely below the detection limit than outside the search area. 
While the small impact parameters are similar to what is observed for GRB-DLAs, the associated star-formation rates are on average lower than seen for GRB host galaxies. 
This is explained by long-duration GRBs being associated with the death of massive stars, hence pinpointing regions of active star formation in the GRB host galaxies. 

Our observations support the suggestion from the literature that ESDLAs could act as blind analogues of GRB-DLAs, probing high column density neutral gas in the heart of high-redshift galaxies, without any prior on the instantaneous star-formation rate.

}

\maketitle

\section{Introduction}
Damped \lya\ absorbers (hereafter DLAs) observed in the spectra of luminous background sources, such as quasars (QSOs) or $\gamma$-ray burst (GRB) afterglows, are defined to be systems with 
column densities, $\log N(\rm \HI)\ge20.3$ \citep{wolfe2005}. DLAs are found to probe most of the neutral gas content of the Universe \citep[see][]{Noterdaeme2009,Prochaska2009}. 

Simulations \citep[e.g.][]{Pontzen2008,RahmatiandSchaye2014, Yajima2012, Altay+13} show that at high redshift, intervening DLAs probe mostly gas in galactic haloes or in the circum-galactic medium (CGM). However, such simulations also indicate that the systems with highest neutral gas column densities ($N(\HI)\sim 10^{22}$~\cmsq) should probe dense gas closer to the centre of the associated galaxy, at impact parameters less than a few kpc. Interestingly, systems at the high column density end of the $N(\HI)$ distribution 
provides strong constraints on models of stellar feedback and formation of H$_2$ in simulations \citep[see e.g.][]{Altay+13, Bird2014}. Nevertheless, the incidence of such intervening systems along quasar lines of sight is rare due to the small projected area of this dense gas.

One way to probe the gas in the central regions of galaxies is then to study damped \lya\ absorbers located at the redshift of long-duration GRBs and observed in their afterglow spectra (hereafter GRB-DLAs\footnote{We use the term `GRB-DLA' to refer to systems {\sl at the GRB redshift}, as opposed to {\sl intervening GRB-DLAs} with $z_{\rm DLA} \ll z_{\rm GRB}$, while `QSO-DLA' refers to {\sl intervening} DLAs along quasar sightlines, as opposed to {\sl proximate QSO-DLAs} with $z_{\rm DLA} \approx z_{\rm QSO}$. This is in agreement with the standard terminology in the literature.}). Long duration GRBs are thought to be associated with the death of massive stars \citep[e.g.][]{Woosley2006}, and are indeed found to be located in the actively star forming regions of their host galaxies \citep[see][]{Fruchter2006, Lyman2017}.
As expected, the $N(\HI)$ distribution of GRB-DLAs is found to be skewed towards high $N(\HI)$ values \citep[see][]{Fynbo2009, Selsing2019}.

However, the main difficulty with studying GRB-DLAs is that high resolution spectroscopic observations of GRB afterglows are difficult to obtain due to the rapid decrease of their luminosity. The situation gets further complicated in the presence of dust in the associated absorber. Dark GRBs have large dust attenuation (A$_{V}\,>\,0.5$\,mag) and hence difficult to detect as the GRB afterglow becomes too faint for optical follow-up \citep[see e.g.][]{Ledoux2009, Bolmer2019}.

With the very large number of quasar spectra available in the Sloan Digital Sky Survey \citep[SDSS;][]{York2000}, it is now possible to build samples of {\sl intervening} absorbers with very high \HI\ column densities. \citet{noterdaeme2014} have found 104 systems with $\log N(\rm \HI)\geq21.7$ ({they coined the name extremely strong DLAs, ESDLAs, for such systems}) in the spectra of 140\,000 quasars from the Baryon Oscillation Spectroscopic Survey (BOSS) of the SDSS-III Data Release 11 \citep{Paris2017}. From \lya\ emission detected in stacked spectra of ESDLAs, \citet{noterdaeme2014} suggest that these systems indeed arise at small impact parameters\footnote{The impact parameter is defined as the projected separation between the centroid of the nebular emission and the quasar sightline.} (typically $\rho<2.5$~kpc), {a result also substantiated by two direct detections at $\rho\sim 1$~kpc reported by \citet{Noterdaeme2012} and \citet{Ranjan+2018}}. Furthermore, ESDLAs are found to have higher density and H$_2$ incidence rate compared to the `normal' DLA population \citep{Noterdaeme2015a, Balashev+18}. In general, all these findings are in line with the expectations based on the Kennicutt-Schmidt law which implies that probing neutral gas at the highest column densities can help reveal molecular gas and star-forming regions originating inside the optical disk of the associated galaxy.

Several works in the literature bring accumulating evidence for a similarity between GRB-DLAs and strong QSO-DLAs. \citet{Guimaraes2012} show that the metallicity and depletion of high N(\HI) QSO-DLAs ($\log N(\rm \HI)\,>\,21.5$) are similar to GRB-DLAs. \citet{Noterdaeme2012} show that the properties of one detected ESDLA galaxy (SDSS J\,1135$-$0010), at very small impact parameter from the quasar line of sight, is similar to some GRB host galaxies. Furthermore, \citet{noterdaeme2014} and \citet{Noterdaeme2015b} discuss the similarity in absorption characteristics of ESDLAs and GRB-DLAs. 

More recently, \citet{Bolmer2019} show that there is no apparent lack of \HH\, in their GRB-DLA sample as compared to QSO-DLAs, supporting the previous findings of the similarity in chemical enrichment as well as providing indirect evidence of the similarity of surrounding UV field between the two absorber sub-sets. The presence of \HH\ indicates that the sight-lines are passing close to the centre of the associated galaxies, where the gas pressure favours the \HI\,-\HH\, transition \citep[see][]{Blitz2006, Balashev+17}. Furthermore, impact parameter measurements of GRB-DLAs by \citet{Lyman2017} and \citet{Arabsalmani2015} give direct evidence for the proximity of the absorbing gas to its associated galaxy.  

It is, however, important to note that ESDLA observations are scarce which has led to very small and heterogeneous samples of ESDLAs in the past.
In order to further study the chemical enrichment, the molecular content and the association with galaxies of ESDLAs, we have therefore started a follow-up campaign of ESDLAs using the Very Large Telescope (VLT) of the European Southern Observatory. 
As part of this campaign, 11 ESDLA systems have been observed with the X-shooter instrument, more than doubling the previous sample size. 
A detailed analysis of one of these systems has already been presented by \citet{Ranjan+2018}. We present here the analysis of the remaining 
10 ESDLAs and compare the properties of ESDLAs in our sample (supplemented with literature measurements) to those of GRB-DLAs. The observations and data reduction 
are presented together with the compilation of literature data in Section~\ref{observations}. The analysis of the absorption lines and dust reddening is described in Section~\ref{absorption_analysis} and that of associated emission lines in Sect.~\ref{emission_analysis}. 
We then discuss our results in Sect.~\ref{discussion} and summarize our findings in Sect.~\ref{Conclusion}.
Column densities are always stated in units of cm$^{-2}$. 
Standard $\Lambda$CDM flat cosmology is used for the paper with 
$\rm H_0 = 67.8~km\,s^{-1} Mpc^{-1}$, $\rm \Omega_{\Lambda} = 0.692$ and $\rm \Omega_{m} = 0.308 $ \citep{Planck2016}.


\section{Observations and data reduction \label{observations} }

Observations of the 11 quasars were carried out in service mode under good seeing conditions (typically 0.7-0.8$\arcsec$) between April, 2015 and July, 2016 under ESO program ID 095.A-0224(A) 
with the multiwavelength, medium-resolution spectrograph X-shooter \citep{Vernet2011} mounted at the Cassegrain focus of the Very Large Telescope (VLT-UT2) at Paranal, Chile. The observations were carried out using a 2-step nodding mode with an offset of 4 arcsec between the two integrations.
Because of the failure of the atmospheric dispersion corrector during that period, we performed all observations with the slit aligned with the parallactic angle at the start of each exposure; the slit angle then remained fixed on the sky during the integration. The parallactic angle changed little between different observations of a given target ($\pm 15^{\circ}$) except for the quasar SDSS~J\,223250.98+124225.29. In the following, we use a short notation for the quasar names, e.g., SDSS~J\,223250.98+124225.29 is simply referred to as J2232+1242.
The log of observations is given in Table~\ref{tab:journal}. 
We reduced the data using the X-shooter pipeline \citep{Modigliani2010} and combined individual exposures weighting each pixel by the inverse of its variance to obtain the combined 2D and 1D spectra used in this paper. For absorption line and dust analysis, we used only the combined 1-D spectra. For the emission-line analysis, we used not only the combined 2D spectra, but also the individual 2D spectra.

\begin{table*}
\centering
\caption{Log of VLT X-shooter observations}
\label{tab:journal}
\begin{tabular}{ccccccc}
\hline \hline
{\Large \strut} Quasar     & Date  & PA\tablefootmark{a}          & Airmass    & Seeing   & Slit widths\tablefootmark{b}   & Exposure times\tablefootmark{b} \\ 
       &                    & (degrees) &         	& (arcsec) & (arcsec)      & (s)         	\\
\hline

SDSS~J001743.88+130739.84   & 2015-12-07 & -165.17                              & 1.297   & 0.69     & 1.6,0.9,1.2                                   & 2$\times$(1480, 1430, 3$\times$480)                         \\
             & 2015-12-08 & -170.14                              & 1.297   & 0.79     & ''                                            & ''                                               \\
             & 2016-07-29 & 173.0                                & 1.268   & 0.60     & ''                                            & ''                                               \\
SDSS~J002503.03+114547.80   & 2015-09-09 & -168.03                              & 1.261   & 0.77     & ''                                            & ''                                               \\
             & 2015-11-07 & 167.88                               & 1.253   & 0.80     & ''                                            & ''                                               \\
             & 2015-12-09 & -167.57                              & 1.270   & 0.54     & ''                                            & ''                                               \\
SDSS~J114347.21+142021.60   & 2016-03-06 & 174.0                                & 1.284   & 0.93     & ''                                            & ''                                               \\
             & 2015-04-09 & 153.48                               & 1.363   & 0.76     & ''                                            & ''                                               \\
SDSS~J125855.41+121250.21   & 2015-04-16 & -170.97                              & 1.259   & 0.58     & ''                                            & ''                                               \\
             & 2016-03-06 & 175.62                               & 1.247   & 0.42     & ''                                            & ''                                               \\
             & 2016-03-12 & -167.61                              & 1.267   & 0.56     & ''                                            & ''                                               \\
SDSS~J134910.45+044819.91   & 2015-04-23 & -179.29                              & 1.147   & 1.13     & ''                                            & ''                                               \\
             & 2015-05-14 & -168.88                              & 1.158   & 0.60     & ''                                            & ''                                               \\
             & 2015-05-21 & 155.39                               & 1.176   & 1.21     & ''                                            & ''                                               \\
SDSS~J141120.51+122935.96   & 2015-07-16 & 179.58                               & 1.253   & 0.77     & ''                                            & ''                                               \\
             & 2016-03-05 & -178.18                              & 1.256   & 0.73     & ''                                            & ''                                               \\
             & 2016-04-09 & -169.96                              & 1.266   & 0.82     & ''                                            & ''                                               \\
SDSS~J151349.52+035211.68   & 2015-04-15 & -179.25                              & 1.138   & 0.89     & ''                                            & ''                                               \\
             & 2015-04-16 & -160.17                              & 1.165   & 0.69     & ''                                            & ''                                               \\
             & 2015-05-14 & -172.22                              & 1.142   & 0.57     & ''                                            & ''                                               \\
             & 2015-05-15 & -163.75                              & 1.153   & 0.60     & ''                                            & ''                                               \\
SDSS~J214043.02$-$032139.29 & 2015-06-15 & 167.23                               & 1.075   & 0.79     & ''                                            & ''                                               \\
             & 2015-06-17 & 141.07                               & 1.125   & 0.74     & ''                                            & ''                                               \\
SDSS~J223250.98+124225.29   & 2016-06-30 & 132.14                               & 1.731   & 0.81     & ''                                            & ''                                               \\
             & 2016-07-13 & 147.06                               & 1.384   & 0.91     & ''                                            & ''                                               \\
             & 2016-07-29 & -175.21                              & 1.262   & 0.67     & ''                                            & ''                                               \\
SDSS~J224621.14+132821.32   & 2015-11-08 & -165.84                              & 1.298   & 0.83     & ''                                            & ''                                               \\
             & 2015-11-15 & -170.89                              & 1.284   & 0.81     & ''                                            & ''                                               \\
             & 2016-07-30 & -168.52                              & 1.290   & 0.70     & ''                                            & ''                                               \\
SDSS~J232207.30+003348.99   & 2015-09-09 & 144.5                                & 1.164   & 0.83     & ''                                            & ''                                               \\
             & 2015-09-18 & 135.82                               & 1.227   & 0.77     & ''                                            & ''                                               \\
             & 2016-09-20 & 150.6                                & 1.140   & 0.77     & ''                                            & ''                                               \\ \hline
\hline
\end{tabular}
\tablefoot{
\tablefoottext{a}{Position angle, North to East.}
\tablefoottext{b}{For UVB, VIS and NIR arms respectively.}

}
\end{table*}

\subsection{Literature Sample \label{sample_selection}}

Given the small size of our sample (11 systems), we include measurements for other ESDLAs selected from the same parent SDSS sample \citep{noterdaeme2014}. The system towards J\,1513$+$0352 has been singled out by \citet{Ranjan+2018} but belongs to the same selection and observational program that we present in this work. Four additional systems (SDSS J0154$+$1395, SDSS J0816$+$1446, SDSS J1456$+$1609, and SDSS J2140$-$0321) have also been selected based on high N(\HI) content from SDSS, but observed with UVES instead of X-shooter \citep{Guimaraes2012, Noterdaeme2015a}. One of them (SDSS J2140$-$0321) is in common with the present X-shooter sample. All these constitute our {\sl homogeneous} (indicating homogeneity in selection based on N(\HI) measurement from SDSS) ESDLA sample (14 systems). 
We furthermore build a `{\sl total}' sample of ESDLAs that meet the N(\HI) criterion but have been selected based on different criteria. In this total sample, we include the aforementioned `homogeneous sample' together with the following targets:

SDSS J0843$+$0221, which was targeted for the clear presence of H$_2$ lines seen in the SDSS spectrum \citep{Balashev2014}, SDSS J1135$-$0010, which was targeted for the presence of \lya\ emission \citep{Noterdaeme2012}, as well as three additional ESDLAs not from the SDSS sample (HE0027$-$1836, Q0458$-$0203, and Q1157$+$0128) but with coverage of the H$_2$ lines \citep{Noterdaeme2015b}.

For the comparison between absorption properties, we use only the {\sl `homogeneous} sample' of ESDLAs. However, for the comparison of impact parameters, we use the `total sample' of ESDLAs as this comparison is mainly qualitative in nature, given the small sample size. In total, 5 emission counterparts have been detected towards ESDLAs: J1135+0352 \citep{Noterdaeme2012,Kulkarni12}; Q0458--0203 \citep{Moller2004, Krogager+12}; J0025+1145, J1143+1420 (both from this work) and J1513+0352 \citep{Ranjan+2018}.

For the purpose of comparing the absorption properties of ESDLAs with GRB-DLAs, we use the GRB-DLA sample listed by \citet{Bolmer2019} as their sample size (22 systems) is similar to ours and their targets have also been observed using VLT/X-Shooter. For the analysis of kinematics ($\Delta v_{90}$), we combine measurements from \citet{Arabsalmani2015} and \citet{Arabsalmani2018} to build a larger sample of GRB-DLAs (13 systems excluding low resolution data). 
Impact parameter measurements for GRB-DLAs have been compiled from the following works: \citet{Castro2003}, \citet{Perley2013}, \citet{Thone2013}, \citet{Arabsalmani2015}, and \citet{Lyman2017}.

\section{Absorption-line analysis \label{absorption_analysis}}

We analysed the absorption lines using standard Voigt-profile fitting, taking the atomic and molecular wavelengths, oscillator strengths and damping constants from 
the source website of VPFIT\footnote{\url{https://www.ast.cam.ac.uk/~rfc/vpfit}} \citep{Carswell2014}, where most records are taken from \citet{Morton2003}. 
We present here the general methodology and measurements for our sample of ESDLAs, but most of the figures showing multicomponent Voigt profile fits to the data are presented in the Appendix to keep the main text concise.
 
\subsection{Neutral atomic hydrogen}

\begin{figure}
    \centering
    \addtolength{\tabcolsep}{-4pt}
    \begin{tabular}{ccc}
   \includegraphics[trim=40 40 55 40,clip,width=0.49\hsize]{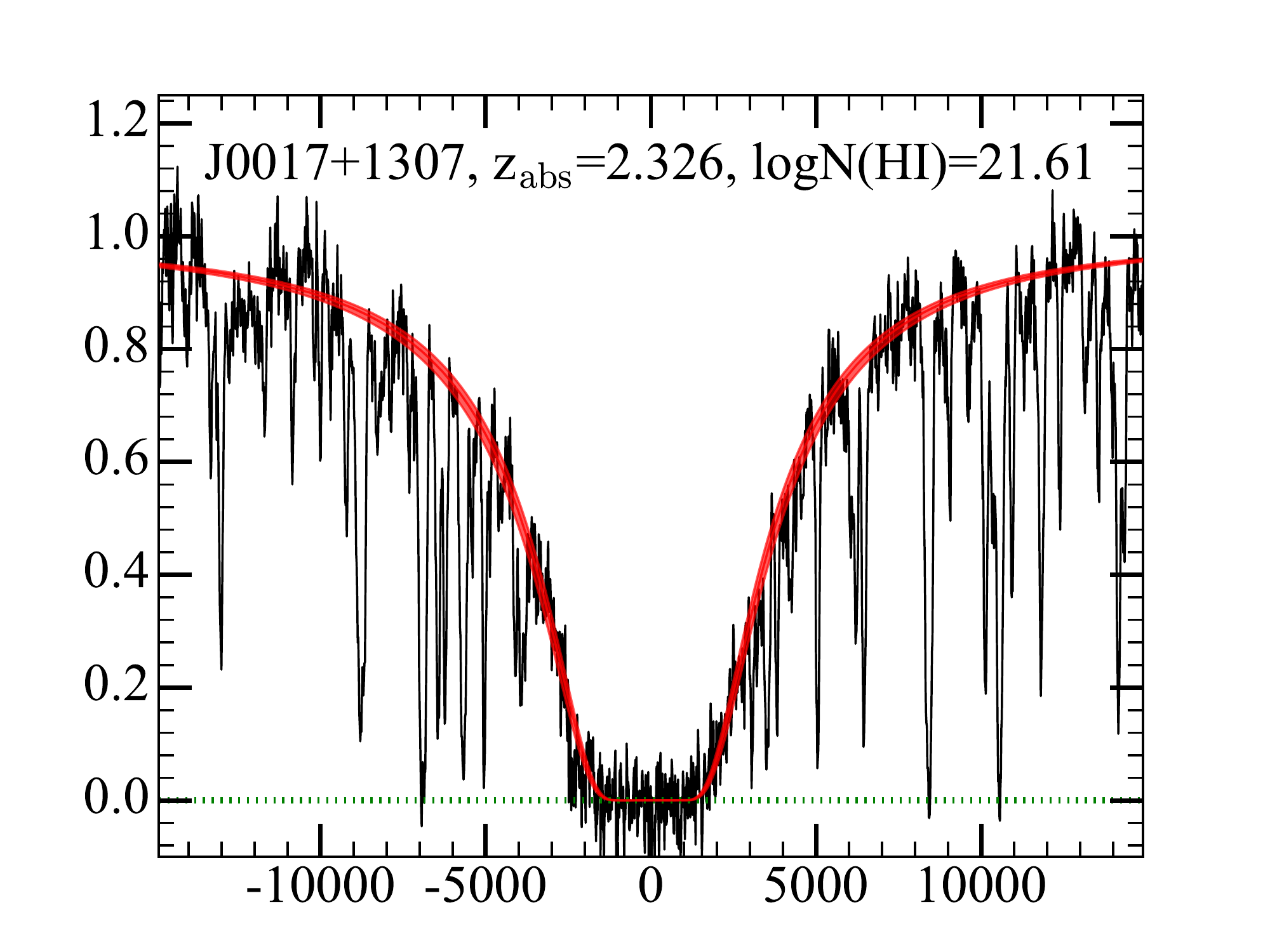} & \includegraphics[trim=40 40 55 40,clip,width=0.49\hsize]{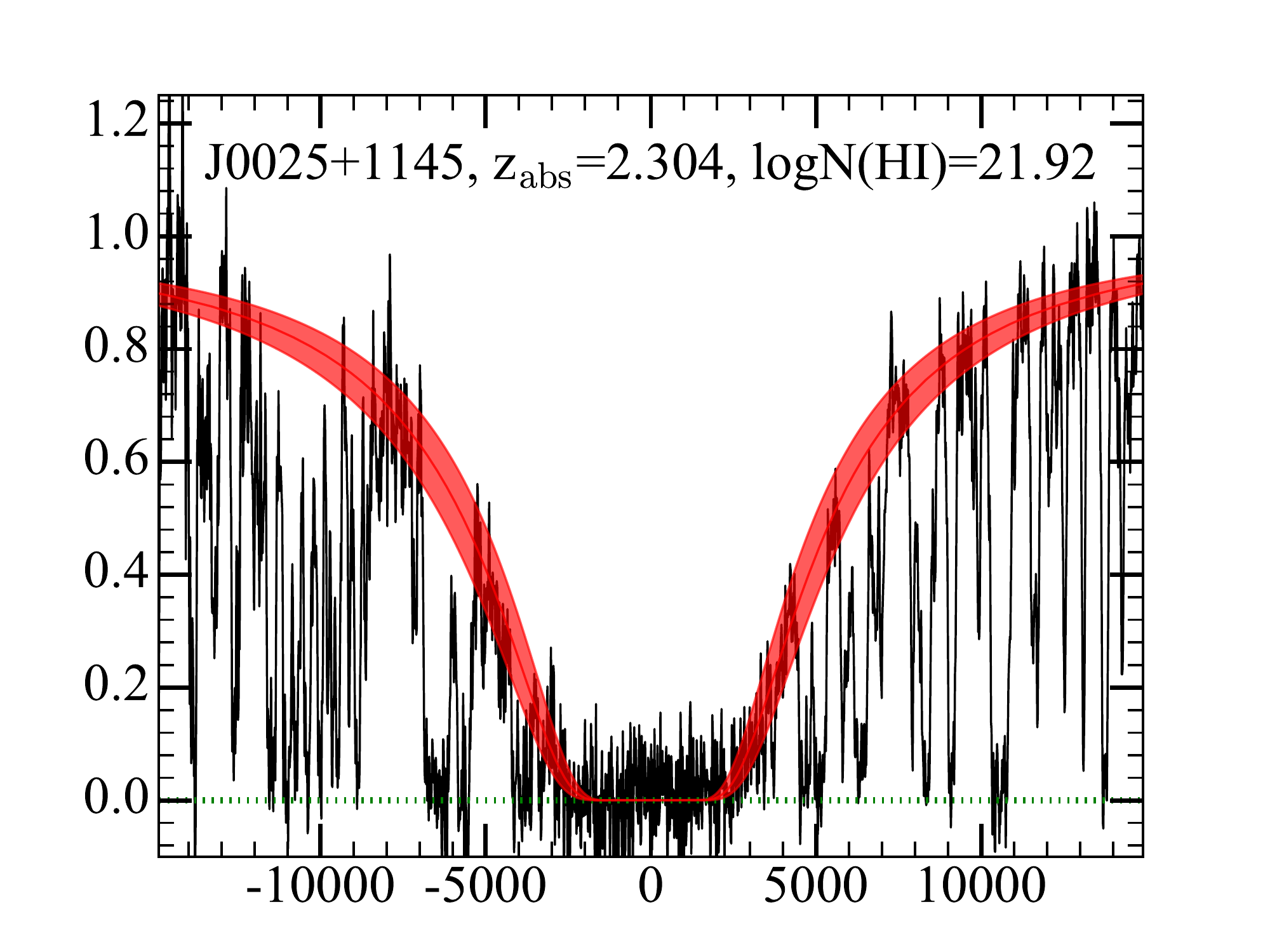} \\ 
   \includegraphics[trim=40 40 55 40,clip,width=0.49\hsize]{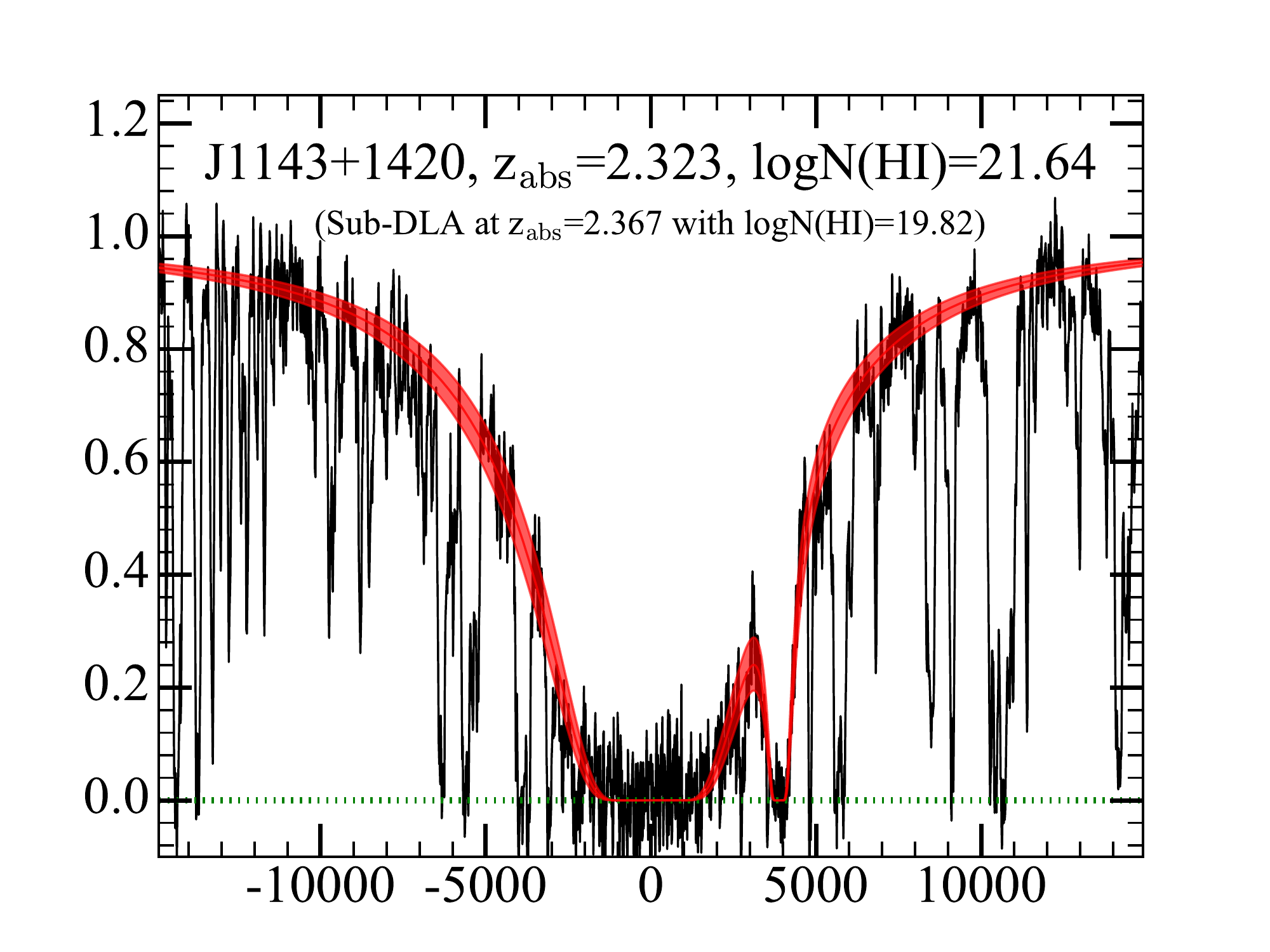} & \includegraphics[trim=40 40 55 40,clip,width=0.49\hsize]{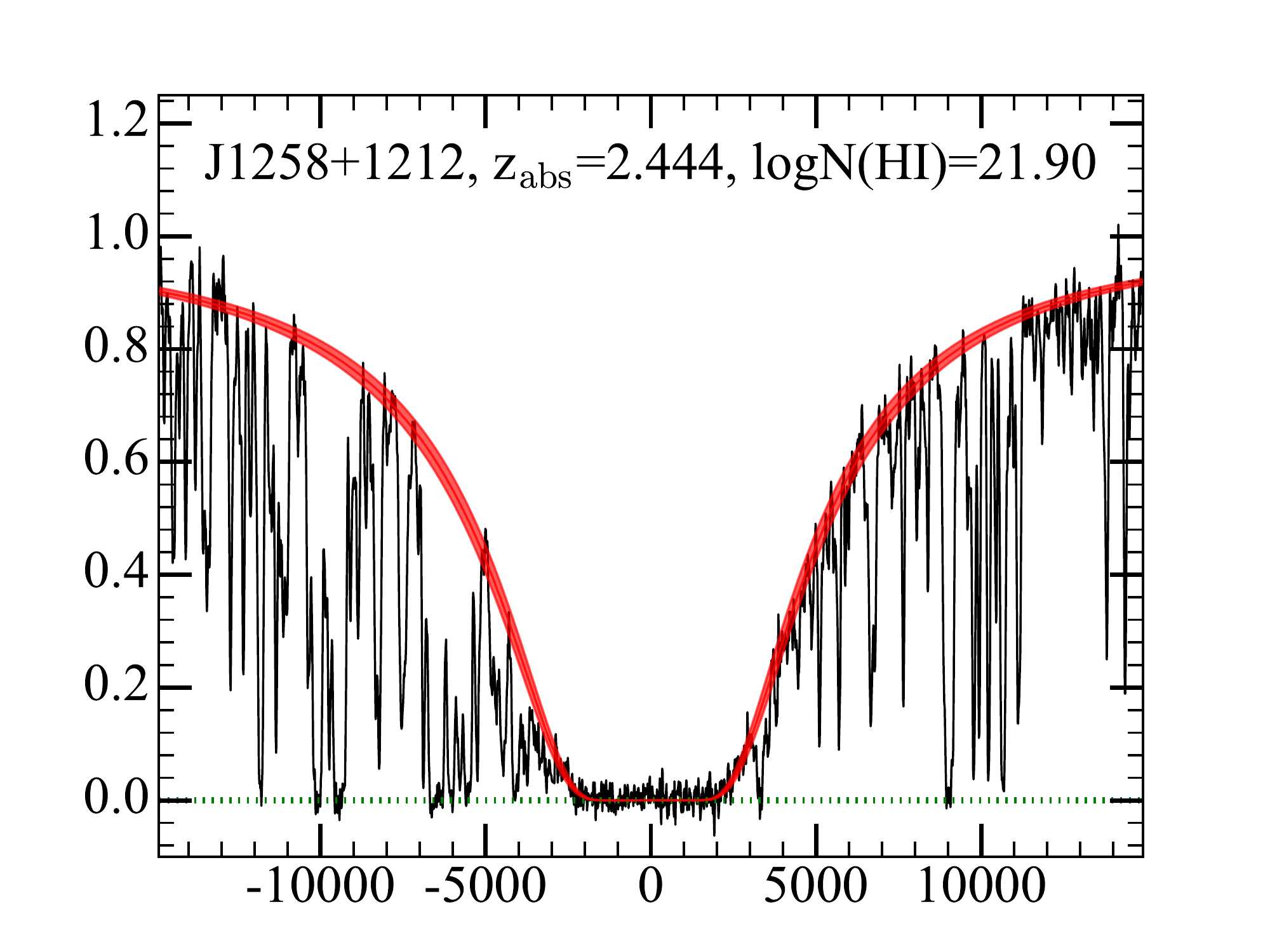} \\ 
   \includegraphics[trim=40 40 55 40,clip,width=0.49\hsize]{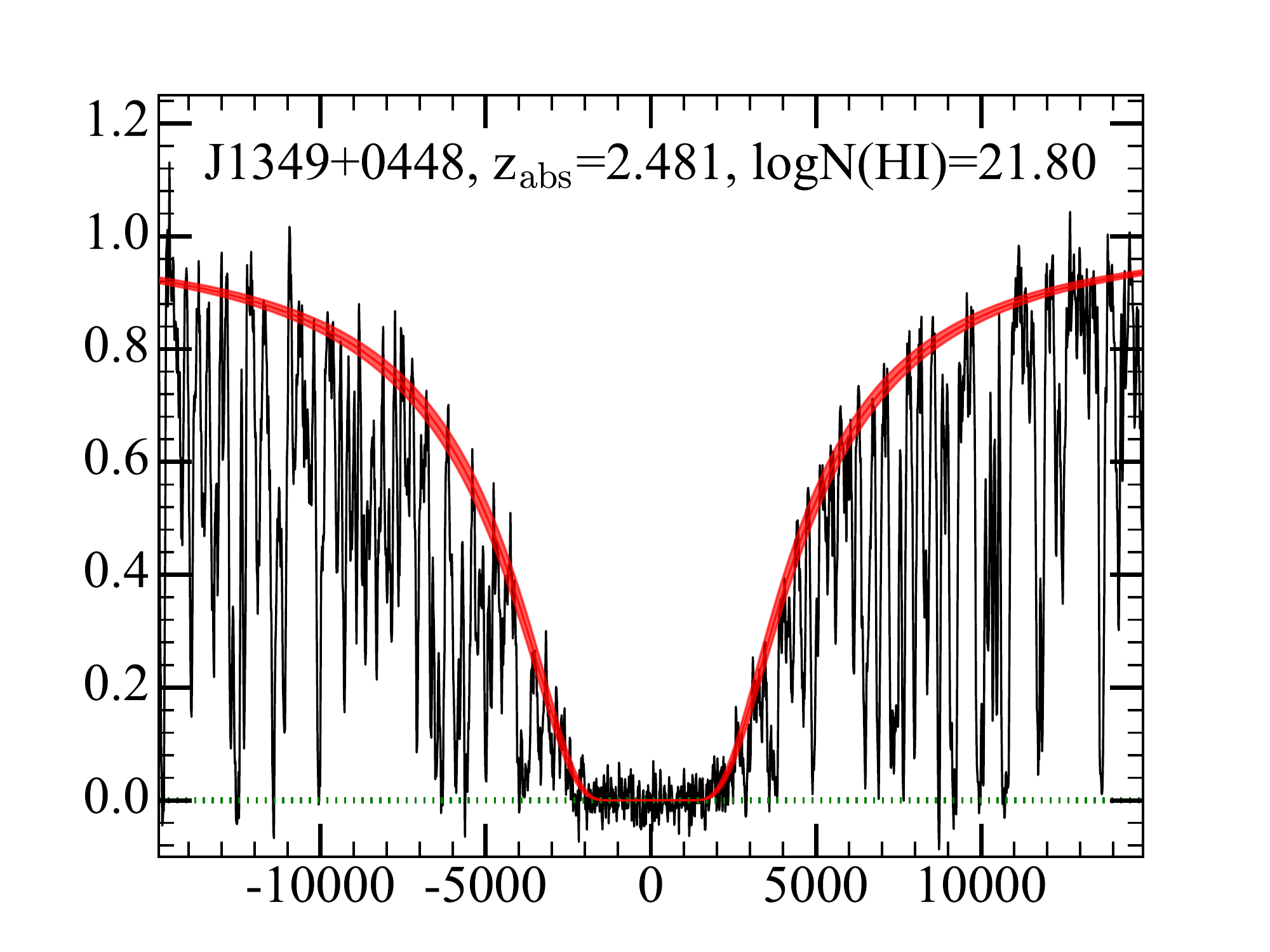} & \includegraphics[trim=40 40 55 40,clip,width=0.49\hsize]{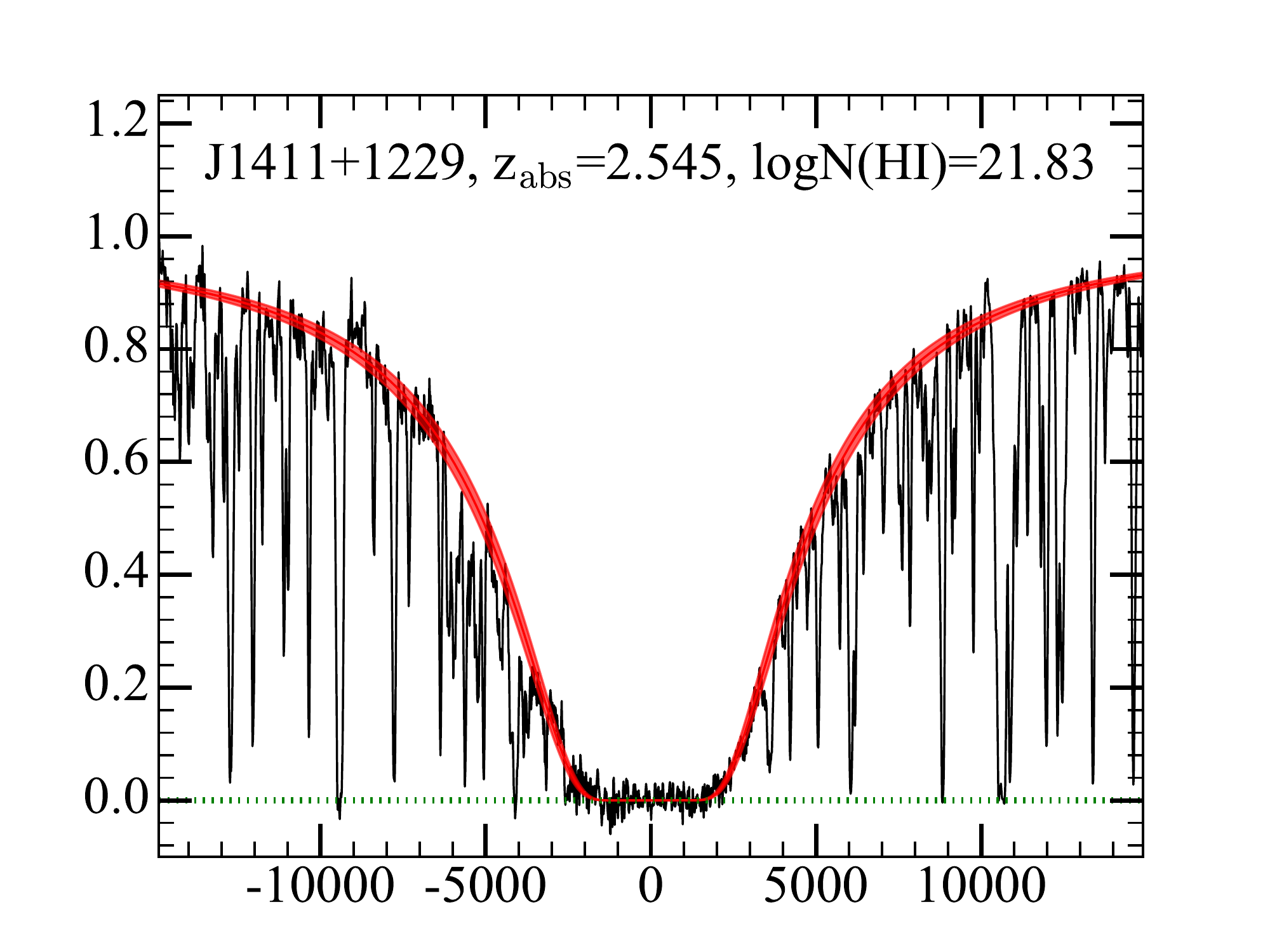} \\ 
   \includegraphics[trim=40 40 55 40,clip,width=0.49\hsize]{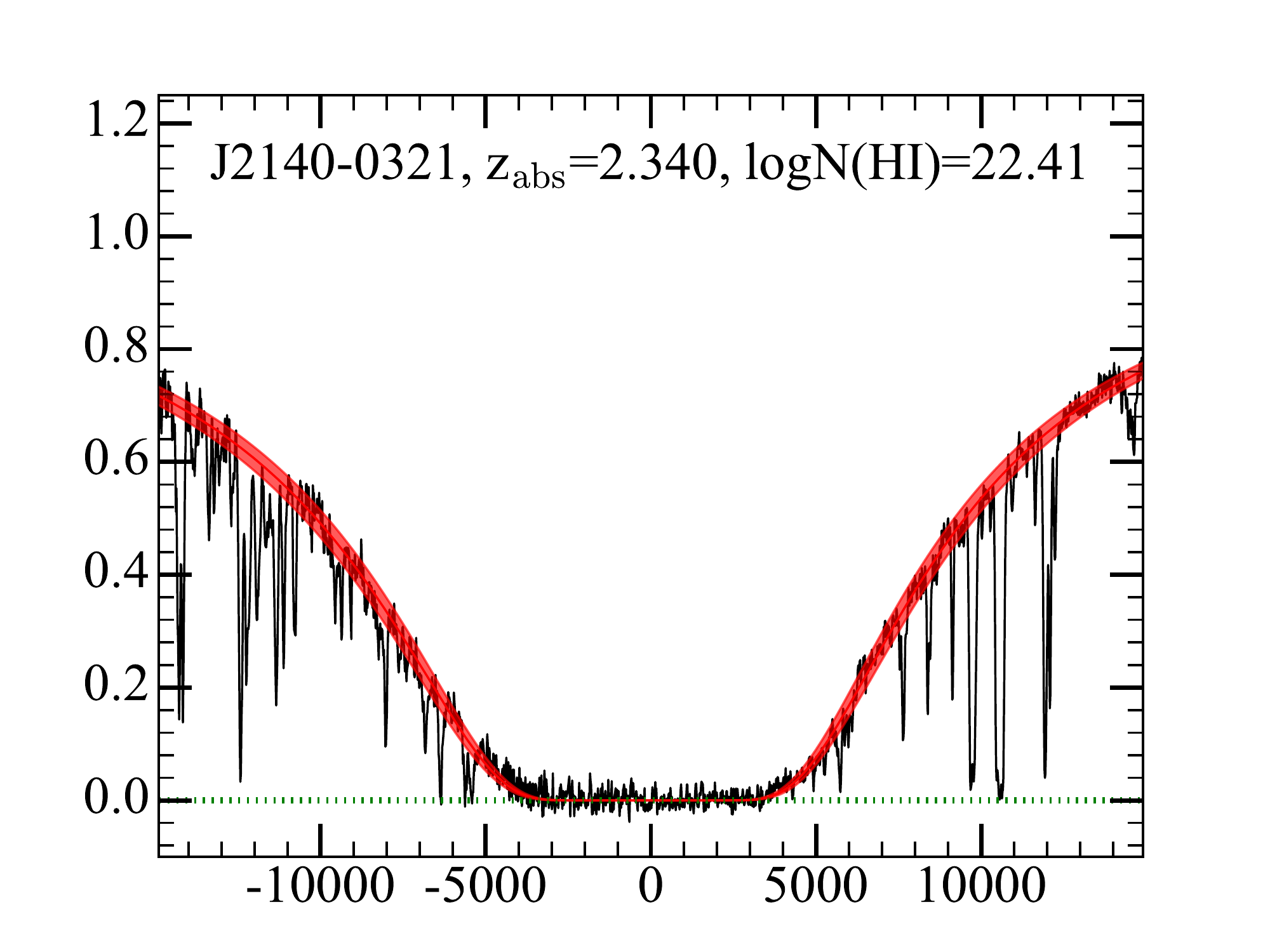} & \includegraphics[trim=40 40 55 40,clip,width=0.49\hsize]{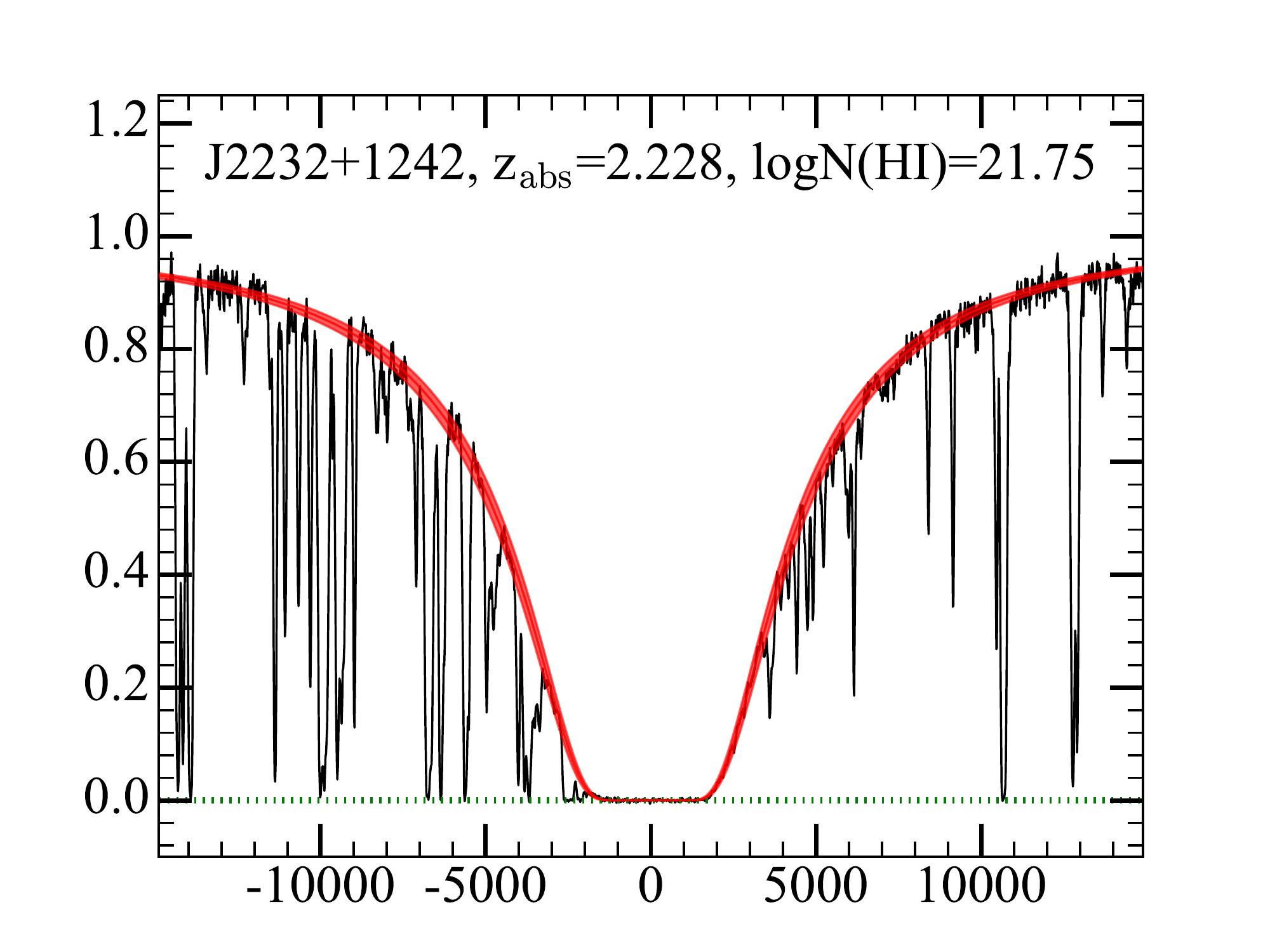} \\ 
   \includegraphics[trim=40 20 55 40,clip,width=0.49\hsize]{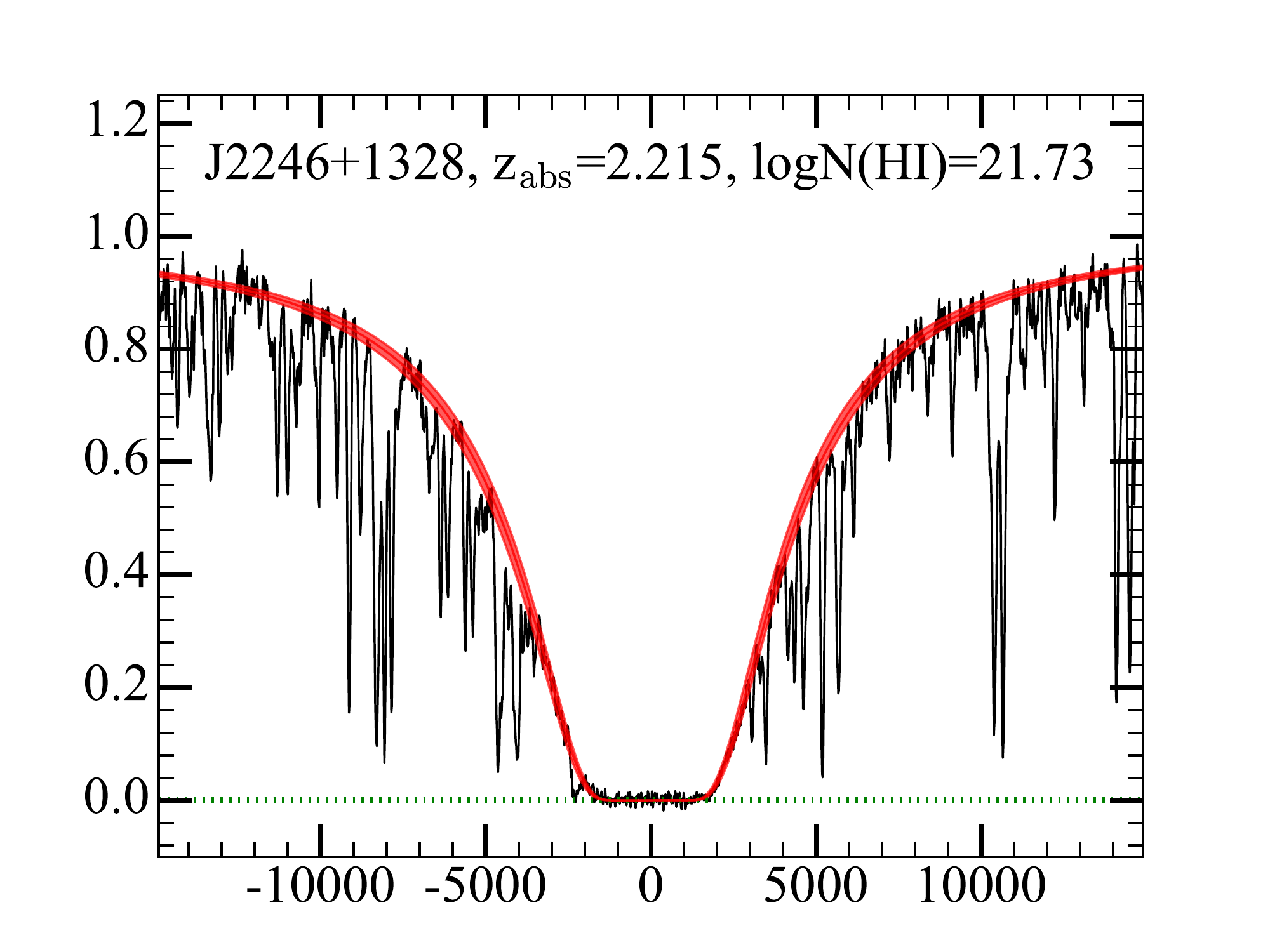} & \includegraphics[trim=40 20 55 40,clip,width=0.49\hsize]{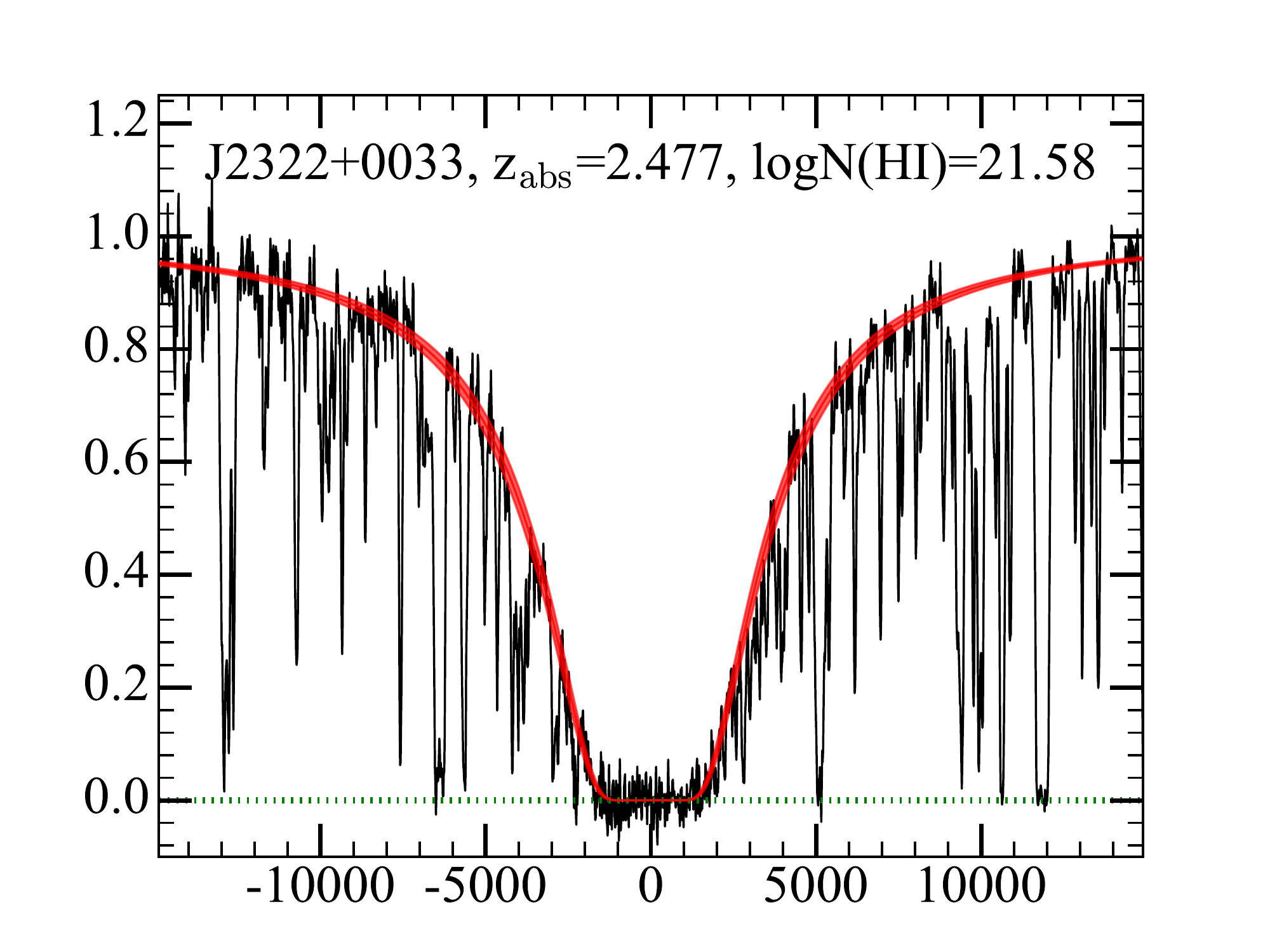} \\ 
   \multicolumn{2}{c}{Relative velocity (\kms)}
    \end{tabular}
 \addtolength{\tabcolsep}{4pt}
    \caption{Fits to the damped Lyman-$\alpha$ lines for the ten systems analysed in this work. The normalised X-shooter spectra are shown in black with the best-fit Voigt profile and over-plotted in red. Shaded regions depict statistical uncertainties. The relative velocity scale is defined with respect to the absorption redshift quoted in each panel.
    }
    \label{fig:HIfits}
\end{figure}

The column density of atomic hydrogen was measured by fitting the damped \lya\ line together with other lines from the \HI\ Lyman series when covered by the spectra.
Because the strong damping wings affect the apparent quasar continuum over a wide range of wavelengths, we first estimated an approximate continuum by eye 
taking reference from the quasar template by \citet{Selsing2016} and then used Chebyshev polynomials to model the remaining fluctuations, while fitting the \HI\ lines simultaneously. {The best fitted Voigt profiles to DLAs are shown in Fig.~\ref{fig:HIfits}}. 

\begin{figure}
    \centering
    \includegraphics[width=0.95\hsize]{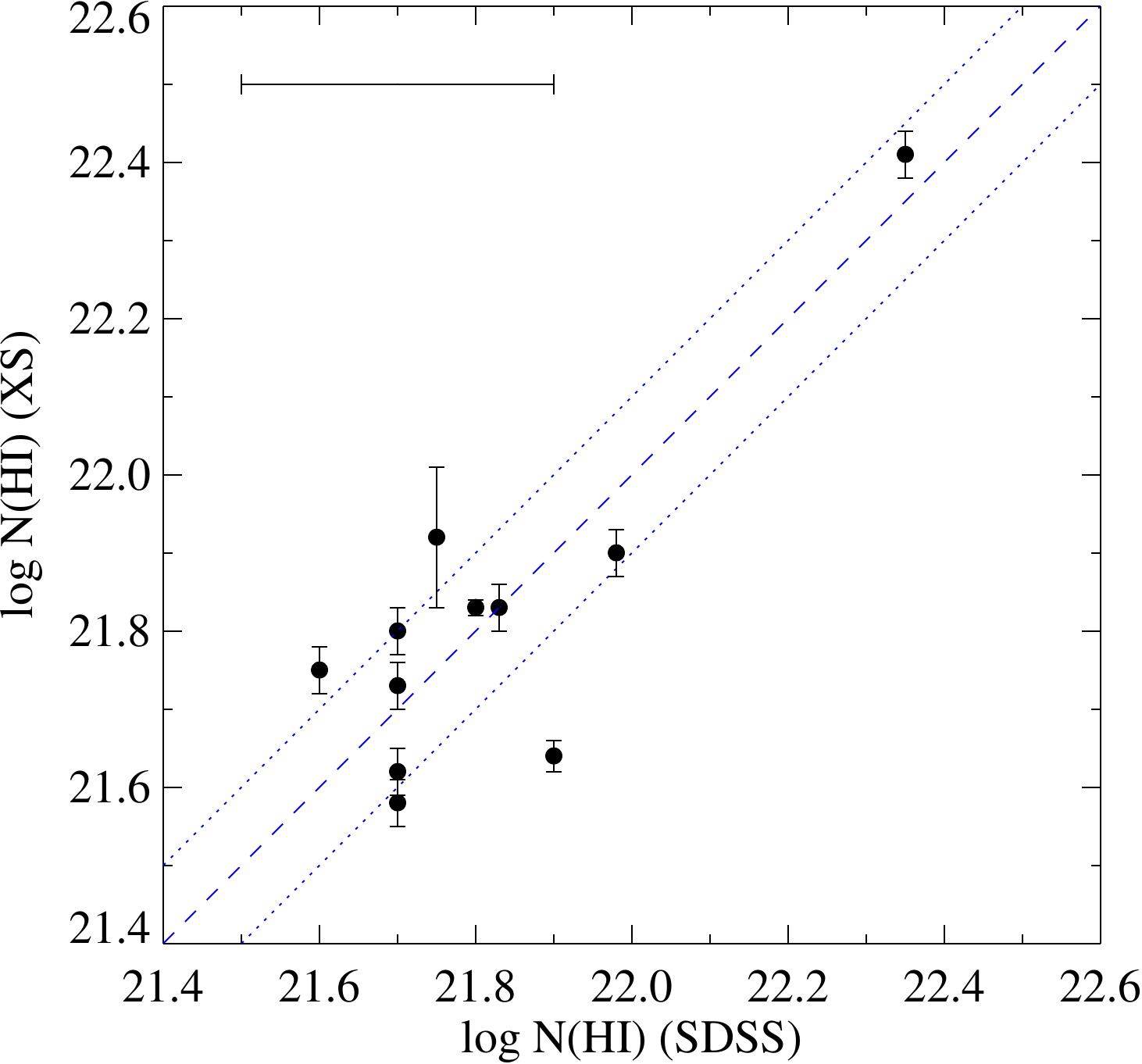}
    
    \caption{Comparison of \HI\ column densities measured in our sample of 11 systems observed with X-shooter to those originally obtained from the low-resolution, low S/N SDSS data. The horizontal bar in the upper left corner shows the typical uncertainty of SDSS measurements. The dashed line shows the one-to-one relation, with dotted lines showing $\pm$0.1~dex around this relation.}
    \label{fig:nhicomp}
\end{figure}

In Fig.~\ref{fig:nhicomp}, we compare the column densities obtained from VLT follow-up spectroscopy with those originally obtained from the low-resolution, low S/N data by 
\citet{noterdaeme2014}. There appears to be no systematic differences between both studies and most values agree to within 0.1~dex. This shows that the SDSS-based column densities are quite robust at the high-end of the \HI\ column density distribution. 
Nevertheless, our X-shooter analysis shows that a few of the systems (towards  J0017$+$1307, J1143+1420 and J2322+0033) have column densities slightly below the ESDLA threshold initially defined by \citet{noterdaeme2014}, $N(\HI)\ge 5 \times 10^{21}\cmsq$, but we will continue to call them ESDLAs in the following, for brevity.

\subsection{Molecular hydrogen}

We searched for H$_2$ absorption lines in the Lyman and Werner bands redshifted to the wavelength range covered by the UVB arm of X-shooter for all systems in our sample. In addition to the systems towards J\,1513+0352 and J\,2140$-$0321, for which the detection of H$_2$ has already been reported by \citet{Ranjan+2018} and \citet{Noterdaeme2015a}, respectively, we also clearly detect H$_2$ absorption lines towards J\,1143+1420 and J\,2232+1242. The spectrum of J\,0025+1145 presents strong absorption at the expected position of \HH, but only one band is available due to a Lyman-break caused by a system at higher redshift. 
Since J\,0025+1145 also presents strong reddening and \CI\ absorption lines, the detection of a high H$_2$ column density is expected and treated as a firm detection throughout the paper. Although there is an indication of damping wings and the data are best fitted with $\log N(\HH)\sim20$, this column density remains highly uncertain.

\begin{figure}
    \centering
    \includegraphics[trim=0 15 10 50,clip,width=0.8\hsize]{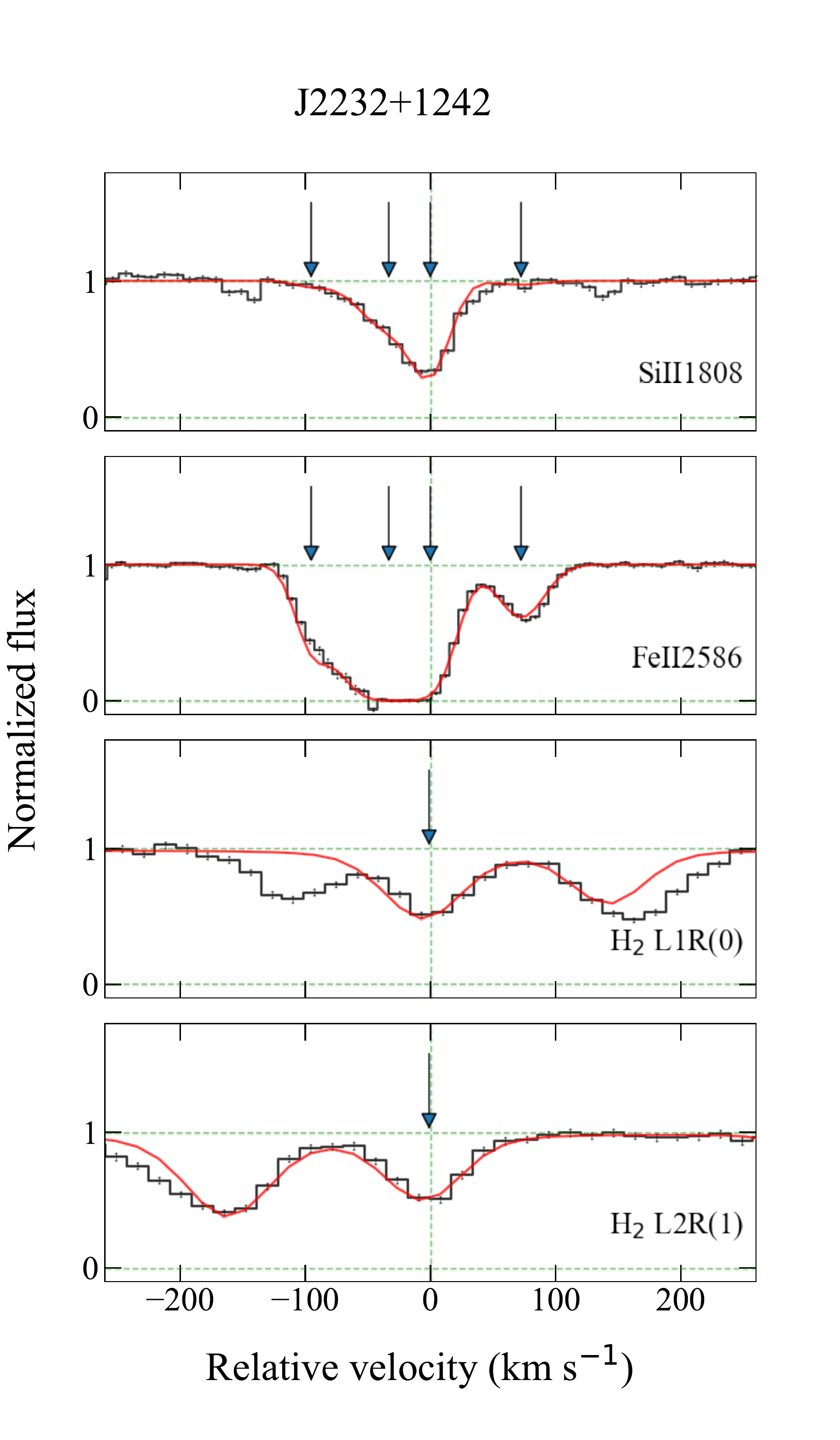}
    \caption{Velocity structure of metal lines tracing the overall neutral gas (top panels) compared with that seen in H$\rm _2$ (bottom panels), where the transition naming 
    is 'L' for Lyman, followed by the band number (vibrational level of the upper state), the branch (corresponding to the selection function) and the rotational level of the lower state 
    in parenthesis. 
    Downwards arrows indicate the positions of the different components in the best-fit model. Note that the absorption seen at 
    $+$150~\kms\  (resp. $-$150~\kms) in the L1R(0) (resp. L2R(1)) panel is not an additional component but absorption from L1R(1) (resp. L2R(0)).  
    The zero of the velocity scale is set relative to the redshift of the strongest metal component.}
    \label{fig:nh2comparison}
\end{figure}

For all these systems, we measured the \HH\ column densities in the lowest rotational levels (typically $J=0,1,2$) by simultaneously fitting the corresponding absorption lines together with the continuum modelled by Chebyshev polynomials. 
The \HH\ redshifts were allowed to vary independently from those of \HI\ given that the observed \HI\ redshift is just indicative of the middle point in the saturated \lya\ profile. 
We find that H$_2$ always corresponds to a component seen in the profile of metals. Fig.~\ref{fig:nh2comparison} shows one such example.  
While for some of the H$_2$-bearing systems, we do see signs of absorption lines from high rotational levels as well, we caution that the column densities could be more uncertain 
than the uncertainties from the best fit suggest because of insufficient spectral resolution. Notwithstanding, in the case of SDSS J\,2140$-$0321, the values derived using X-shooter agree surprisingly well (within 0.2~dex) with those derived using available UVES data by \citet{Noterdaeme2015a} for all rotational levels up to J=4 (see Table~\ref{NH2J2140}), although the statistical errors obtained here from the fitting of the X-shooter data appear to be underestimated. This is due to the need to assume a single $b$-parameter for all $J$-levels when fitting X-shooter data, while this was left free when fitting the UVES data.

\begin{table}
\caption{Column densities of molecular hydrogen at $\zabs=2.34$ towards J\,2140-0321 derived 
in this work (X-Shooter) and by \citet{Noterdaeme2015a} (UVES). \label{NH2J2140}}
    \centering
   \begin{tabular}{ccc}
\hline \hline
{\Large \strut}Rotational level ($J$) & \multicolumn{2}{c}{$\log N(\HH,J)$} \\
                       & X-Shooter & UVES \\
\hline
0        &      19.95 $\pm$ 0.01 & 19.84 $\pm$ 0.09\\
1        &      19.78 $\pm$ 0.01 & 19.81 $\pm$ 0.04\\
2        &      17.81  $\pm$ 0.08   & 17.96 $\pm$ 0.14\\
3        &      17.67  $\pm$ 0.09   & 17.76 $\pm$ 0.40\\
4        &      16.04  $\pm$ 0.11   & 15.88 $\pm$ 0.26\\
5        &      15.55  $\pm$ 0.09   & 15.17 $\pm$ 0.16\\
Total    &      20.18 $\pm$ 0.01  & 20.13 $\pm$ 0.07\\
\hline
\end{tabular}
\end{table}

For the six remaining systems, the presence of H$_2$ is difficult to ascertain due to blending with the Lyman-$\alpha$ forest at the spectral resolution of X-shooter and/or a small number of covered transitions. Instead, we measured conservative upper limits by identifying the highest column density model still consistent with the observed spectra. To do this, we created synthetic H$_2$ absorption models with a fixed excitation temperature, T$_{ex}=100$~K as typically seen in H$_2$-bearing DLAs, a corresponding low Doppler parameter $b=1~\kms$, and the increasing total $N(\HH)$ value {convolved with the appropriate instrumental resolution to mimic the X-shooter spectrum}. We also tested redshifts within a 150~\kms\ window centred on the strongest metal component. For each model, we compared the distribution of positive residuals only (negative residuals are mostly due to absorption from the \lya\ forest) with the expected distribution given the error spectrum and assuming intrinsically Gaussian-distributed uncertainties. The maximum allowed column density is obtained when the fraction of positive residuals above 1\,$\sigma$ and 2\,$\sigma$ both exceeds 32\% and 4.5\% level, respectively. Our conservative upper limit is then given by the model with the highest column density still consistent with the data. The upper-limits obtained range from $\log N(\HH)\sim 16$ to 18.3.

\subsection{Metal column densities}

Absorption lines from metals in various ionisation stages are seen associated to the ESDLAs. Since here we are mostly interested in 
the chemical enrichment in the neutral and the molecular gas phase, we focus on the low-ionisation species, \ion{Fe}{ii}, \ion{Si}{ii}, \ion{Cr}{ii} 
and \ion{Zn}{ii} that have first ionisation energy below 13.6~eV but second ionisation energy above this value.

These species are fitted simultaneously assuming a common velocity structure, i.e. the velocities and Doppler parameters for various elements are tied together.
We also included neutral magnesium, since \ZnII$\lambda2026$ is easily blended with \MgI$\lambda2026$. While being fitted simultaneously with 
the above-mentioned species, we allowed \MgI\ to have a different velocity structure since it does not correspond to the main ionisation stage of magnesium, which is mostly \MgII.

The number and location of the velocity components were first identified visually from the numerous \ion{Fe}{ii} and \ion{Si}{ii} lines to serve as an initial guess. 
We then fitted the metal species, adjusting the number of components if necessary. During the fitting process, we noted that the main \SiII\ lines were intrinsically saturated, even the relatively weak \SiII$\lambda1808$ because of the very high column densities involved. However, weaker lines such as \ion{Si}{ii}~$\lambda2335$ remain undetectable at the achieved S/N with X-shooter. The resulting \SiII\ column densities are therefore highly uncertain.

\subsection{Neutral Carbon \label{neutral_carbon}}

We searched for the presence of neutral carbon lines in our X-shooter spectra of ESDLAs since this species is known to be a good tracer of molecular hydrogen \citep[e.g.][]{srianand2005,Noterdaeme2018}. \CI\ lines are detected in four systems of our sample, towards J\,0025$+$1145, J\,1258$+$1212, J\,1513+0352 and J\,2140$-$0321. All of them have clear H$_2$ lines except J\,1258$+$1212, for which we could only obtain a conservative limit of $\log N(\HH)<18.3$. 
We note that while the spectra J1143+1420 and J2232+1242 feature clear H$_2$ lines,
we do not detect \CI\ lines.
However, the presence of detectable \CI\ lines at the achieved spectral resolution and 
S/N is not necessarily expected. Indeed, the presence of \CI\ is also dependent on the metallicity \citep{Noterdaeme2018,Zou2018,Heintz2019} and strong H$_2$ systems with low \CI\ column densities have been identified in the literature \citep[e.g.][]{Balashev+17}.  

Whenever detected, we simultaneously fitted the \CI\ lines in various fine-structure levels of the electronic ground state, denoted as \ion{C}{i} ($J=0$), \ion{C}{i}* ($J=1$) and \ion{C}{i}** ($J=2$). 
The velocities and Doppler parameters for each component are assumed to be the same over the different fine-structure levels, consistent with what is usually seen in \CI\ absorbers, even at high spectral resolution. However, the fitting procedure did not allow us to obtain accurate Doppler parameters because the \CI\ lines are intrinsically much narrower than the instrumental line spread function. The resulting column densities are thus to be considered with care and may well represent lower limits to the actual values.

\subsection{Dust extinction \label{extinction_measurement}}

The extinction of the quasar light by dust in the absorbing system was estimated by fitting a quasar composite spectrum to the data with various amounts of reddening, $A_V$, applied in the rest-frame of the absorber. For this purpose, we used the quasar template spectrum by \citet{Selsing2016}. Since the appropriate reddening law is unknown, we fitted various reddening laws for each target in our sample. The extinction laws for the average Small Magellanic Cloud (SMC), the average Large Magellanic Cloud (LMC), and the average LMC super-shell (LMC2) were described using the parameters obtained for different environments by \citet{Gordon2003}.
For each target, we shifted the template spectrum to the given quasar redshift and re-sampled the template onto the observed wavelength grid. The template was smoothed with a 20 pixel top-hat filter in order to avoid the fit being biased by noise peaks in the empirical template spectrum. We then applied the reddening in the rest-frame of the absorber with a variable amount of rest-frame $V$-band extinction, $A_V$. This is the only free parameter in the fit, since the absolute flux-scaling is obtained by normalizing the reddened template to match the observed spectrum. The best-fitted values of $A_V$ were obtained by a standard $\chi^2$ minimization and the resulting best-fitted values are provided in Table~\ref{absorption_table}.

The main uncertainty for the extinction measurement using this method comes from the fact that the intrinsic spectral shape of the quasar is not known and is in fact highly degenerate with the amount of dust extinction. We quantify this systematic uncertainty based on the observed variations of quasar spectral indices of 0.19~dex \citep{Krawczyk2015}. This translates into an uncertainty of 0.033~mag on $E(B-V)$ which should then be scaled by $R_V$ to get the systematic uncertainty on $A_V$ for the different extinction curves, corresponding to roughly 0.1~mag in $A_V$. These systematic uncertainties are given in Table~\ref{absorption_table}. For two targets, we were not able to fit the reddening as the spectra are bluer than the observed template, which would yield a nonphysical, negative value for $A_V$. Therefore for these targets, we just report upper limits on $A_V < 0.1$ in Table~\ref{absorption_table} with no extinction law.

\section{Emission-line analysis\label{emission_analysis}}

We searched for emission counterparts of the ESDLAs by looking for \lya\ emission in the UVB spectra and nebular emission lines in the NIR spectra: [\ion{O}{II}]\,$\lambda\lambda$ 3727, 3729 doublet, the [\ion{O}{III}]\,$\lambda\lambda$ 4959, 5007 doublet, \Ha\ and \Hb.
In the case of \lya\, the quasar light is naturally removed by the intervening DLA, so that we can directly look for emission in the DLA core. For emission lines in the NIR spectra, we first need to subtract the quasar continuum emission in order to search for the weak lines from the ESDLA counterparts. The quasar trace in the 2D data was subtracted using a procedure similar to that described by \citet{Moller2000}, where we used a local estimate of the spectral point spread function (SPSF) instead of relying on a parametrized model. This empirical SPSF was constructed using a median combination of the SPSF from the quasar dominated trace on either side of the expected location of the emission line. We discarded the SPSF from columns in the spectrum within 40 pixels around the line centroid in wavelength-space.
For each quasar, we looked for emission using not only the individual 2D spectra but also the combined spectrum. While the latter is composed of individual 2D spectra with different PAs (hence the spatial axis loses some meaning), the position angles differed by only 24 deg. The combined 2D spectra, therefore, allow us to obtain a higher signal-to-noise ratio.

We detected emission lines associated to 3 ESDLAs in our sample: J0025+1145, J1143+1420 and J1513+0352. The details regarding the emission counterpart of J1513+0352 detected in \lya\ are presented by \citet{Ranjan+2018} and the details regarding the 2 new detections are presented in the following subsections.
In these cases, we measured the line fluxes in the quasar-subtracted NIR spectra. For each detection, we extracted a 1D spectrum in an aperture centred at the position of the emission line with a width of twice the FWHM of the spatial extent of the emission. Based on the extracted 1D spectra we fitted the emission line with a Gaussian profile in order to obtain the integrated line flux and the observed line width. The uncertainty on the line flux is estimated by randomly sampling the Gaussian profile according to the best-fit values and their uncertainties. We double check that the estimated uncertainties are consistent with the noise in the residuals from the best-fit.

For systems where no emission lines were detected, we estimated upper limits to the line flux based on 100 random apertures in the quasar subtracted frame. In all cases, we used an aperture size in velocity-space of 300\,km\,s$^{-1}$ corresponding to twice the average line-width expected for such emission lines \citep[e.g.][]{Krogager2013}. In the spatial direction, we used an aperture size of twice the measured FWHM of the trace. The noise in the aperture is the added in quadrature and the resulting 3\,$\sigma$ limits are given in Table~\ref{tab:emission}.

\subsection{J0025+1145 \label{J0025+1145_Detection_of_NIR_emission_lines}}

\begin{figure*}
    \centering
    \includegraphics[width=1.0\hsize]{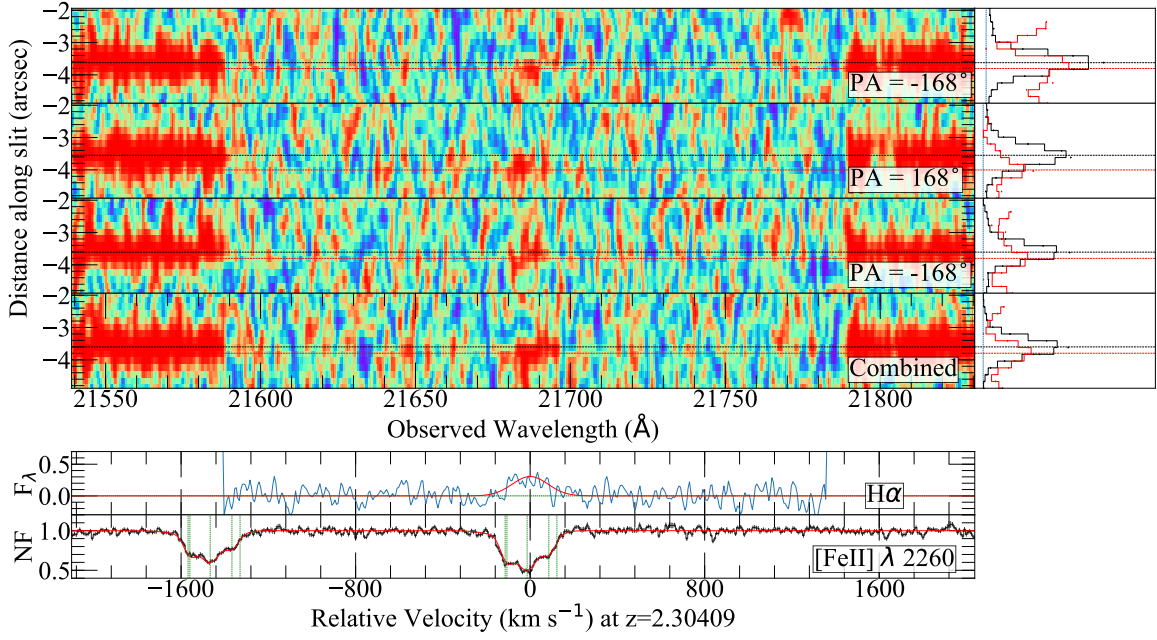}\\
    \caption{Detection of \Ha\ emission associated to the $\zabs=2.304$ ESDLA towards J0025+1145. The top panels represent the 3 individual 2D spectra and the combination of all of them. The quasar trace has been removed over the central region ($\Delta v\sim 3000~\kms$). The \Ha\ emission corresponds to the remaining blob in the centre of the images, and is quite evident in the combined 2D spectrum.  The side figures show the spatial extent of the trace (black) and of the \Ha\ emission (red) obtained by collapsing the 2D data along the wavelength axis. The location of their centroid (horizontal dotted lines) provides a measure of the impact parameter along the slit direction. 
    The fifth panel shows the quasar-subtracted 1D data (in blue) summed over the spatial axis together with a Gaussian fit (in red) to the emission line. The normalised absorption profile of \FeII$\lambda$2260 is shown in the bottom panel for comparison. Note that the absorption seen at $v\sim-1600~\kms$ is due to \FeII$\lambda$2249.}
    \label{J0025+1145_NIR_1}
\end{figure*}

\begin{figure*}
    \centering
    \includegraphics[width=1.0\hsize]{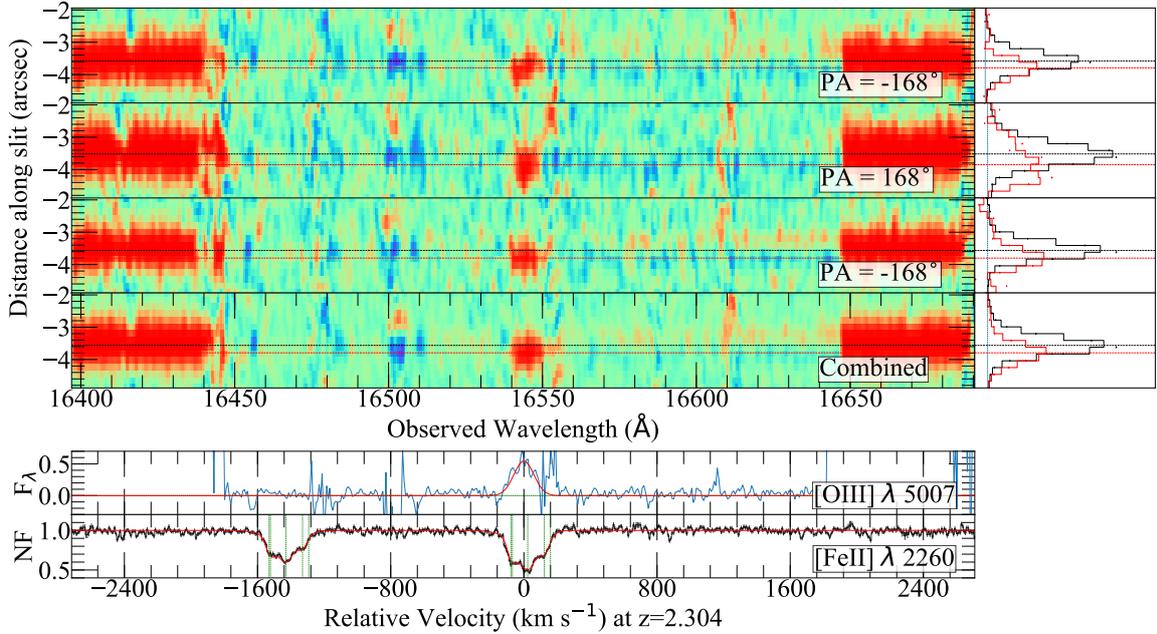}
    \caption{Same as Fig.~\ref{J0025+1145_NIR_1} for the [\OIII]$\lambda$5007 emission line at the redshift of the $\zabs=2.304$ DLA towards J0025+1145. }
    \label{J0025+1145_NIR_2}
\end{figure*}

We detected both \Ha\ and [\OIII]$\lambda$5007 in all individual 2D spectra as well as in the combined 2D spectrum, see Figs.~\ref{J0025+1145_NIR_1} and \ref{J0025+1145_NIR_2}.

The \Ha\ line flux in the combined frame was measured to be $(4.63\pm1.43)\times 10^{-17}$\,erg\,s$^{-1}$\,cm$^{-2}$ at an impact parameter of $0.23 \pm 0.05$~arcsec.

The line flux of [\OIII]\,$\lambda$ 5007 was measured to be $(4.16 \pm 0.48) \times 10^{-17}$~erg~s$^{-1}$~cm$^{-2}$ in the combined 2D spectrum. The other emission line of the doublet, namely [\OIII]\,$\lambda$ 4959, falls exactly on top of a sky emission line; hence, we were not able to detect this line. We measure the impact parameter along the slit to be $\rho_{\parallel}=1.8\pm0.2$~kpc for the PA = $-168$\,deg observations and $2.8\pm0.8$~kpc for PA = 168\,deg. Since the two position angles do not differ by much, the true impact parameter and relative position angle are not constrained from triangulation. In turn, we used the average impact parameter as measured from the combined 2D data, $\rho = 1.9 \pm 0.1$~kpc. This value also corresponds to the most likely impact parameter when assuming a uniformly distributed random angle between the slit and quasar-galaxy direction \citep[see][]{Ranjan+2018}.

\subsection{J1143+1420 \label{J1143+1420_Detection_of_emission_lines}}

\begin{figure*}
    \centering
    \includegraphics[width=1.0\hsize]{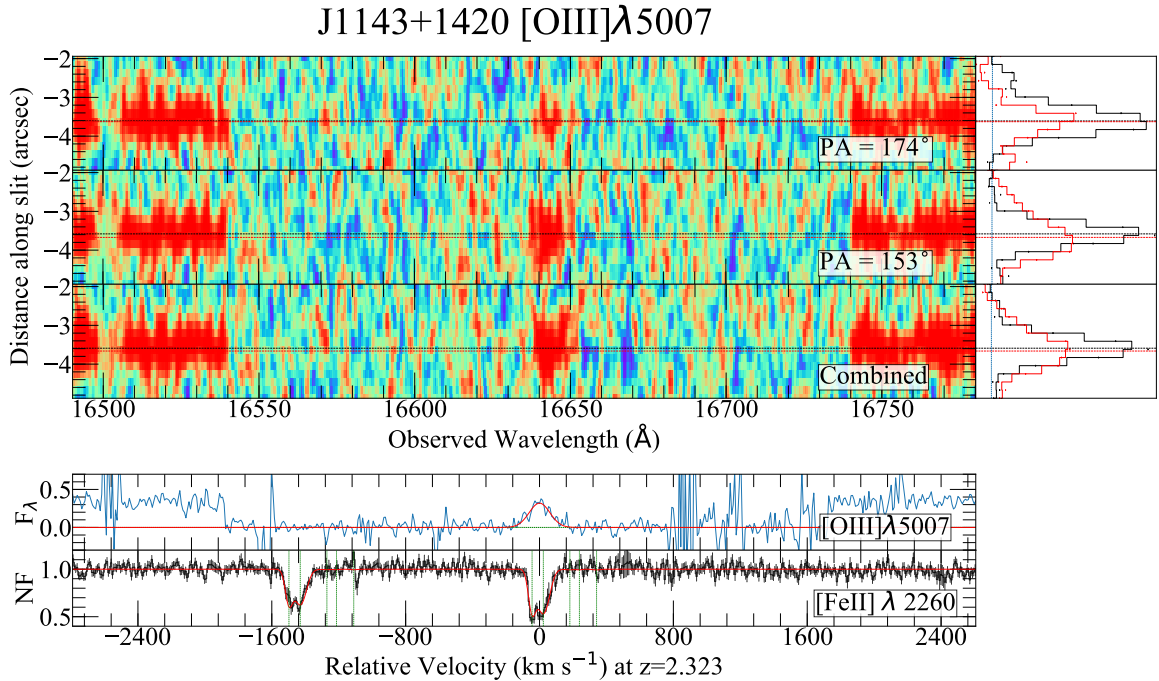}
        \caption{Same as Fig.~\ref{J0025+1145_NIR_1} for the [\OIII]$\lambda$5007 emission line at the redshift of the $\zabs=2.323$ DLA towards J1143+1420.}
    \label{J1143+1420_NIR_1}
\end{figure*}

For J1143+1420, we detected the [\ion{O}{III}]\,$\lambda$ 5007 emission line in the 2D NIR spectra as shown in Fig.~\ref{J1143+1420_NIR_1}. The integrated flux in the combined spectrum is $(2.74 \pm 0.35) \times 10^{-17}$~erg~s$^{-1}$~cm$^{-2}$ with the emission centred at $z = 2.3232$. Since the slit position angles of the two observations differed by only 21\,deg, it is not possible to triangulate the exact sky position relative to the quasar. However, we find that in both spectra the emission is observed at a very small impact parameter of $\rho$ = 0.7 $\pm$ 0.3 kpc.

\begin{table*}
        \caption{Emission properties of ESDLAs from our X-shooter observing programme. All measurements are from this work except J1513$+$0352 \citep[from][]{Ranjan+2018}.}
    \centering
    \begin{tabular}{c c c c c c c}
    \hline\hline
    {\Large \strut}Quasar     & $z_{\rm ESDLA}$ &  \multicolumn{3}{c}{Line luminosity ($10^{40}$erg\,s$^{-1}$)} & $\rho$ \\
               &                 & \lya\                   & [\OIII]$\lambda5007$ & \Ha\   & (kpc) \\
    \hline
    J0017+1307 &  2.326          & $<$27.2                                 & $<$43.9                  & $<$127.7          &  \ldots    \\
    J0025+1145 &  2.304          & $<$16.3                                 & 178 $\pm$ 21                 & 198 $\pm$ 61          &  1.9 $\pm$ 0.1     \\
    J1143+1420 &  2.323          & $<$30.2                                 & 120 $\pm$ 15                 &  $<$351         &  0.6 $\pm$ 0.3     \\
    J1258+1212 &  2.444          & $<$11.9                                 & $<$249                  & $<$334          &  \ldots     \\
    J1349+0448 &  2.482          & $<$9.2                                 & $<$92                  & $<$402          &  \ldots    \\
    J1411+1229 &  2.545          & $<$12                                 & $<$63                  & $<$274          &  \ldots    \\
    J1513+0352 &  2.464          & 9.6 $\pm$ 3.5                                & $<$69                 & $<$131          &  1.4 $\pm$ 0.9    \\
  J2140$-$0321 &  2.339          & $<$12                                 & $<$130                  & $<$76          &  \ldots     \\
    J2232+1242 &  2.230          & $<$8.7                                 & $<$137                  & $<$72          &  \ldots     \\
    J2246+1328 &  2.215          & $<$7.4                                 & $<$36                  & $<$88          &  \ldots     \\
    J2322+0033 &  2.477          & $<$14                                 & $<$46                  & $<$317          &  \ldots     \\
    \hline
    \end{tabular}
    \label{tab:emission}
    \tablefoot{The upper limits correspond to the mean detection limit in the spatial area covered by the slit (see text). 
    }
\end{table*}

\section{Discussion \label{discussion} } 

In this section, we discuss the absorption and emission properties
of ESDLAs and compare them to those of GRB-DLAs. The samples used for the comparison are described in detail in Section~\ref{sample_selection}.

\subsection{Gas properties}

In the following, the gas-phase abundance of a species (X) is expressed relative to the Solar value as:

\begin{equation}
\rm {[X/H]} = \log\left(\frac{N({X})}{N({H_{tot}})}\right) - \log\left(\frac{X}{H}\right)_{\odot},
\end{equation}

\noindent where $\rm N({H_{tot}})=N(\HI)$ ($\rm +2N(H_2)$ if available) is the total hydrogen column density.
Note that the abundances are calculated as an average over the entire absorption profile because we cannot do a component-wise estimation of neutral hydrogen from saturated \lya\ profiles (nor from other Lyman-series absorption).
We adopted the same Solar abundances as in \citet{DeCia2016}, i.e., from \citet{Asplund+09}, following the recommendations of \citet{lodders2009} about the choice of photospheric, meteoritic, or average values. 
We use zinc as a reference element for metallicity for all the systems in our sample since this species is known to be volatile and little depleted into dust grains. In turn, iron is known to deplete heavily and we, therefore, use the iron-to-zinc ratio, [Fe/Zn$]=\log N(\ZnII)/N(\FeII) - \log ({\rm Zn/Fe})_{\odot}$, as a measure of depletion, as usually done in the literature.

\begin{table*}[]
\caption{Main absorption properties of extremely strong DLAs with VLT spectroscopic follow-up.} 
\label{absorption_table}
\begin{tabular}{c c c c c c c c c c}
\hline\hline
Quasar         & $z\rm _{QSO}$ & \zabs  & $\log N(\HI)$    & Inst.\tablefootmark{a}  & $\log N($H$_2)$         & $\log Z$\tablefootmark{b}     & [Fe/Zn]\tablefootmark{b}   & A$_V$\tablefootmark{c} & Refs. \\
\hline
\multicolumn{10}{l}{\sl Homogeneous ESDLA sample:}\\
J0017$+$1307  & 2.594  & 2.326  & 21.62$\pm$0.03  & XS      &  \textless{}18.3                  & -1.50$\pm$0.09     & -0.20$\pm$0.09  &  0.36 (LMC2) & 1   \\
J0025$+$1145  & 2.961  & 2.304  & 21.92$\pm$0.09  & XS      &  $\sim 20$                        & -0.53$\pm$0.11     & -0.90$\pm$0.12  &    0.51 (SMC)   & 1   \\
J0154$+$1935  & 2.513  & 2.251  & 21.75$\pm$0.15  & UVES    &  $\sim$18.0                       & -0.72$\pm$0.15     & -0.42$\pm$0.03  & \ldots  & 2   \\
J0816$+$1446  & 3.846  & 3.287  & 22.00$\pm$0.10  & UVES    &  18.6$\pm$0.4                     & -1.10$\pm$0.10     & -0.48$\pm$0.02  & \ldots  & 3   \\
J1143$+$1420  & 2.583  & 2.323  & 21.64$\pm$0.06  & XS      &  18.3$\pm$0.1                     & -0.80$\pm$0.06     & -0.54$\pm$0.03  & 0.23 (SMC)  & 1   \\
J1258$+$1212  & 3.055  & 2.444  & 21.90$\pm$0.03  & XS      &  \textless{}18.3                  & -1.43$\pm$0.04     & -0.50$\pm$0.04  &  0.04 (SMC) & 1   \\
J1349$+$0448  & 3.353  & 2.482  & 21.80$\pm$0.01  & XS      &  \textless{}18.1                  & -1.35$\pm$0.06     & -0.34$\pm$0.08  &  0.07(SMC) & 1   \\
J1411$+$1229  & 2.713  & 2.545  & 21.83$\pm$0.03  & XS      &  \textless{}15.9                  & -1.59$\pm$0.08     & -0.34$\pm$0.11  &  0.09 (SMC) & 1   \\
J1456$+$1609  & 3.683  & 3.352  & 21.70$\pm$0.10  & UVES    &  17.10$\pm$0.09                   & -1.39$\pm$0.11     & -0.40$\pm$0.07  &  \ldots & 2   \\
J1513$+$0352  & 2.680  & 2.464  & 21.83$\pm$0.01  & XS      &  21.31$\pm$0.01                   & -0.84$\pm$0.23     & -1.22$\pm$0.23  & 0.43 (LMC)  & 4   \\
J2140$-$0321  & 2.479  & 2.339  & 22.41$\pm$0.03  & XS/UVES &  20.13$\pm$0.07\tablefootmark{d}  & -1.52$\pm$0.08     & -0.70$\pm$0.10  &  0.12 (SMC) & 1,2 \\
J2232$+$1242  & 2.299  & 2.230  & 21.75$\pm$0.03  & XS      &  18.56$\pm$0.02                   & -1.48$\pm$0.05     & -0.21$\pm$0.04  &  0.01 (SMC) & 1   \\
J2246$+$1328  & 2.514  & 2.215  & 21.73$\pm$0.03  & XS      &  \textless{}16.3                  & -1.82$\pm$0.63     & -0.17$\pm$0.64  &  $<0.1$\tablefootmark{f}    & 1   \\
J2322$+$0033  & 2.693  & 2.477  & 21.58$\pm$0.03  & XS      &  \textless{}16.0                  & -1.71$\pm$0.13     & -0.10$\pm$0.14  &  $<0.1$\tablefootmark{f} & 1   \\
\hline
\multicolumn{10}{l}{\sl Additional ESDLAs:}\\
HE0027$-$1836 & 2.560  & 2.402  & 21.75$\pm$0.10  & UVES    &  17.43$\pm$0.02                   & -1.59$\pm$0.10     & -0.67$\pm$0.03  & \ldots  & 5,6 \\
Q0458$-$0203  & 2.290  & 2.040  & 21.70$\pm$0.10  & UVES    & \textless{}14.9                   & -1.26$\pm$0.10     & -0.67$\pm$0.04  &  \ldots & 7   \\
J0843$+$0221   & 2.920  & 2.786  & 21.82$\pm$0.11  & UVES    &  21.21$\pm$0.02                   & -1.59$\pm$0.10     & -0.91$\pm$0.13  & 0.09 (SMC)  & 8   \\
J1135$-$0010  & 2.890  & 2.207  & 22.10$\pm$0.05  & XS/UVES &  NA\tablefootmark{e}              & -1.13$\pm$0.05     & -0.61$\pm$0.03  & 0.11 (SMC)  & 9   \\  
Q1157$+$0128  & 1.990  & 1.944  & 21.80$\pm$0.10  & UVES    &  \textless{}14.7                  & -1.40$\pm$0.10     & -0.41$\pm$0.01  &  \ldots & 7   \\
\hline
\end{tabular}
\tablefoot{
\tablefoottext{a}{Instrument used for the observations: X-shooter (XS) or UVES.} 
\tablefoottext{b}{Metallicities and depletion factors from the literature have been corrected to use our adopted Solar abundances.}
\tablefoottext{c}{Extinction measurements are available for systems observed with X-shooter only, except for J0843+0221 for which an estimate has been obtained
directly from the SDSS spectrum \citep{Balashev+17}. The best-fit extinction law is indicated in parenthesis. Uncertainties are dominated by systematics due to intrinsic quasar-shape variations and hence are all of the order of 0.1~mag.}
\tablefoottext{d}{UVES-based measurement from \citet{Noterdaeme2015a}.}
\tablefoottext{e}{No H$_2$ line is covered for this system due to the Lyman break of a system at higher redshift; see \citet{Noterdaeme2012} and \citet{Kulkarni12}.}
\tablefoottext{f}{These quasars are bluer than the composite spectrum (see Sect.~\ref{extinction_measurement}).}
}
\tablebib{(1)~This Work; (2) \citet{Noterdaeme2015a}; (3) \citet{Guimaraes2012}; (4) \citet{Ranjan+2018}; (5) \citet{Noterdaeme2007b}; (6) \citet{Rahmani2013};  (7) \citet{noterdaeme2008}; (8) \citet{Balashev+17}; (9) \citet{Noterdaeme2012}.}
\end{table*}

\subsubsection*{Chemical enrichment \label{neutral_gas_to_metallicity}}

The average metallicity of ESDLAs in our homogeneous sample is $\mean{\rm [Zn/H]} = -1.30 \pm 0.05$. A similar value is obtained for the `total sample' of ESDLAs (see Sect.~\ref{sample_selection}). This confirms the typical ESDLA value estimated by \citet{noterdaeme2014} ([Zn/H] $\sim -1.3$), even though the latter was obtained from analyzing a composite spectrum of SDSS data with low resolution and low S/N.

\begin{figure*}[!t]
 \centering
 \addtolength{\tabcolsep}{-3pt}
 \begin{tabular}{ccc}
    \includegraphics[trim=10 10 10 0,clip,width=0.32\hsize]{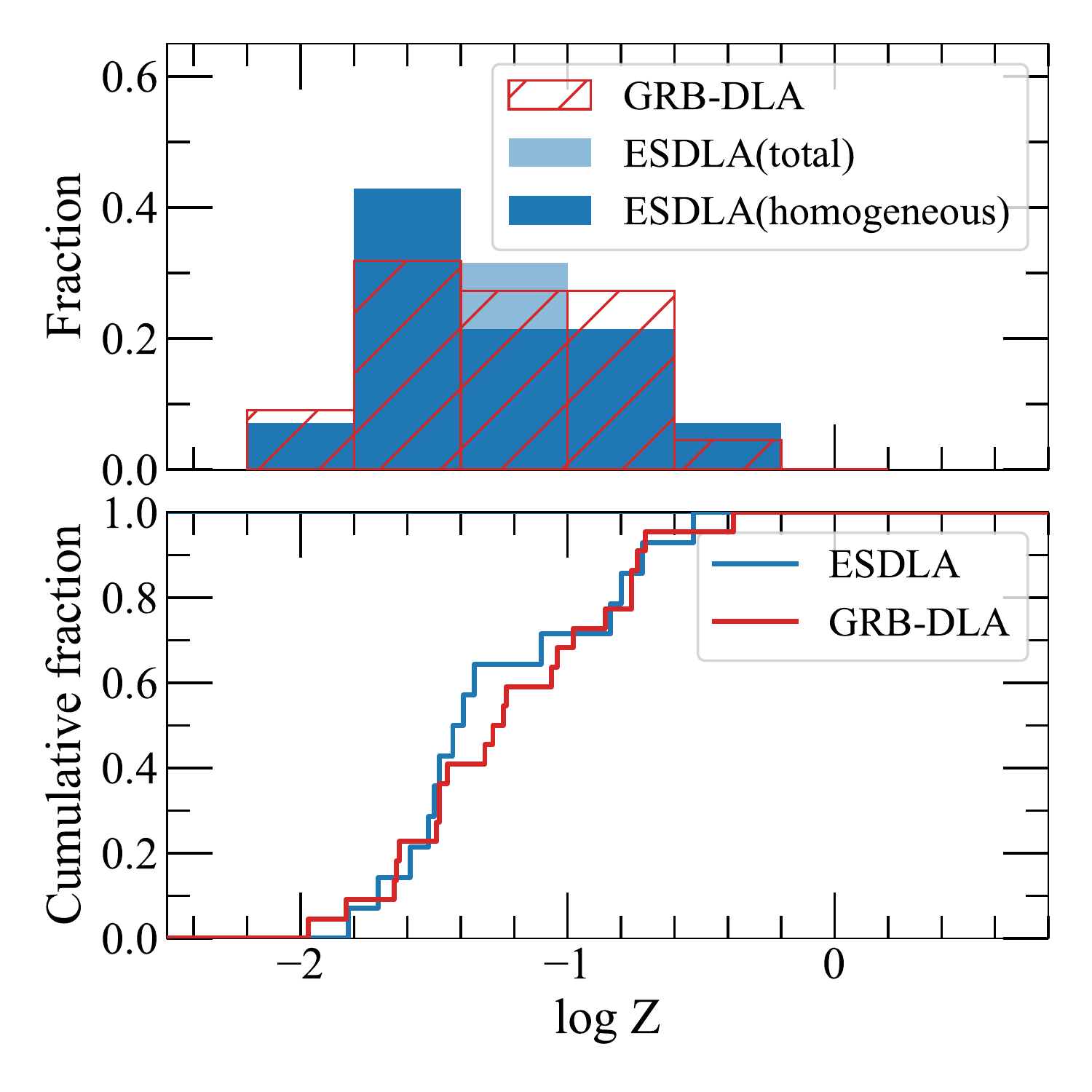} & 
    \includegraphics[trim=10 10 10 0,clip,width=0.32\hsize]{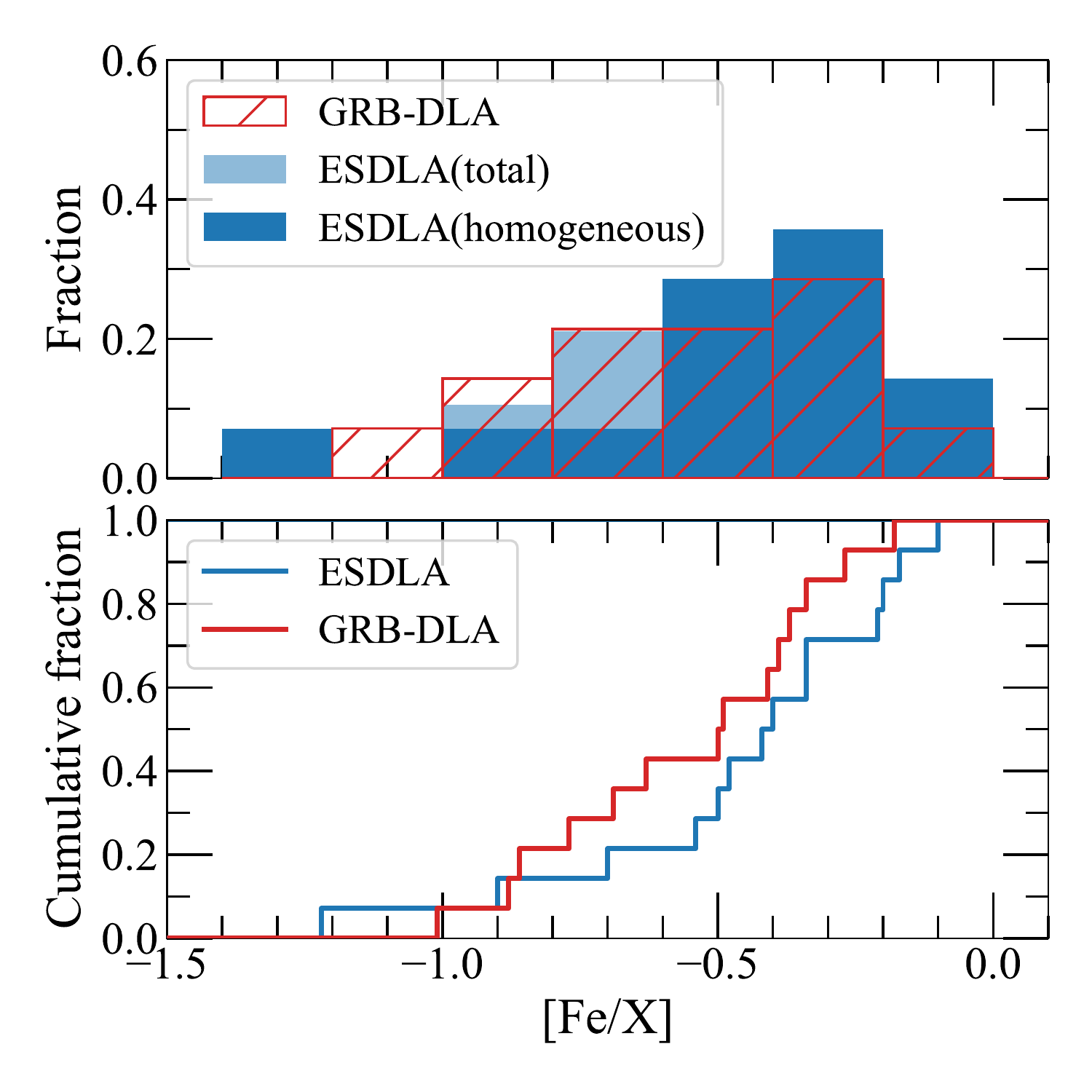} &
    \includegraphics[trim=10 10 10 0,clip,width=0.32\hsize]{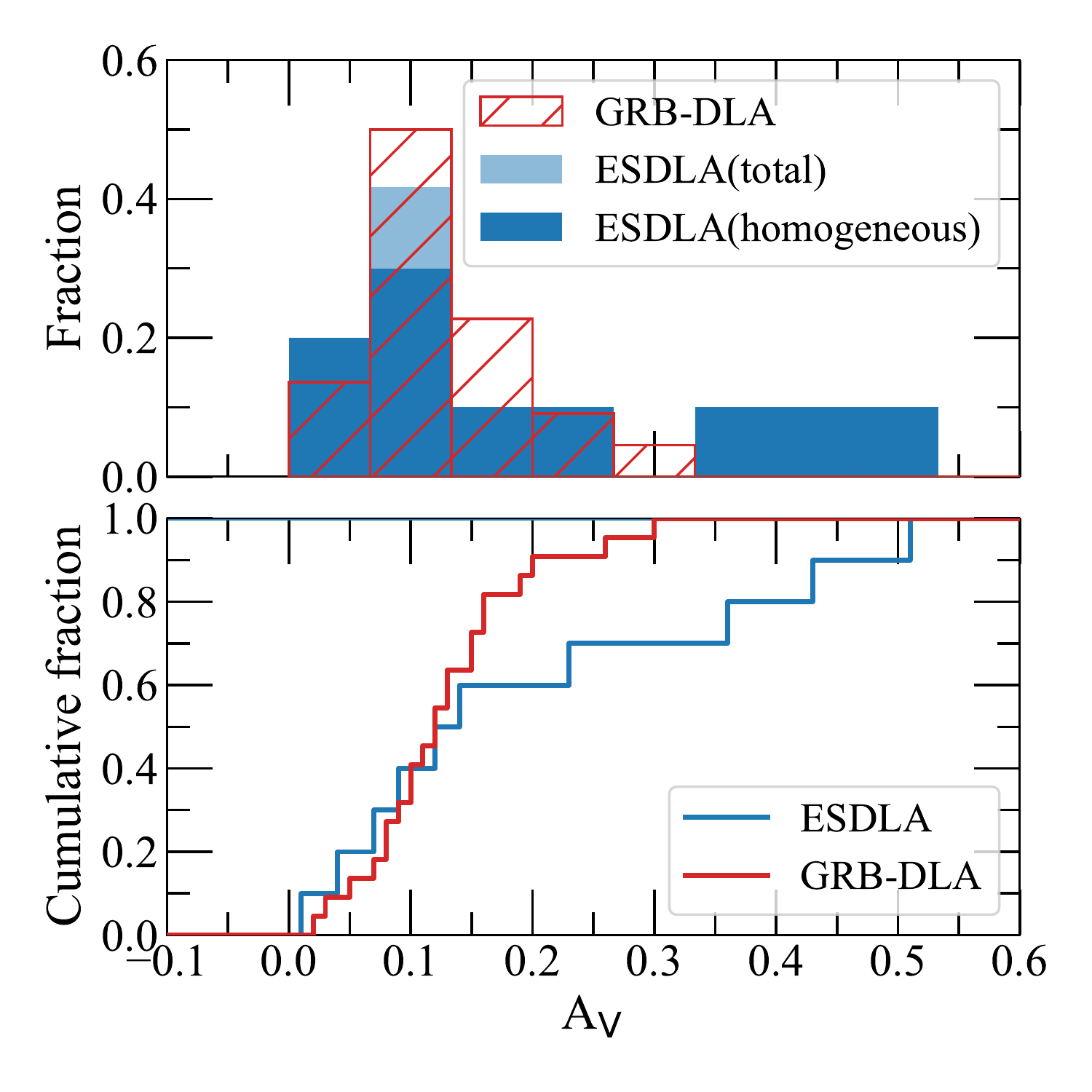}\\
 \end{tabular}
  \addtolength{\tabcolsep}{+3pt}
  \caption{Left: Metallicity distribution from different samples of DLAs (top). ESDLAs are plotted in blue with darker blue for the {\sl homogeneous} sample; GRB-DLAs are over-plotted as red hashed histogram.
  The bottom panel shows the cumulative fractions for the same distributions (including only homogeneous ESDLAs for comparison). 
  Middle and right panels: Same for the depletion factor [Fe/X] (where X is an undepleted species) and for the measured dust-extinction ($A_v$), respectively.
  }
  \label{metallicity_distribution_esdla}
  \label{depletion_distribution_esdla}
  \label{dust_distribution_esdla}
\end{figure*}

In the left panel of Fig.~\ref{metallicity_distribution_esdla}, we compare the distribution of metallicities for ESDLA to that for GRB-DLAs.
The metallicity distribution of ESDLAs is found to be very similar to that of GRB-DLAs.
The mean GRB-DLA metallicity is $\mean{\rm [Zn/H]} = -1.23 \pm 0.02$ and the two-sided KS test between ESDLAs and GRB-DLAs metallicities give $P=0.73$. This strengthens 
the similarity between ESDLAs and GRB-DLAs already discussed in the literature, in particular in the recent work by \citet{Bolmer2019}.

The distribution of depletion factors in our ESDLA sample is also compared to that of GRB-DLAs in Fig.~\ref{depletion_distribution_esdla} (middle panel). The two distributions are found to be similar with average values of $\mean{\rm [Fe/Zn]} = -0.50 \pm 0.05$ (ESDLAs)
and $\mean{\rm [Fe/Zn]} = -0.56 \pm 0.04$ (GRB-DLAs). The two-sided KS test gives $P=0.95$. Therefore, we cannot reject the null hypothesis that the two distributions are the same.

\subsubsection*{Dust reddening}

With both similar column densities and chemical enrichment, we can expect the dust reddening induced by ESDLAs to be similar to that induced by GRB-DLAs. 
The two distribution are compared in the right panel of Fig~\ref{dust_distribution_esdla}. While the two distributions overlap (the KS test gives $P=0.45$), ESDLAs tend to have higher dust reddening than GRB-DLAs, with average values of 
$\mean{\rm A_{V}} = 0.2$ and $\mean{\rm A_{V}} = 0.1$, respectively. Differences between the measured extinctions arising from GRB-DLAs and ESDLAs, if any, 
are likely not of physical origin. Instead, these can be explained by differences in the measurement methods and observational biases.
About the extinction measurements themselves, we note that the intrinsic spectral shape of quasars is more complex than those of GRB afterglows and can 
vary strongly from one quasar to another. In addition, there is a degeneracy between dust at the quasar redshift and dust in the absorber, while for GRB-DLAs, 
the light source and the absorber virtually have the same redshift. The higher reddening measured for ESDLAs compared to GRB-DLAs is also mostly 
driven by the absence of GRB-DLAs with $A_{V}>0.3$ in the sample from \citet{Bolmer2019}. 
As discussed by these authors (and previously by \citealt{Ledoux2009}), the extinction by dust in so-called dark bursts (with $A_{V}>0.5$) is high enough so that the optical afterglow becomes too faint for spectroscopic follow-up, implying a bias against dimmed systems in the spectral follow-up campaigns of GRB afterglows.

In principle, a similar bias could also apply to quasars, when identified from their colours in a flux-limited sample, as demonstrated by \citet{Krogager19} for the SDSS-II. However, the following SDSS stages are likely less inclined to dust bias thanks to improved selection
methods and deeper observations. In addition, unlike GRB afterglows, quasars are not transient and hence do not suffer from a strong targeting bias affecting the decision for follow-up. 
In conclusion, within the current limitations and taking into account the uncertainty on the dust measurements of ESDLAs and the absence of dark GRB-DLAs, there is no evidence for an intrinsic difference between the dust extinction from ESDLAs and that due to GRB-DLAs.

\subsubsection*{Molecular hydrogen} 
Theoretical \HI-H$_2$ transition models show that the presence of molecular hydrogen is mostly dependent on three parameters: the total gas column density, 
the abundance of dust, and the UV flux to density ratio \citep{Krumholz2012, Sternberg2014, Bialy2016}. We have already shown that ESDLAs and GRB-DLAs have 
similar column densities, chemical enrichment, and dust properties. Comparing the molecular content of both populations could then provide hints about any possible 
difference in the physical conditions. For example, from an early small sample of GRB-DLAs without H$_2$, \citet{Tumlinson2007} have suggested that the 
proximity to the explosion-site in the case of GRB-DLAs could influence the abundance of H$_2$. \citet{Ledoux2009} estimate a distance of 0.5~kpc between the absorber and the GRB explosion by photoexcitation modelling of \FeII. They also show that, at such distances, the GRB afterglow photons cannot influence the abundance of \HH\, and that X-Shooter observations could possibly find molecular hydrogen in GRB-DLAs. Indeed, \citet{Bolmer2019} show that there is no apparent lack of \HH\ in GRB-DLAs, with 6 firm detections in their sample of 22. Moreover, since dusty systems are more likely to harbour \HH, the absence of dark burst systems may lead to an underestimate of the actual \HH\ incidence rate in GRB-DLAs. Quantifying the number for the supposed dark bursts, \citet{Bolmer2019} conclude that the \HH\, detection rate in GRB-DLAs could be as high as 41\%. 

In our total sample of ESDLAs (Table~\ref{absorption_table}), H$_2$ is detected in 10 out of 18 systems (no limit at all could be put towards J\,1135$-$0010),
which is consistent with the detection rate in GRB-DLAs given the small number statistics and caveats about detection bias in GRB-DLAs. However, the detection limits are different from one system to another. For example, H$_2$ is firmly detected with $\log N(\HH)\sim 17.4$ towards HE\,0027$-$1836, that has extensively been observed with UVES \citep[see][]{Noterdaeme2007b,Rahmani2013}, while we could only put a weak limit ($N(\HH)< \mbox{a few} \times 10^{18}$~\cmsq) for several systems observed with X-shooter. A fair comparison, therefore, cannot be done considering the detection rates at face value. Restricting to systems with $\log N(\HH)>18.3$ (our weakest detection limit), there are 6/14 ($\sim 43\%$) strong H$_2$ systems in our homogeneous sample of ESDLAs (7/18 ($\sim$39\%) for the total sample). These values agree well with the $\sim$35 per cent incidence rate of high H$_2$ column densities among ESDLAs as derived from a stack of SDSS spectra \citep{Balashev+18}. Regarding the GRB-DLA sample, four (possibly five, if considering that GRB\,151021A as an H$_2$ detection) are seen in the sample of 22 GRB-DLAs by \citeauthor{Bolmer2019}, i.e., a lower rate of $\sim$20\%. However, we note that while most GRB-DLAs tend to have very high $N(\HI)$ content, there are still 8 GRB-DLAs with $\log N(\HI)<21.5$. Restricting the comparison to only those GRB-DLAs with $\log N(\HI) > 21.5$, we find that GRB-DLAs have a strong-H$_2$ detection rate of 4--5 out of 14, i.e., about 30\%. This is consistent with the ESDLA detection rate given the low number statistics.

\begin{figure}
\centering
   \includegraphics[trim=18 10 10 10,clip,width=0.95\hsize]{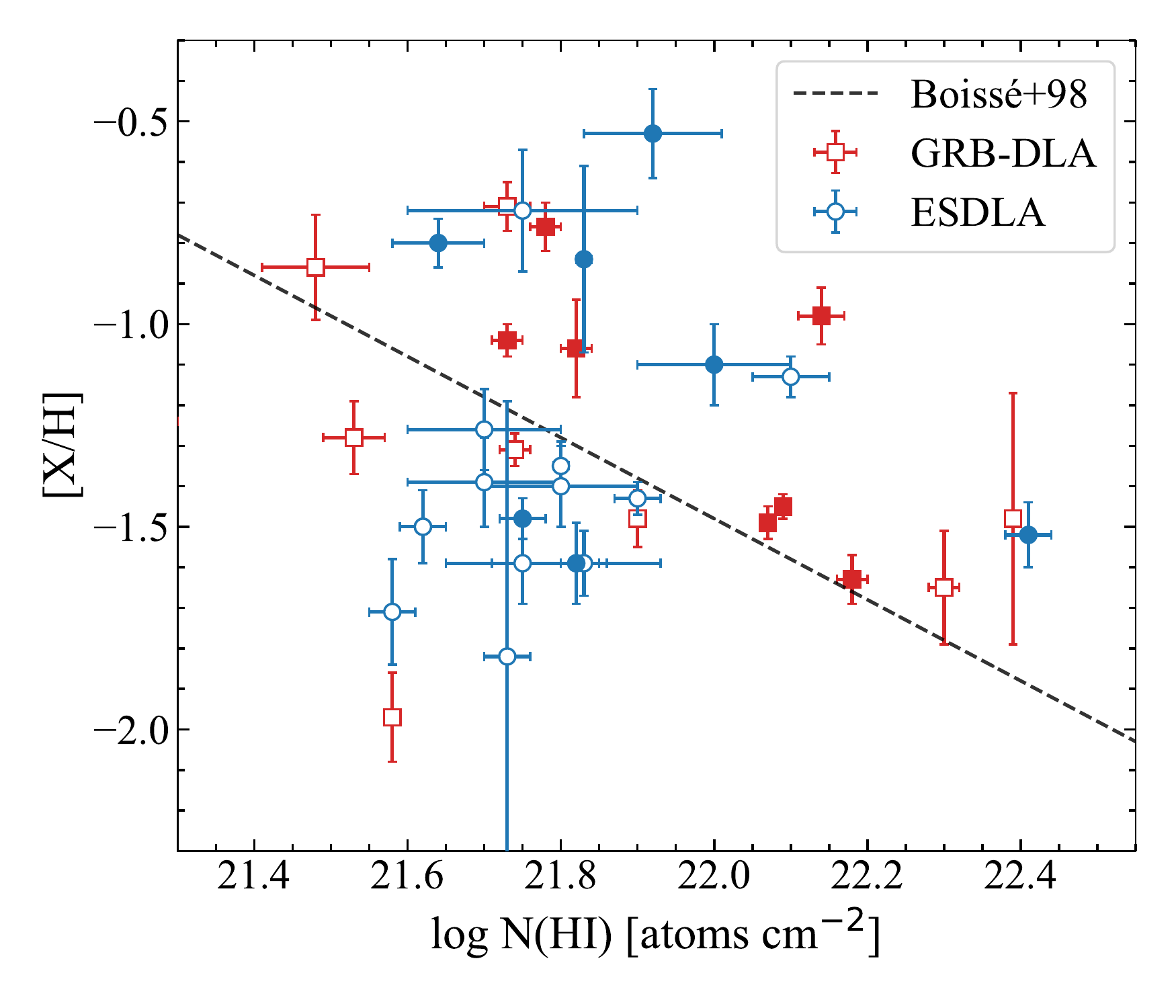}
    \caption{Metallicity vs \HI\ column density for the sample of ESDLAs (homogeneous, blue) and GRB-DLAs (red). Filled symbols represent the systems with firm detection of strong H$_2$ lines ($\log N(\HH)\ge 18.3)$. The dashed line shows the constant total metal column density corresponding to $\log N(\ZnII)=13.15$ from \citet{Boisse1998}.
   }
    \label{nh1_vs_Z}
\end{figure}

To illustrate this further, we plot the location of the GRB-DLAs and ESDLAs in the column density versus metallicity space, identifying those systems with strong H$_2$ lines, see Fig.~\ref{nh1_vs_Z}. The two populations seem to be indistinguishable, with most of the strong H$_2$ bearing systems being located in the upper right region of the figure (high $N(\HI)$ and high metallicity, beyond the limit originally proposed by \citet{Boisse1998} as being due to dust bias).
The similarity between the location of \HH\ and non-\HH\ bearing ESDLAs with respect to GRB-DLAs strengthens the conclusions by \citet{Bolmer2019} and \citet{Ledoux2009} that the effect of the GRB explosion itself is not strong enough to significantly dissociate H$_2$ in the observed GRB-DLAs. In other words, GRB-DLAs are most likely arising from gas located within the same galaxy as that of the GRB itself, yet they are far enough away from the UV radiation of the explosion site or the associated star-forming region (with distances, d$>$0.5kpc) to form \HH.

Finally, we note that \CI\ lines are detected in about 36\% of our sample, while \citet{Heintz2019} observe about 25\% in GRB-DLAs. However, this difference is not significant given the small number statistics of both samples and, even more importantly, the detection of \CI\ depends strongly on the resolution and the exact S/N ratio achieved.

\subsubsection*{Kinematics} 
Lastly, we compare the kinematics of the ESDLA and GRB-DLA samples.
We used a common way to quantify kinematics of the DLAs by means of the velocity width, $\Delta v_{90}$, defined as the velocity interval comprising 90\% of the line optical depth \citep[see][]{Prochaska_and_wolfe_1997}.
We have ignored the low-resolution (110 $<$ FWHM $<$ 480 km s$^{-1}$) data for GRB-DLAs, due to the fact that low spectral resolution leads to a broadening of the absorption lines, which in turn leads to an over-estimation of $\Delta v_{90}$. \citet{Arabsalmani2015} show that QSO-DLAs and GRB-DLAs are indistinguishable in the $\Delta v_{90}$--metallicity plane. The authors also show that, as opposed to the general spatial distribution of QSO-DLAs around their host galaxy, GRB-DLA sightlines pass from a deeper part of the dark matter halo potential well indicating close proximity to the centre of their associated galaxy. We proceed to perform a similar comparison of GRB-DLAs with our subset of ESDLAs.

\begin{figure}
\centering
   \includegraphics[width=0.95\hsize]{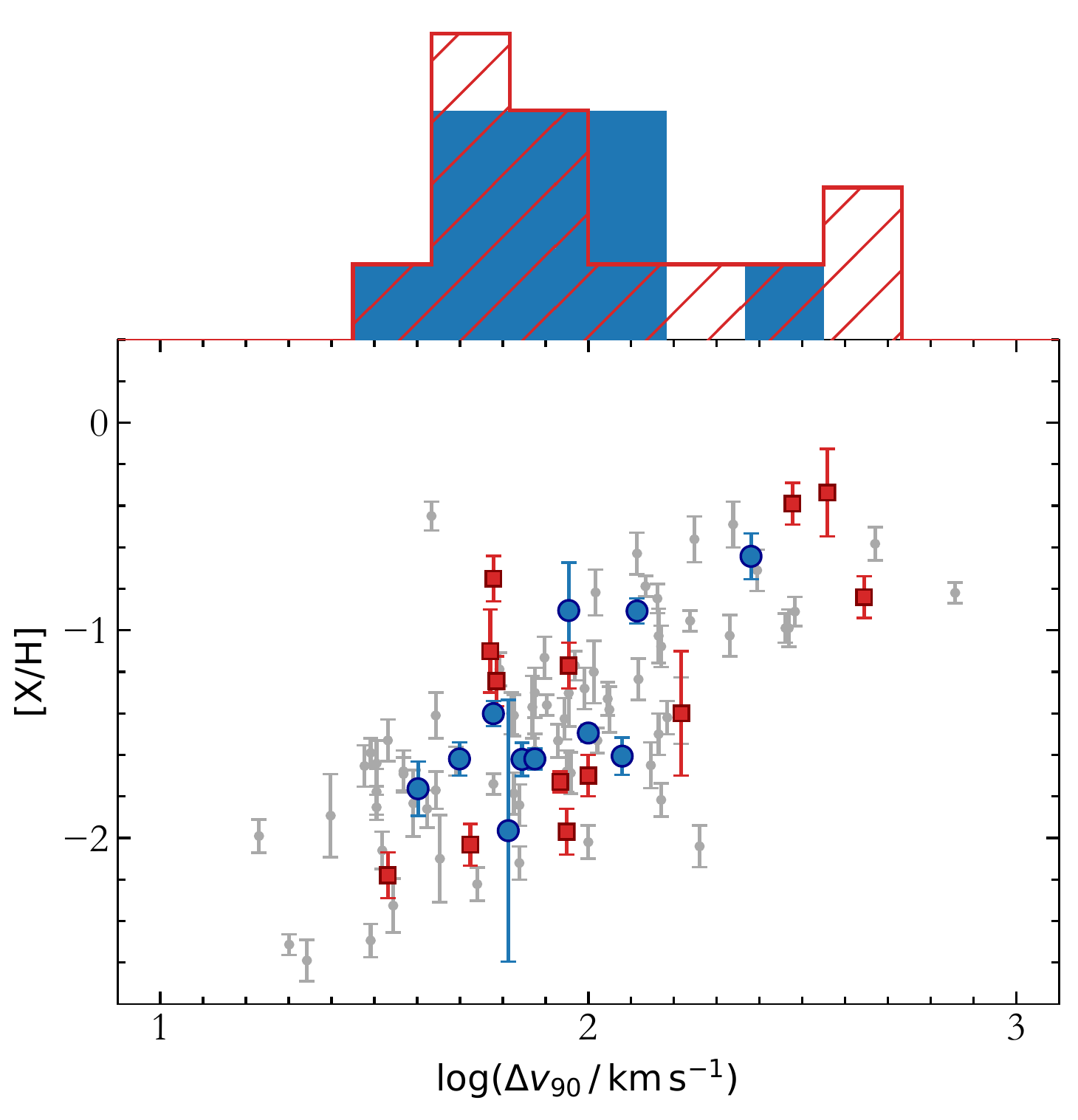}
    \caption{Comparison of the metallicities and $\Delta v_{90}$ values of ESDLAs (blue circles) and GRB-DLAs (red squares) observed with X-shooter. 
    For comparison, the grey points show the values for the overall intervening QSO-DLA population from \citet{Ledoux2006}. 
    The histograms in the top panel provide the $\Delta v_{90}$ distributions of the two samples using the same colour coding as in Fig.~\ref{nh1_vs_Z}.}
    \label{deltav90}
\end{figure}

For our ESDLA sample, we estimated $\Delta v_{90}$ using the metal absorption lines with peaked absorption strength $\sim0.3$ of the normalized flux. The exact lines used for the measurements are given in Table.~\ref{delta_v90_table} in the Appendix. At the spectral resolution of X-shooter, this corresponds to lines that are in principle neither saturated nor weak. We used the same correction for the spectral resolution as applied by \citet{Arabsalmani2015}. The comparison of these samples is shown in Fig.~\ref{deltav90}, along with the values for the overall QSO-DLA population by \citet{Ledoux2006}. One can see that the ESDLA sample is located in the same region of the $\Delta v_{90}$--metallicity plane as the GRB- and QSO-DLA samples. The distribution of the values of $\Delta v_{90}$ for ESDLAs and GRB-DLAs are shown on the top panel of Fig.~\ref{deltav90}.  The two-sided KS test with the null hypothesis that these two samples are drawn from the same distribution gives $P=0.86$.

\subsection{Associated galaxies  \label{associated_galaxies_discussion}}

The similarity between the gas properties of GRB-DLAs and ESDLAs (i.e., metallicities, depletion, H$_2$ content, \HI\ column densities, and kinematics) suggests that the two classes of absorbers probe a similar galaxy population and similar environments within their respective host galaxies.
Since the long-duration GRBs (which we consider here) are linked to the death of a massive star \citep[e.g.][]{Woosley2006}, they are expected to arise at relatively small impact parameters in their host galaxies where young, massive stars are formed. 
This is indeed confirmed by observations \citep[e.g.][]{Fruchter2006, Arabsalmani2015, Lyman2017}. 

QSO-DLAs, on the other hand, are observed to span a large range of impact parameters \citep{Fynbo2009,Krogager+17}.  However, both observations and simulations indicate an 
anti-correlation between the typical impact parameter and the hydrogen column density \citep[e.g.,][]{RahmatiandSchaye2014}. It is therefore expected that ESDLAs and GRB-DLAs 
should probe a similar range of impact parameters.
Nonetheless, due to the low incidence of the very highest column density QSO-DLAs (i.e. ESDLAs), the impact parameters of ESDLAs and the overlap with GRB-DLAs has been poorly studied until now.

\begin{figure*}
    \centering
    \includegraphics[trim=80 0 80 0,clip,width=\hsize]{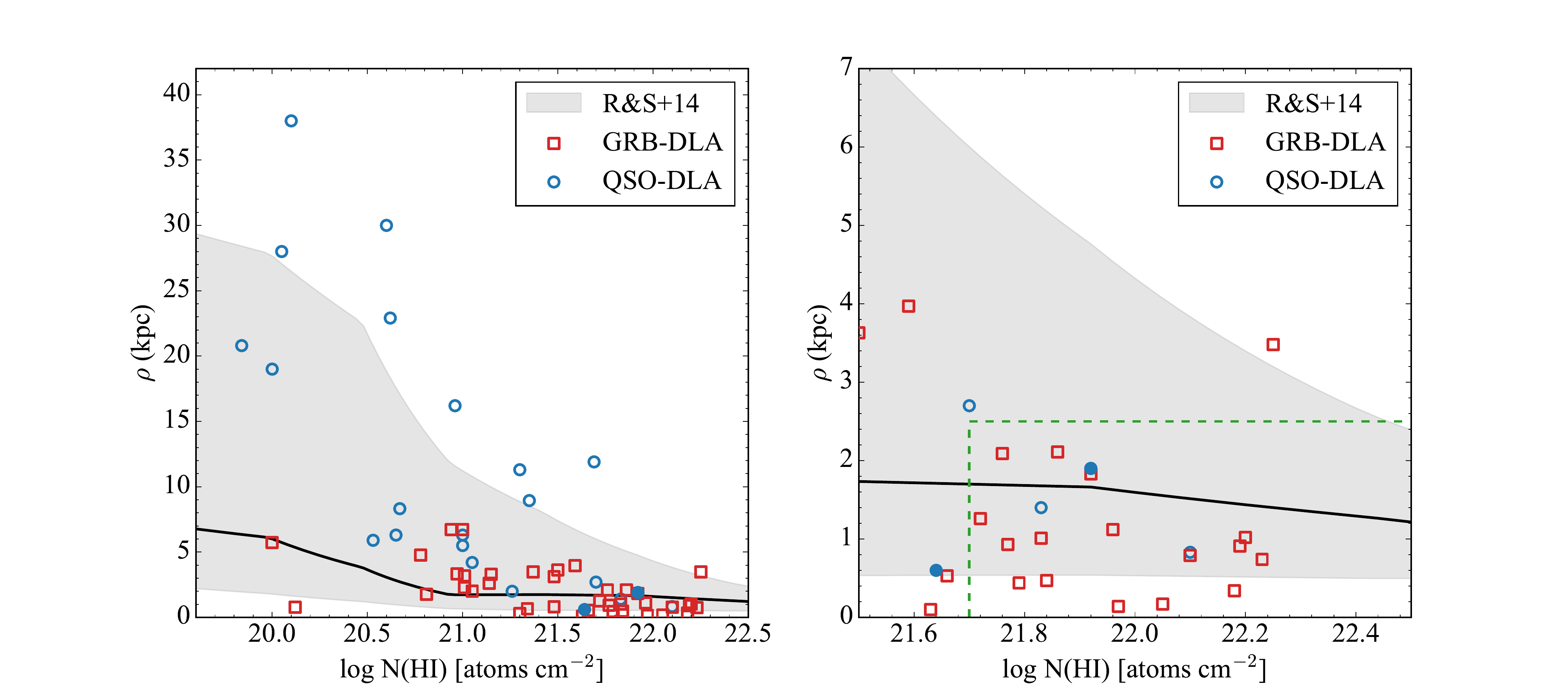}
    \caption{Impact parameter, $\rho$, versus the neutral hydrogen column density, $\log N(\HI)$, measured along the line-of-sight. Systems associated to the generic population of QSO-DLAs are plotted as blue circles (from  
    \citealt{Bouche2013,Fynbo2011,Fynbo2013,Hartoog2015,Kashikawa+14,Krogager+12,Krogager2013,Krogager+17, Ma2018, Moller+93, Moller2004, Neeleman2018, Noterdaeme2012, srianand2016, Ranjan+2018, Rudie2017, Warren2001,Weatherley+05,Zafar2017}), with the two new detections presented in this paper
    as filled circles. Values corresponding to GRB-DLAs are plotted as red squares.
    The grey line with the shaded region shows the predicted median and the 68\% region from simulations by \cite{RahmatiandSchaye2014}. The right panel shows a zoom-in around the high column density region.
    The extent of the y-axis corresponds to the impact parameter out to which we uniformly cover the area around the quasar, determined by the slit-width used in our observations.
 The dashed green rectangle shows the region of typical impact parameters derived by \citet{noterdaeme2014} for $\log N(\HI)>21.7$ systems (see text).  
   }
    \label{nhi_vs_rho}
\end{figure*}

To test this expectation further, we compare the impact parameters seen for our sample of ESDLAs to those observed for GRB-DLAs, see Fig.~\ref{nhi_vs_rho}. For this comparison, we disregard those ESDLAs that have only been observed with UVES, as the UVES data do not provide useful constraints on the associated emission due to the fact that the UVES slit is not fixed on the sky during integration.
In total, emission counterparts are detected for 5 ESDLAs (see Sect.~\ref{sample_selection}). For the remaining systems, we only obtain upper limits to the line luminosities {\sl within the effective slit aperture.} While we cannot formally reject the possibility that the emission from associated galaxies falls outside the region covered by the slit, we note that for all detections, the impact parameters are very small (<2~kpc). 
This is well within the range of impact parameters for which we have uniform coverage, determined by half the slit width (i.e. $\approx$6.6~kpc, for the median redshift of this sample). However, since the slit is much longer, we do cover impact parameters out to $\sim$40~kpc along the random slit direction. We, therefore, conclude that the non-detections are most likely due to the associated emission being below the detection limit \citep[see also][]{Krogager+17}. This is further bolstered by the fact that all 5 detections have $\rho<3$~kpc and none has impact parameters in the range $3<\rho<6.6$~kpc, where our effective sky coverage is still complete.
Lastly, our findings are also consistent with the suggestion by \citet{noterdaeme2014} that ESDLAs typically probe impact parameters $\rho \lesssim$2.5~kpc, as evidenced by the detection of \lya\ emission in a stack of SDSS fiber spectra and the similarity between the emission properties and \lya\ emitting galaxies.

In Fig.~\ref{nhi_vs_rho}, we also show the results from simulations by \citet{RahmatiandSchaye2014} in which the very high $N(\HI)$ regime is found at very small impact parameters. This is in excellent agreement with our observations. 
This is however not the case at lower column densities, where most of the observed data points are located well above the expected median values. This is likely the result of biasing towards high-metallicity systems, which were shown to be statistically located at larger impact parameters compared to low-metallicity systems \citep{Krogager+17}, due to the correlation between galaxy masses and sizes with metallicity. It is also possible that many of the absorber--galaxy associations suffer from an identification bias in which the absorber is associated to the brightest galaxy in a group. As we are detecting systems at very small impact parameters, such a bias is much less likely to occur in our ESDLA sample. 

Considering indeed that the non-detections of galactic emission for several systems is more likely due to the line fluxes being lower than our detection limits rather than the galaxy being located outside the area covered by the slit, we can estimate upper limits on the star-formation rates using the calibration by \citet{Kennicutt1998}:

\begin{equation}
    {\rm SFR\ /\ M_{\odot}\, yr^{-1}}\, =\, 7.9\,\times\,10^{-42}\, L(\Ha)\ /\ {\rm erg\, s^{-1}}~.
\end{equation}
\noindent

The median upper limit to the SFR from \Ha\ ranges from 6 to 30~M$_{\rm \odot}$\,yr$^{-1}$. Using \lya\ instead and assuming standard case-B recombination theory \citep{Brocklehurst1971, Osterbrock1989}, assuming a low \lya\ escape fraction of 5\% \citep{Hayes2011}, we obtain more stringent limits of typically 2$-$5~M$_{\rm \odot}$\,yr$^{-1}$, which are lower than typically seen for high-redshift GRB-DLA host galaxies. This is however not surprising since GRBs are, by selection, associated with active star formation \citep[see discussion by][]{Vergani2015}.

\section{Summary \label{Conclusion}}

We have presented the analysis of a sample of eleven intervening quasar absorption-line systems selected for their very high \HI\ column densities and observed with the X-shooter spectrograph at the Very Large Telescope. 
One system has been presented by \citet{Ranjan+2018} and in this work we have measured the metal abundances, molecular and dust content, and gas kinematics, and searched for the associated galaxy counterparts for the remaining ten systems. 

We clearly detect molecular hydrogen in five out of eleven systems, three of them being reported here for the first time. 
For the remaining systems, the presence of H$_2$ is hard to ascertain and we instead provide conservative upper limits from the maximum H$_2$ column density still consistent with the X-shooter data. We supplement our ESDLA sample with measurements from the literature and compare their absorption properties with GRB-DLAs.

We find that the two populations are indistinguishable in terms of chemical enrichment (metallicity and dust depletion) and there is no marked difference in the conditions for the presence of H$_2$. This supports the suggestion by \citet{noterdaeme2014} that ESDLAs are likely arising from similar environments as GRB-DLAs and the suggestion by \citet{Bolmer2019} that GRB-DLAs are likely located far enough from the explosion site for their H$_2$ content to be little affected by the GRB explosion itself.

The similarity between ESDLAs and GRB-DLAs is further bolstered by the similar absorption-line kinematics and by the observed impact parameters. 
Indeed, we detect emission lines associated to 3 systems out of our sample of 11 systems, two of them being reported here for the first time. This brings the number of emission counterparts obtained at the very high $N(\HI)$-end of the quasar-DLA column-density distribution to a total of five.
All emission counterparts of ESDLAs are located at very small impact parameters ($\rho <3$~kpc), when much larger impact parameters are generally reported for the overall population of DLAs. Since our observations are complete out to $\rho \sim$ 7~kpc, we argue that the counterparts for the remaining ESDLAs are more likely below the detection limit than located outside the area covered by the slit.

Long-duration $\gamma$-ray bursts have been shown to be tracers of star formation over the history of the Universe \citep{Vergani2015}. While GRB-DLAs then probe neutral gas inside high-redshift, highly star-forming galaxies, ESDLAs may act as completely blind analogues, probing neutral gas in the heart of high-redshift galaxies without any prior on the instantaneous star-formation rate.
In the near future, blind radio absorption line surveys such as the MeerKAT Absorption Line Survey (MALS, \citealt{GuptaMALS}) will further explore the population of neutral gas absorbers without the effects of dust biasing which currently affects DLA samples.

\begin{acknowledgements}
We thank the anonymous referee for useful comments and suggestions that helped to improve the clarity of the manuscript.
The authors are grateful to the European Southern Observatory (ESO) and in particular the Paranal observatory's staff for carrying out their observations in service mode.
AR, PN, PPJ, RS, and NG gratefully acknowledge the support of the Indo-French Centre for the Promotion of Advanced Research (Centre Franco-Indien pour la Promotion de la Recherche Avanc\'ee) under contract no. 5504-2 (PIs Gupta and Noterdaeme). We also thank the Institut d'Astrophysique de Paris (IAP) and the Inter-University Centre for Astronomy and Astrophysics (IUCAA) for hospitality during visits to these institutes.
The research leading to these results has received funding from the French {\sl Agence Nationale de la Recherche} under grant no ANR-17-CE31-0011-01 (project "HIH2" - PI Noterdaeme). PN and JKK thank the ESO Science Visitor program for hospitality and support during their visit to the ESO headquarters in Chile. SAB is supported by RFBR (grant No. 18-02-00596). SAB also thanks the Basis foundation and the IAP for hospitality and financial support during the time part of this work was performed.

\end{acknowledgements}

\bibliographystyle{aa}
\bibliography{bibliography.bib}

\begin{thebibliography}{83}
\expandafter\ifx\csname natexlab\endcsname\relax\def\natexlab#1{#1}\fi

\bibitem[{{Altay} {et~al.}(2013){Altay}, {Theuns}, {Schaye}, {Booth}, \& {Dalla
  Vecchia}}]{Altay+13}
{Altay}, G., {Theuns}, T., {Schaye}, J., {Booth}, C.~M., \& {Dalla Vecchia}, C.
  2013, \mnras, 436, 2689

\bibitem[{{Arabsalmani} {et~al.}(2015){Arabsalmani}, {M{\o}ller}, {Fynbo},
  {Christensen}, {Freudling}, {Savaglio}, \& {Zafar}}]{Arabsalmani2015}
{Arabsalmani}, M., {M{\o}ller}, P., {Fynbo}, J.~P.~U., {et~al.} 2015, \mnras,
  446, 990

\bibitem[{{Arabsalmani} {et~al.}(2018){Arabsalmani}, {M{\o}ller}, {Perley},
  {Freudling}, {Fynbo}, {Le Floc'h}, {Zwaan}, {Schulze}, {Tanvir},
  {Christensen}, {Levan}, {Jakobsson}, {Malesani}, {Cano}, {Covino}, {D'Elia},
  {Goldoni}, {Gomboc}, {Heintz}, {Sparre}, {de Ugarte Postigo}, \&
  {Vergani}}]{Arabsalmani2018}
{Arabsalmani}, M., {M{\o}ller}, P., {Perley}, D.~A., {et~al.} 2018, \mnras,
  473, 3312

\bibitem[{{Asplund} {et~al.}(2009){Asplund}, {Grevesse}, {Sauval}, \&
  {Scott}}]{Asplund+09}
{Asplund}, M., {Grevesse}, N., {Sauval}, A.~J., \& {Scott}, P. 2009, \araa, 47,
  481

\bibitem[{{Balashev} {et~al.}(2014){Balashev}, {Klimenko}, {Ivanchik},
  {Varshalovich}, {Petitjean}, \& {Noterdaeme}}]{Balashev2014}
{Balashev}, S.~A., {Klimenko}, V.~V., {Ivanchik}, A.~V., {et~al.} 2014, \mnras,
  440, 225

\bibitem[{{Balashev} \& {Noterdaeme}(2018)}]{Balashev+18}
{Balashev}, S.~A. \& {Noterdaeme}, P. 2018, \mnras, 478, L7

\bibitem[{{Balashev} {et~al.}(2017){Balashev}, {Noterdaeme}, {Rahmani},
  {Klimenko}, {Ledoux}, {Petitjean}, {Srianand}, {Ivanchik}, \&
  {Varshalovich}}]{Balashev+17}
{Balashev}, S.~A., {Noterdaeme}, P., {Rahmani}, H., {et~al.} 2017, \mnras, 470,
  2890

\bibitem[{{Bialy} \& {Sternberg}(2016)}]{Bialy2016}
{Bialy}, S. \& {Sternberg}, A. 2016, \apj, 822, 83

\bibitem[{{Bird} {et~al.}(2014){Bird}, {Vogelsberger}, {Haehnelt}, {Sijacki},
  {Genel}, {Torrey}, {Springel}, \& {Hernquist}}]{Bird2014}
{Bird}, S., {Vogelsberger}, M., {Haehnelt}, M., {et~al.} 2014, \mnras, 445,
  2313

\bibitem[{{Blitz} \& {Rosolowsky}(2006)}]{Blitz2006}
{Blitz}, L. \& {Rosolowsky}, E. 2006, \apj, 650, 933

\bibitem[{{Boiss\'e} {et~al.}(1998){Boiss\'e}, {Le Brun}, {Bergeron}, \&
  {Deharveng}}]{Boisse1998}
{Boiss\'e}, P., {Le Brun}, V., {Bergeron}, J., \& {Deharveng}, J.-M. 1998,
  \aap, 333, 841

\bibitem[{{Bolmer} {et~al.}(2019){Bolmer}, {Ledoux}, {Wiseman}, {De Cia},
  {Selsing}, {Schady}, {Greiner}, {Savaglio}, {Burgess}, {D'Elia}, {Fynbo},
  {Goldoni}, {Hartmann}, {Heintz}, {Jakobsson}, {Japelj}, {Kaper}, {Tanvir},
  {Vreeswijk}, \& {Zafar}}]{Bolmer2019}
{Bolmer}, J., {Ledoux}, C., {Wiseman}, P., {et~al.} 2019, \aap, 623, A43

\bibitem[{{Bouch{\'e}} {et~al.}(2013){Bouch{\'e}}, {Murphy}, {Kacprzak},
  {P{\'e}roux}, {Contini}, {Martin}, \& {Dessauges-Zavadsky}}]{Bouche2013}
{Bouch{\'e}}, N., {Murphy}, M.~T., {Kacprzak}, G.~G., {et~al.} 2013, Science,
  341, 50

\bibitem[{{Brocklehurst}(1971)}]{Brocklehurst1971}
{Brocklehurst}, M. 1971, \mnras, 153, 471

\bibitem[{{Carswell} \& {Webb}(2014)}]{Carswell2014}
{Carswell}, R.~F. \& {Webb}, J.~K. 2014, {VPFIT: Voigt profile fitting
  program}, Astrophysics Source Code Library

\bibitem[{{Castro} {et~al.}(2003){Castro}, {Galama}, {Harrison}, {Holtzman},
  {Bloom}, {Djorgovski}, \& {Kulkarni}}]{Castro2003}
{Castro}, S., {Galama}, T.~J., {Harrison}, F.~A., {et~al.} 2003, \apj, 586, 128

\bibitem[{{De Cia} {et~al.}(2016){De Cia}, {Ledoux}, {Mattsson}, {Petitjean},
  {Srianand}, {Gavignaud}, \& {Jenkins}}]{DeCia2016}
{De Cia}, A., {Ledoux}, C., {Mattsson}, L., {et~al.} 2016, \aap, 596, A97

\bibitem[{{Fruchter} {et~al.}(2006){Fruchter}, {Levan}, {Strolger},
  {Vreeswijk}, {Thorsett}, {Bersier}, {Burud}, {Castro Cer{\'o}n},
  {Castro-Tirado}, {Conselice}, {Dahlen}, {Ferguson}, {Fynbo}, {Garnavich},
  {Gibbons}, {Gorosabel}, {Gull}, {Hjorth}, {Holland}, {Kouveliotou}, {Levay},
  {Livio}, {Metzger}, {Nugent}, {Petro}, {Pian}, {Rhoads}, {Riess}, {Sahu},
  {Smette}, {Tanvir}, {Wijers}, \& {Woosley}}]{Fruchter2006}
{Fruchter}, A.~S., {Levan}, A.~J., {Strolger}, L., {et~al.} 2006, \nat, 441,
  463

\bibitem[{{Fynbo} {et~al.}(2009){Fynbo}, {Jakobsson}, {Prochaska}, {Malesani},
  {Ledoux}, {de Ugarte Postigo}, {Nardini}, {Vreeswijk}, {Wiersema}, {Hjorth},
  {Sollerman}, {Chen}, {Th{\"o}ne}, {Bj{\"o}rnsson}, {Bloom}, {Castro-Tirado},
  {Christensen}, {De Cia}, {Fruchter}, {Gorosabel}, {Graham}, {Jaunsen},
  {Jensen}, {Kann}, {Kouveliotou}, {Levan}, {Maund}, {Masetti},
  {Milvang-Jensen}, {Palazzi}, {Perley}, {Pian}, {Rol}, {Schady}, {Starling},
  {Tanvir}, {Watson}, {Xu}, {Augusteijn}, {Grundahl}, {Telting}, \&
  {Quirion}}]{Fynbo2009}
{Fynbo}, J.~P.~U., {Jakobsson}, P., {Prochaska}, J.~X., {et~al.} 2009, \apjs,
  185, 526

\bibitem[{{Fynbo} {et~al.}(2013){Fynbo}, {Krogager}, {Venemans}, {Noterdaeme},
  {Vestergaard}, {M{\o}ller}, {Ledoux}, \& {Geier}}]{Fynbo2013}
{Fynbo}, J.~P.~U., {Krogager}, J.-K., {Venemans}, B., {et~al.} 2013, \apjs,
  204, 6

\bibitem[{{Fynbo} {et~al.}(2011){Fynbo}, {Ledoux}, {Noterdaeme}, {Christensen},
  {M{\o}ller}, {Durgapal}, {Goldoni}, {Kaper}, {Krogager}, {Laursen}, {Maund},
  {Milvang-Jensen}, {Okoshi}, {Rasmussen}, {Thorsen}, {Toft}, \&
  {Zafar}}]{Fynbo2011}
{Fynbo}, J.~P.~U., {Ledoux}, C., {Noterdaeme}, P., {et~al.} 2011, \mnras, 413,
  2481

\bibitem[{{Gordon} {et~al.}(2003){Gordon}, {Clayton}, {Misselt}, {Landolt}, \&
  {Wolff}}]{Gordon2003}
{Gordon}, K.~D., {Clayton}, G.~C., {Misselt}, K.~A., {Landolt}, A.~U., \&
  {Wolff}, M.~J. 2003, \apj, 594, 279

\bibitem[{{Guimar{\~a}es} {et~al.}(2012){Guimar{\~a}es}, {Noterdaeme},
  {Petitjean}, {Ledoux}, {Srianand}, {L{\'o}pez}, \& {Rahmani}}]{Guimaraes2012}
{Guimar{\~a}es}, R., {Noterdaeme}, P., {Petitjean}, P., {et~al.} 2012, \aj,
  143, 147

\bibitem[{{Gupta} {et~al.}(2016){Gupta}, {Srianand}, {Baan}, {Baker},
  {Beswick}, {Bhatnagar}, {Bhattacharya}, {Bosma}, {Carilli}, {Cluver},
  {Combes}, {Cress}, {Dutta}, {Fynbo}, {Heald}, {Hilton}, {Hussain}, {Jarvis},
  {Jozsa}, {Kamphuis}, {Kembhavi}, {Kerp}, {Kloeckner}, {Krogager}, {Kulkarni},
  {Ledoux}, {Mahabal}, {Mauch}, {Moodley}, {Momjian}, {Morganti}, {Noterdaeme},
  {Oosterloo}, {Petitjean}, {Schroeder}, {Serra}, {Sievers}, {Spekkens},
  {Vaisanen}, {van der Hulst}, {Vivek}, {Wang}, {Wong}, \& {Zungu}}]{GuptaMALS}
{Gupta}, N., {Srianand}, R., {Baan}, W., {et~al.} 2016, in Proceedings of
  MeerKAT Science: On the Pathway to the SKA. 25-27 May, 2016 Stellenbosch,
  South Africa (MeerKAT2016)., 14

\bibitem[{{Hartoog} {et~al.}(2015){Hartoog}, {Fynbo}, {Kaper}, {De Cia}, \&
  {Bagdonaite}}]{Hartoog2015}
{Hartoog}, O.~E., {Fynbo}, J.~P.~U., {Kaper}, L., {De Cia}, A., \&
  {Bagdonaite}, J. 2015, \mnras, 447, 2738

\bibitem[{{Hayes} {et~al.}(2011){Hayes}, {Schaerer}, {{\"O}stlin}, {Mas-Hesse},
  {Atek}, \& {Kunth}}]{Hayes2011}
{Hayes}, M., {Schaerer}, D., {{\"O}stlin}, G., {et~al.} 2011, \apj, 730, 8

\bibitem[{{Heintz} {et~al.}(2019){Heintz}, {Ledoux}, {Fynbo}, {Jakobsson},
  {Noterdaeme}, {Krogager}, {Bolmer}, {M{\o}ller}, {Vergani}, {Watson},
  {Zafar}, {De Cia}, {Tanvir}, {Malesani}, {Japelj}, {Covino}, \&
  {Kaper}}]{Heintz2019}
{Heintz}, K.~E., {Ledoux}, C., {Fynbo}, J.~P.~U., {et~al.} 2019, \aap, 621, A20

\bibitem[{{Kashikawa} {et~al.}(2014){Kashikawa}, {Misawa}, {Minowa}, {Okoshi},
  {Hattori}, {Toshikawa}, {Ishikawa}, \& {Onoue}}]{Kashikawa+14}
{Kashikawa}, N., {Misawa}, T., {Minowa}, Y., {et~al.} 2014, \apj, 780, 116

\bibitem[{{Kennicutt}(1998)}]{Kennicutt1998}
{Kennicutt}, Jr., R.~C. 1998, \araa, 36, 189

\bibitem[{{Krawczyk} {et~al.}(2015){Krawczyk}, {Richards}, {Gallagher},
  {Leighly}, {Hewett}, {Ross}, \& {Hall}}]{Krawczyk2015}
{Krawczyk}, C.~M., {Richards}, G.~T., {Gallagher}, S.~C., {et~al.} 2015, \aj,
  149, 203

\bibitem[{{Krogager} {et~al.}(2013){Krogager}, {Fynbo}, {Ledoux},
  {Christensen}, {Gallazzi}, {Laursen}, {M{\o}ller}, {Noterdaeme},
  {P{\'e}roux}, {Pettini}, \& {Vestergaard}}]{Krogager2013}
{Krogager}, J.-K., {Fynbo}, J.~P.~U., {Ledoux}, C., {et~al.} 2013, \mnras, 433,
  3091

\bibitem[{{Krogager} {et~al.}(2012){Krogager}, {Fynbo}, {M{\o}ller}, {Ledoux},
  {Noterdaeme}, {Christensen}, {Milvang-Jensen}, \& {Sparre}}]{Krogager+12}
{Krogager}, J.-K., {Fynbo}, J.~P.~U., {M{\o}ller}, P., {et~al.} 2012, \mnras,
  424, L1

\bibitem[{{Krogager} {et~al.}(2019){Krogager}, {Fynbo}, {M{\o}ller},
  {Noterdaeme}, {Heintz}, \& {Pettini}}]{Krogager19}
{Krogager}, J.-K., {Fynbo}, J. P.~U., {M{\o}ller}, P., {et~al.} 2019, \mnras,
  486, 4377

\bibitem[{{Krogager} {et~al.}(2017){Krogager}, {M{\o}ller}, {Fynbo}, \&
  {Noterdaeme}}]{Krogager+17}
{Krogager}, J.-K., {M{\o}ller}, P., {Fynbo}, J.~P.~U., \& {Noterdaeme}, P.
  2017, \mnras, 469, 2959

\bibitem[{{Krumholz}(2012)}]{Krumholz2012}
{Krumholz}, M.~R. 2012, \apj, 759, 9

\bibitem[{{Kulkarni} {et~al.}(2012){Kulkarni}, {Meiring}, {Som}, {P{\'e}roux},
  {York}, {Khare}, \& {Lauroesch}}]{Kulkarni12}
{Kulkarni}, V.~P., {Meiring}, J., {Som}, D., {et~al.} 2012, \apj, 749, 176

\bibitem[{{Ledoux} {et~al.}(2006){Ledoux}, {Petitjean}, {Fynbo}, {M{\o}ller},
  \& {Srianand}}]{Ledoux2006}
{Ledoux}, C., {Petitjean}, P., {Fynbo}, J.~P.~U., {M{\o}ller}, P., \&
  {Srianand}, R. 2006, \aap, 457, 71

\bibitem[{{Ledoux} {et~al.}(2009){Ledoux}, {Vreeswijk}, {Smette}, {Fox},
  {Petitjean}, {Ellison}, {Fynbo}, \& {Savaglio}}]{Ledoux2009}
{Ledoux}, C., {Vreeswijk}, P.~M., {Smette}, A., {et~al.} 2009, \aap, 506, 661

\bibitem[{Lodders {et~al.}(2009)Lodders, Palme, \& Gail}]{lodders2009}
Lodders, K., Palme, H., \& Gail, H. 2009, JE Tr{\"u}mper, 4, 44

\bibitem[{{Lyman} {et~al.}(2017){Lyman}, {Levan}, {Tanvir}, {Fynbo}, {McGuire},
  {Perley}, {Angus}, {Bloom}, {Conselice}, {Fruchter}, {Hjorth}, {Jakobsson},
  \& {Starling}}]{Lyman2017}
{Lyman}, J.~D., {Levan}, A.~J., {Tanvir}, N.~R., {et~al.} 2017, \mnras, 467,
  1795

\bibitem[{{Ma} {et~al.}(2018){Ma}, {Brammer}, {Ge}, {Prochaska}, \&
  {Lundgren}}]{Ma2018}
{Ma}, J., {Brammer}, G., {Ge}, J., {Prochaska}, J.~X., \& {Lundgren}, B. 2018,
  \apjl, 857, L12

\bibitem[{{Modigliani} {et~al.}(2010){Modigliani}, {Goldoni}, {Royer},
  {Haigron}, {Guglielmi}, {Fran{\c c}ois}, {Horrobin}, {Bristow}, {Vernet},
  {Moehler}, {Kerber}, {Ballester}, {Mason}, \& {Christensen}}]{Modigliani2010}
{Modigliani}, A., {Goldoni}, P., {Royer}, F., {et~al.} 2010, in \procspie, Vol.
  7737, Observatory Operations: Strategies, Processes, and Systems III, 773728

\bibitem[{{M{\o}ller}(2000)}]{Moller2000}
{M{\o}ller}, P. 2000, The Messenger, 99, 31

\bibitem[{{M{\o}ller} {et~al.}(2004){M{\o}ller}, {Fynbo}, \&
  {Fall}}]{Moller2004}
{M{\o}ller}, P., {Fynbo}, J.~P.~U., \& {Fall}, S.~M. 2004, \aap, 422, L33

\bibitem[{{M{\o}ller} \& {Warren}(1993)}]{Moller+93}
{M{\o}ller}, P. \& {Warren}, S.~J. 1993, \aap, 270, 43

\bibitem[{{Morton}(2003)}]{Morton2003}
{Morton}, D.~C. 2003, \apjs, 149, 205

\bibitem[{{Neeleman} {et~al.}(2018){Neeleman}, {Kanekar}, {Prochaska},
  {Christensen}, {Dessauges-Zavadsky}, {Fynbo}, {M{\o}ller}, \&
  {Zwaan}}]{Neeleman2018}
{Neeleman}, M., {Kanekar}, N., {Prochaska}, J.~X., {et~al.} 2018, \apjl, 856,
  L12

\bibitem[{{Noterdaeme} {et~al.}(2012){Noterdaeme}, {Laursen}, {Petitjean},
  {Vergani}, {Maureira}, {Ledoux}, {Fynbo}, {L{\'o}pez}, \&
  {Srianand}}]{Noterdaeme2012}
{Noterdaeme}, P., {Laursen}, P., {Petitjean}, P., {et~al.} 2012, \aap, 540, A63

\bibitem[{Noterdaeme {et~al.}(2008)Noterdaeme, Ledoux, Petitjean, \&
  Srianand}]{noterdaeme2008}
Noterdaeme, P., Ledoux, C., Petitjean, P., \& Srianand, R. 2008, Astronomy \&
  Astrophysics, 481, 327

\bibitem[{{Noterdaeme} {et~al.}(2018){Noterdaeme}, {Ledoux}, {Zou},
  {Petitjean}, {Srianand}, {Balashev}, \& {L{\'o}pez}}]{Noterdaeme2018}
{Noterdaeme}, P., {Ledoux}, C., {Zou}, S., {et~al.} 2018, \aap, 612, A58

\bibitem[{{Noterdaeme} {et~al.}(2009){Noterdaeme}, {Petitjean}, {Ledoux}, \&
  {Srianand}}]{Noterdaeme2009}
{Noterdaeme}, P., {Petitjean}, P., {Ledoux}, C., \& {Srianand}, R. 2009, \aap,
  505, 1087

\bibitem[{Noterdaeme {et~al.}(2014)Noterdaeme, Petitjean, P{\^a}ris, Cai,
  Finley, Ge, Pieri, \& York}]{noterdaeme2014}
Noterdaeme, P., Petitjean, P., P{\^a}ris, I., {et~al.} 2014, Astronomy \&
  Astrophysics, 566, A24

\bibitem[{{Noterdaeme} {et~al.}(2015{\natexlab{a}}){Noterdaeme}, {Petitjean},
  \& {Srianand}}]{Noterdaeme2015b}
{Noterdaeme}, P., {Petitjean}, P., \& {Srianand}, R. 2015{\natexlab{a}}, \aap,
  578, L5

\bibitem[{{Noterdaeme} {et~al.}(2007){Noterdaeme}, {Petitjean}, {Srianand},
  {Ledoux}, \& {Le Petit}}]{Noterdaeme2007b}
{Noterdaeme}, P., {Petitjean}, P., {Srianand}, R., {Ledoux}, C., \& {Le Petit},
  F. 2007, \aap, 469, 425

\bibitem[{{Noterdaeme} {et~al.}(2015{\natexlab{b}}){Noterdaeme}, {Srianand},
  {Rahmani}, {Petitjean}, {P{\^a}ris}, {Ledoux}, {Gupta}, \&
  {L{\'o}pez}}]{Noterdaeme2015a}
{Noterdaeme}, P., {Srianand}, R., {Rahmani}, H., {et~al.} 2015{\natexlab{b}},
  \aap, 577, A24

\bibitem[{{Osterbrock}(1989)}]{Osterbrock1989}
{Osterbrock}, D.~E. 1989, {Astrophysics of gaseous nebulae and active galactic
  nuclei}

\bibitem[{{P{\^a}ris} {et~al.}(2017){P{\^a}ris}, {Petitjean}, {Ross}, {Myers},
  {Aubourg}, {Streblyanska}, {Bailey}, {Armengaud}, {Palanque-Delabrouille},
  {Y{\`e}che}, {Hamann}, {Strauss}, {Albareti}, {Bovy}, {Bizyaev}, {Niel
  Brandt}, {Brusa}, {Buchner}, {Comparat}, {Croft}, {Dwelly}, {Fan},
  {Font-Ribera}, {Ge}, {Georgakakis}, {Hall}, {Jiang}, {Kinemuchi},
  {Malanushenko}, {Malanushenko}, {McMahon}, {Menzel}, {Merloni}, {Nandra},
  {Noterdaeme}, {Oravetz}, {Pan}, {Pieri}, {Prada}, {Salvato}, {Schlegel},
  {Schneider}, {Simmons}, {Viel}, {Weinberg}, \& {Zhu}}]{Paris2017}
{P{\^a}ris}, I., {Petitjean}, P., {Ross}, N.~P., {et~al.} 2017, \aap, 597, A79

\bibitem[{{Perley} {et~al.}(2013){Perley}, {Levan}, {Tanvir}, {Cenko}, {Bloom},
  {Hjorth}, {Kr{\"u}hler}, {Filippenko}, {Fruchter}, {Fynbo}, {Jakobsson},
  {Kalirai}, {Milvang-Jensen}, {Morgan}, {Prochaska}, \&
  {Silverman}}]{Perley2013}
{Perley}, D.~A., {Levan}, A.~J., {Tanvir}, N.~R., {et~al.} 2013, \apj, 778, 128

\bibitem[{{Planck Collaboration} {et~al.}(2016){Planck Collaboration}, {Ade},
  {Aghanim}, {Arnaud}, {Ashdown}, {Aumont}, {Baccigalupi}, {Banday},
  {Barreiro}, {Bartlett}, \& et~al.}]{Planck2016}
{Planck Collaboration}, {Ade}, P.~A.~R., {Aghanim}, N., {et~al.} 2016, \aap,
  594, A13

\bibitem[{{Pontzen} {et~al.}(2008){Pontzen}, {Governato}, {Pettini}, {Booth},
  {Stinson}, {Wadsley}, {Brooks}, {Quinn}, \& {Haehnelt}}]{Pontzen2008}
{Pontzen}, A., {Governato}, F., {Pettini}, M., {et~al.} 2008, \mnras, 390, 1349

\bibitem[{{Prochaska} {et~al.}(2009){Prochaska}, {Sheffer}, {Perley}, {Bloom},
  {Lopez}, {Dessauges-Zavadsky}, {Chen}, {Filippenko}, {Ganeshalingam}, {Li},
  {Miller}, \& {Starr}}]{Prochaska2009}
{Prochaska}, J.~X., {Sheffer}, Y., {Perley}, D.~A., {et~al.} 2009, \apjl, 691,
  L27

\bibitem[{{Prochaska} \& {Wolfe}(1997)}]{Prochaska_and_wolfe_1997}
{Prochaska}, J.~X. \& {Wolfe}, A.~M. 1997, \apj, 487, 73

\bibitem[{{Rahmani} {et~al.}(2013){Rahmani}, {Wendt}, {Srianand}, {Noterdaeme},
  {Petitjean}, {Molaro}, {Whitmore}, {Murphy}, {Centurion}, {Fathivavsari},
  {D'Odorico}, {Evans}, {Levshakov}, {Lopez}, {Martins}, {Reimers}, \&
  {Vladilo}}]{Rahmani2013}
{Rahmani}, H., {Wendt}, M., {Srianand}, R., {et~al.} 2013, \mnras, 435, 861

\bibitem[{{Rahmati} \& {Schaye}(2014)}]{RahmatiandSchaye2014}
{Rahmati}, A. \& {Schaye}, J. 2014, \mnras, 438, 529

\bibitem[{{Ranjan} {et~al.}(2018){Ranjan}, {Noterdaeme}, {Krogager},
  {Petitjean}, {Balashev}, {Bialy}, {Srianand}, {Gupta}, {Fynbo}, {Ledoux}, \&
  {Laursen}}]{Ranjan+2018}
{Ranjan}, A., {Noterdaeme}, P., {Krogager}, J.-K., {et~al.} 2018, \aap, 618,
  A184

\bibitem[{{Rudie} {et~al.}(2017){Rudie}, {Newman}, \& {Murphy}}]{Rudie2017}
{Rudie}, G.~C., {Newman}, A.~B., \& {Murphy}, M.~T. 2017, \apj, 843, 98

\bibitem[{{Selsing} {et~al.}(2016){Selsing}, {Fynbo}, {Christensen}, \&
  {Krogager}}]{Selsing2016}
{Selsing}, J., {Fynbo}, J.~P.~U., {Christensen}, L., \& {Krogager}, J.-K. 2016,
  \aap, 585, A87

\bibitem[{{Selsing} {et~al.}(2019){Selsing}, {Malesani}, {Goldoni}, {Fynbo},
  {Kr{\"u}hler}, {Antonelli}, {Arabsalmani}, {Bolmer}, {Cano}, {Christensen},
  {Covino}, {D'Avanzo}, {D'Elia}, {De Cia}, {de Ugarte Postigo}, {Flores},
  {Friis}, {Gomboc}, {Greiner}, {Groot}, {Hammer}, {Hartoog}, {Heintz},
  {Hjorth}, {Jakobsson}, {Japelj}, {Kann}, {Kaper}, {Ledoux}, {Leloudas},
  {Levan}, {Maiorano}, {Melandri}, {Milvang-Jensen}, {Palazzi}, {Palmerio},
  {Perley}, {Pian}, {Piranomonte}, {Pugliese}, {S{\'a}nchez-Ram{\'\i}rez},
  {Savaglio}, {Schady}, {Schulze}, {Sollerman}, {Sparre}, {Tagliaferri},
  {Tanvir}, {Th{\"o}ne}, {Vergani}, {Vreeswijk}, {Watson}, {Wiersema},
  {Wijers}, {Xu}, \& {Zafar}}]{Selsing2019}
{Selsing}, J., {Malesani}, D., {Goldoni}, P., {et~al.} 2019, \aap, 623, A92

\bibitem[{{Srianand} {et~al.}(2016){Srianand}, {Hussain}, {Noterdaeme},
  {Petitjean}, {Kr{\"u}hler}, {Japelj}, {P{\^a}ris}, \&
  {Kashikawa}}]{srianand2016}
{Srianand}, R., {Hussain}, T., {Noterdaeme}, P., {et~al.} 2016, \mnras, 460,
  634

\bibitem[{Srianand {et~al.}(2005)Srianand, Petitjean, Ledoux, Ferland, \&
  Shaw}]{srianand2005}
Srianand, R., Petitjean, P., Ledoux, C., Ferland, G., \& Shaw, G. 2005, Monthly
  Notices of the Royal Astronomical Society, 362, 549

\bibitem[{{Sternberg} {et~al.}(2014){Sternberg}, {Le Petit}, {Roueff}, \& {Le
  Bourlot}}]{Sternberg2014}
{Sternberg}, A., {Le Petit}, F., {Roueff}, E., \& {Le Bourlot}, J. 2014, \apj,
  790, 10

\bibitem[{{Th{\"o}ne} {et~al.}(2013){Th{\"o}ne}, {Fynbo}, {Goldoni}, {de Ugarte
  Postigo}, {Campana}, {Vergani}, {Covino}, {Kr{\"u}hler}, {Kaper}, {Tanvir},
  {Zafar}, {D'Elia}, {Gorosabel}, {Greiner}, {Groot}, {Hammer}, {Jakobsson},
  {Klose}, {Levan}, {Milvang-Jensen}, {Nicuesa}, {Palazzi}, {Piranomonte},
  {Tagliaferri}, {Watson}, {Wiersema}, \& {Wijers}}]{Thone2013}
{Th{\"o}ne}, C.~C., {Fynbo}, J.~P.~U., {Goldoni}, P., {et~al.} 2013, \mnras,
  428, 3590

\bibitem[{{Tumlinson} {et~al.}(2007){Tumlinson}, {Prochaska}, {Chen},
  {Dessauges-Zavadsky}, \& {Bloom}}]{Tumlinson2007}
{Tumlinson}, J., {Prochaska}, J.~X., {Chen}, H.-W., {Dessauges-Zavadsky}, M.,
  \& {Bloom}, J.~S. 2007, \apj, 668, 667

\bibitem[{{Vergani} {et~al.}(2015){Vergani}, {Salvaterra}, {Japelj}, {Le
  Floc'h}, {D'Avanzo}, {Fernandez-Soto}, {Kr{\"u}hler}, {Melandri}, {Boissier},
  {Covino}, {Puech}, {Greiner}, {Hunt}, {Perley}, {Petitjean}, {Vinci},
  {Hammer}, {Levan}, {Mannucci}, {Campana}, {Flores}, {Gomboc}, \&
  {Tagliaferri}}]{Vergani2015}
{Vergani}, S.~D., {Salvaterra}, R., {Japelj}, J., {et~al.} 2015, \aap, 581,
  A102

\bibitem[{{Vernet} {et~al.}(2011){Vernet}, {Dekker}, {D'Odorico}, {Kaper},
  {Kjaergaard}, {Hammer}, {Randich}, {Zerbi}, {Groot}, {Hjorth}, {Guinouard},
  {Navarro}, {Adolfse}, {Albers}, {Amans}, {Andersen}, {Andersen}, {Binetruy},
  {Bristow}, {Castillo}, {Chemla}, {Christensen}, {Conconi}, {Conzelmann},
  {Dam}, {de Caprio}, {de Ugarte Postigo}, {Delabre}, {di Marcantonio},
  {Downing}, {Elswijk}, {Finger}, {Fischer}, {Flores}, {Fran{\c c}ois},
  {Goldoni}, {Guglielmi}, {Haigron}, {Hanenburg}, {Hendriks}, {Horrobin},
  {Horville}, {Jessen}, {Kerber}, {Kern}, {Kiekebusch}, {Kleszcz}, {Klougart},
  {Kragt}, {Larsen}, {Lizon}, {Lucuix}, {Mainieri}, {Manuputy}, {Martayan},
  {Mason}, {Mazzoleni}, {Michaelsen}, {Modigliani}, {Moehler}, {M{\o}ller},
  {Norup S{\o}rensen}, {N{\o}rregaard}, {P{\'e}roux}, {Patat}, {Pena}, {Pragt},
  {Reinero}, {Rigal}, {Riva}, {Roelfsema}, {Royer}, {Sacco}, {Santin},
  {Schoenmaker}, {Spano}, {Sweers}, {Ter Horst}, {Tintori}, {Tromp}, {van
  Dael}, {van der Vliet}, {Venema}, {Vidali}, {Vinther}, {Vola}, {Winters},
  {Wistisen}, {Wulterkens}, \& {Zacchei}}]{Vernet2011}
{Vernet}, J., {Dekker}, H., {D'Odorico}, S., {et~al.} 2011, \aap, 536, A105

\bibitem[{{Warren} {et~al.}(2001){Warren}, {M{\o}ller}, {Fall}, \&
  {Jakobsen}}]{Warren2001}
{Warren}, S.~J., {M{\o}ller}, P., {Fall}, S.~M., \& {Jakobsen}, P. 2001,
  \mnras, 326, 759

\bibitem[{{Weatherley} {et~al.}(2005){Weatherley}, {Warren}, {M{\o}ller},
  {Fall}, {Fynbo}, \& {Croom}}]{Weatherley+05}
{Weatherley}, S.~J., {Warren}, S.~J., {M{\o}ller}, P., {et~al.} 2005, \mnras,
  358, 985

\bibitem[{Wolfe {et~al.}(2005)Wolfe, Gawiser, \& Prochaska}]{wolfe2005}
Wolfe, A.~M., Gawiser, E., \& Prochaska, J.~X. 2005, Annu. Rev. Astron.
  Astrophys., 43, 861

\bibitem[{{Woosley} \& {Bloom}(2006)}]{Woosley2006}
{Woosley}, S.~E. \& {Bloom}, J.~S. 2006, \araa, 44, 507

\bibitem[{{Yajima} {et~al.}(2012){Yajima}, {Choi}, \& {Nagamine}}]{Yajima2012}
{Yajima}, H., {Choi}, J.-H., \& {Nagamine}, K. 2012, \mnras, 427, 2889

\bibitem[{{York} {et~al.}(2000){York}, {Adelman}, {Anderson}, {Anderson},
  {Annis}, {Bahcall}, {Bakken}, {Barkhouser}, {Bastian}, {Berman}, {Boroski},
  {Bracker}, {Briegel}, {Briggs}, {Brinkmann}, {Brunner}, {Burles}, {Carey},
  {Carr}, {Castander}, {Chen}, {Colestock}, {Connolly}, {Crocker}, {Csabai},
  {Czarapata}, {Davis}, {Doi}, {Dombeck}, {Eisenstein}, {Ellman}, {Elms},
  {Evans}, {Fan}, {Federwitz}, {Fiscelli}, {Friedman}, {Frieman}, {Fukugita},
  {Gillespie}, {Gunn}, {Gurbani}, {de Haas}, {Haldeman}, {Harris}, {Hayes},
  {Heckman}, {Hennessy}, {Hindsley}, {Holm}, {Holmgren}, {Huang}, {Hull},
  {Husby}, {Ichikawa}, {Ichikawa}, {Ivezi{\'c}}, {Kent}, {Kim}, {Kinney},
  {Klaene}, {Kleinman}, {Kleinman}, {Knapp}, {Korienek}, {Kron}, {Kunszt},
  {Lamb}, {Lee}, {Leger}, {Limmongkol}, {Lindenmeyer}, {Long}, {Loomis},
  {Loveday}, {Lucinio}, {Lupton}, {MacKinnon}, {Mannery}, {Mantsch}, {Margon},
  {McGehee}, {McKay}, {Meiksin}, {Merelli}, {Monet}, {Munn}, {Narayanan},
  {Nash}, {Neilsen}, {Neswold}, {Newberg}, {Nichol}, {Nicinski}, {Nonino},
  {Okada}, {Okamura}, {Ostriker}, {Owen}, {Pauls}, {Peoples}, {Peterson},
  {Petravick}, {Pier}, {Pope}, {Pordes}, {Prosapio}, {Rechenmacher}, {Quinn},
  {Richards}, {Richmond}, {Rivetta}, {Rockosi}, {Ruthmansdorfer}, {Sandford},
  {Schlegel}, {Schneider}, {Sekiguchi}, {Sergey}, {Shimasaku}, {Siegmund},
  {Smee}, {Smith}, {Snedden}, {Stone}, {Stoughton}, {Strauss}, {Stubbs},
  {SubbaRao}, {Szalay}, {Szapudi}, {Szokoly}, {Thakar}, {Tremonti}, {Tucker},
  {Uomoto}, {Vanden Berk}, {Vogeley}, {Waddell}, {Wang}, {Watanabe},
  {Weinberg}, {Yanny}, {Yasuda}, \& {SDSS Collaboration}}]{York2000}
{York}, D.~G., {Adelman}, J., {Anderson}, Jr., J.~E., {et~al.} 2000, \aj, 120,
  1579

\bibitem[{{Zafar} {et~al.}(2017){Zafar}, {M{\o}ller}, {P{\'e}roux}, {Quiret},
  {Fynbo}, {Ledoux}, \& {Deharveng}}]{Zafar2017}
{Zafar}, T., {M{\o}ller}, P., {P{\'e}roux}, C., {et~al.} 2017, \mnras, 465,
  1613

\bibitem[{{Zou} {et~al.}(2018){Zou}, {Petitjean}, {Noterdaeme}, {Ledoux},
  {Krogager}, {Fathivavsari}, {Srianand}, \& {L{\'o}pez}}]{Zou2018}
{Zou}, S., {Petitjean}, P., {Noterdaeme}, P., {et~al.} 2018, \aap, 616, A158

\end{thebibliography}

\begin{appendix}

\section{Details about column density measurements\label{comments_individual} }

\begin{figure}
    \centering
    \includegraphics[trim=0 0 0 0,clip,width=\hsize]{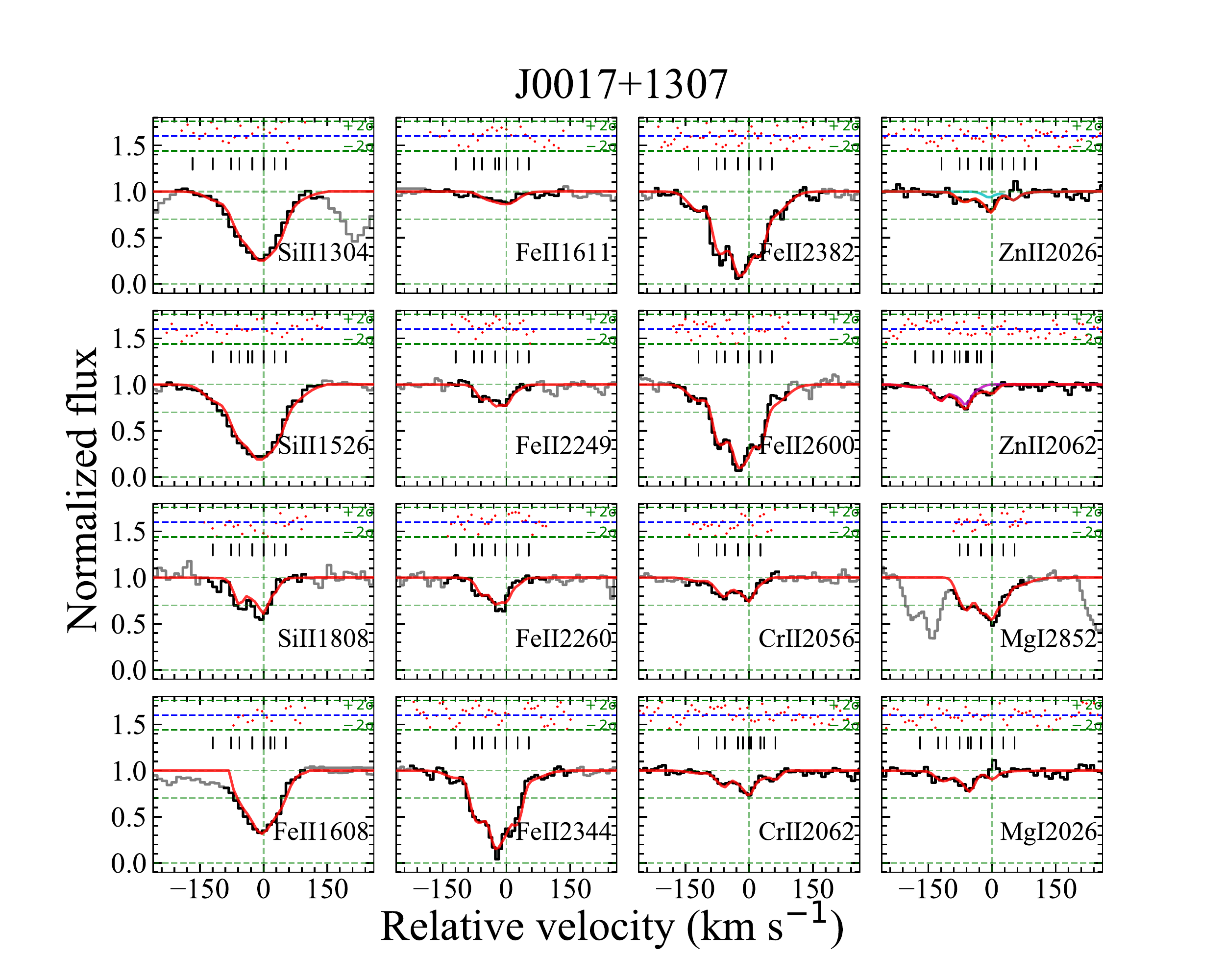}
    \caption{Low-ionisation metal lines associated to the $\zabs=2.326$ system towards J0017+1307. The normalised X-shooter 
    spectrum is shown in grey with the best-fit multi-component Voigt profile over-plotted in red. Dark points show the 
    data actually constraining the fit. Short vertical marks show the location of the different velocity components. The panels showing \ion{Zn}{II}$\lambda$2026 and \ion{Zn}{II}$\lambda$2062 are overplotted with cyan and magenta line showing contributions from \ion{Mg}{I}$\lambda$2026 and \ion{Cr}{II}$\lambda$2062 respectively. Finally, the 
    residuals are shown above each line in units of the standard deviation (from the error spectrum). 
    }
    \label{J0017met}
\end{figure}

\begin{figure}
    \centering
    \includegraphics[trim=0 0 0 0,clip,width=\hsize]{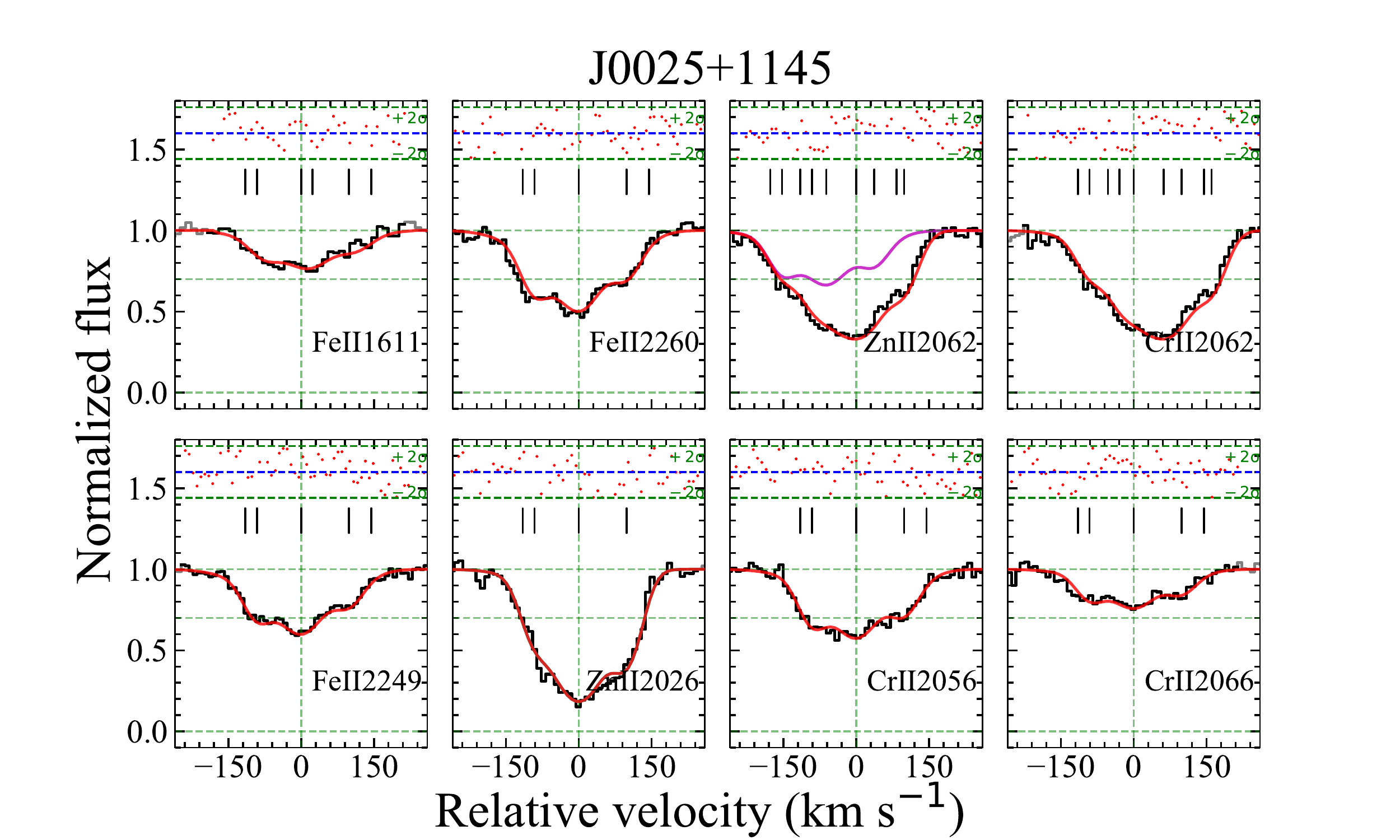}
    \caption{Same as Fig.~\ref{J0017met} for the $\zabs=2.304$ system towards J0025+1145.}
    \label{J0025met}
\end{figure}

\begin{figure}
    \centering
    \includegraphics[trim=0 0 0 0,clip,width=\hsize]{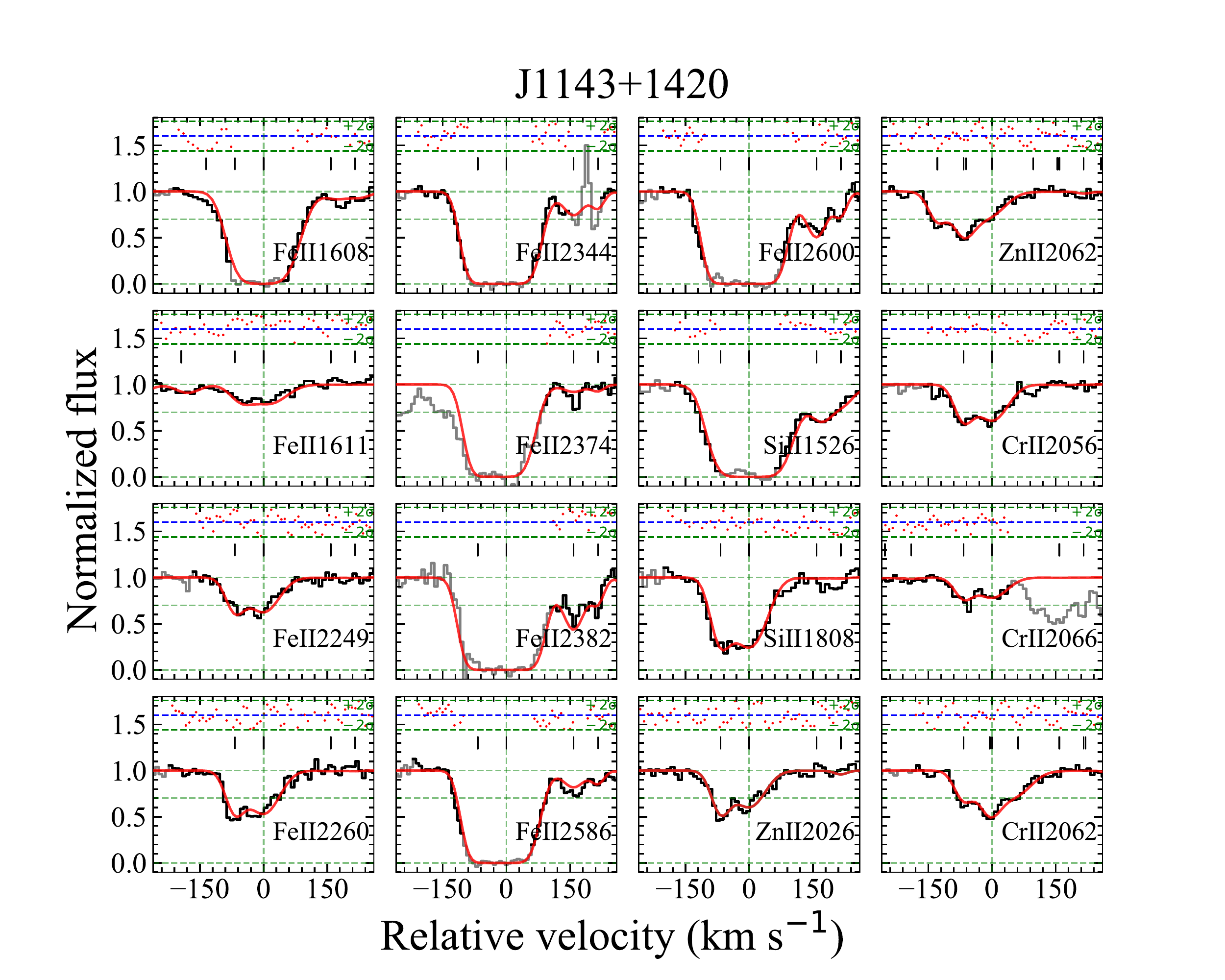}
    \caption{Same as Fig.~\ref{J0017met} for the $\zabs=2.323$ system towards J1143+1420. }
    \label{J1143met}
\end{figure}

\begin{figure}
    \centering
    \includegraphics[trim=0 0 0 0,clip,width=\hsize]{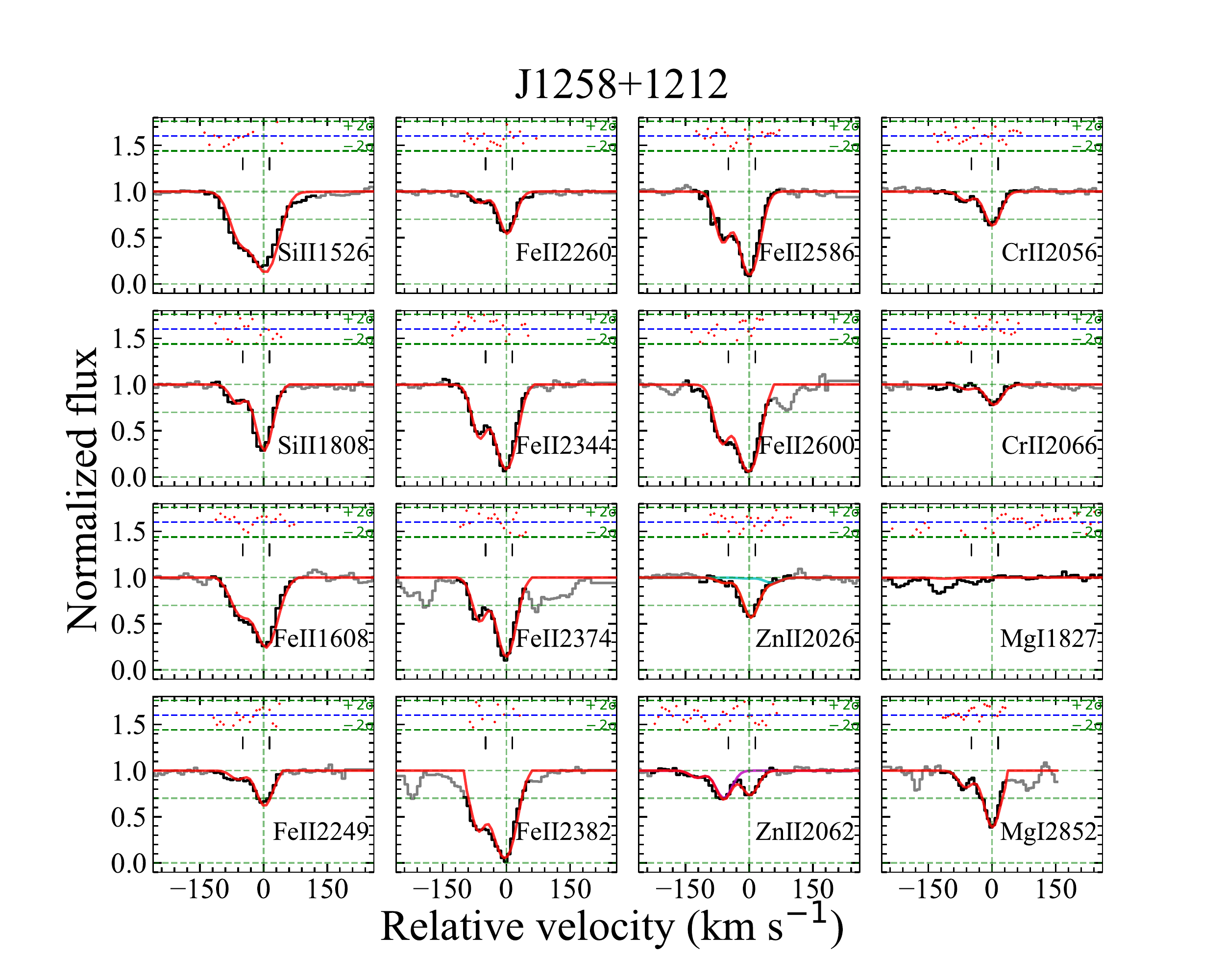}
    \caption{Same as Fig.~\ref{J0017met} for the $\zabs=2.444$ system towards J1258+1212.}
    \label{J1258met}
\end{figure}

\begin{figure}
    \centering
    \includegraphics[trim=0 0 0 0,clip,width=\hsize]{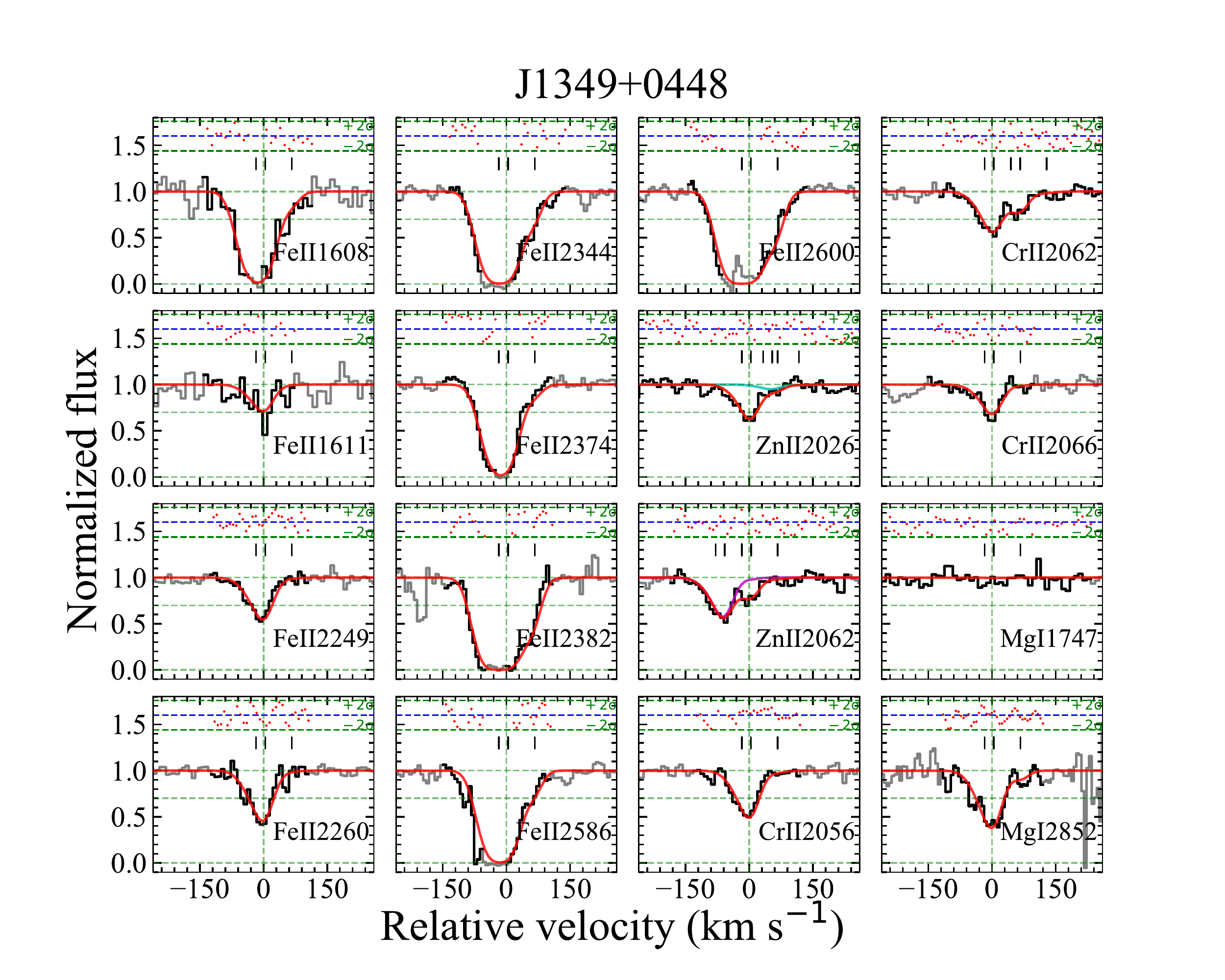}
    \caption{Same as Fig.~\ref{J0017met} for the $\zabs=2.482$ system towards J1349+0448.}
    \label{J1349met}
\end{figure}

\begin{figure}
    \centering
    \includegraphics[trim=0 0 0 0,clip,width=\hsize]{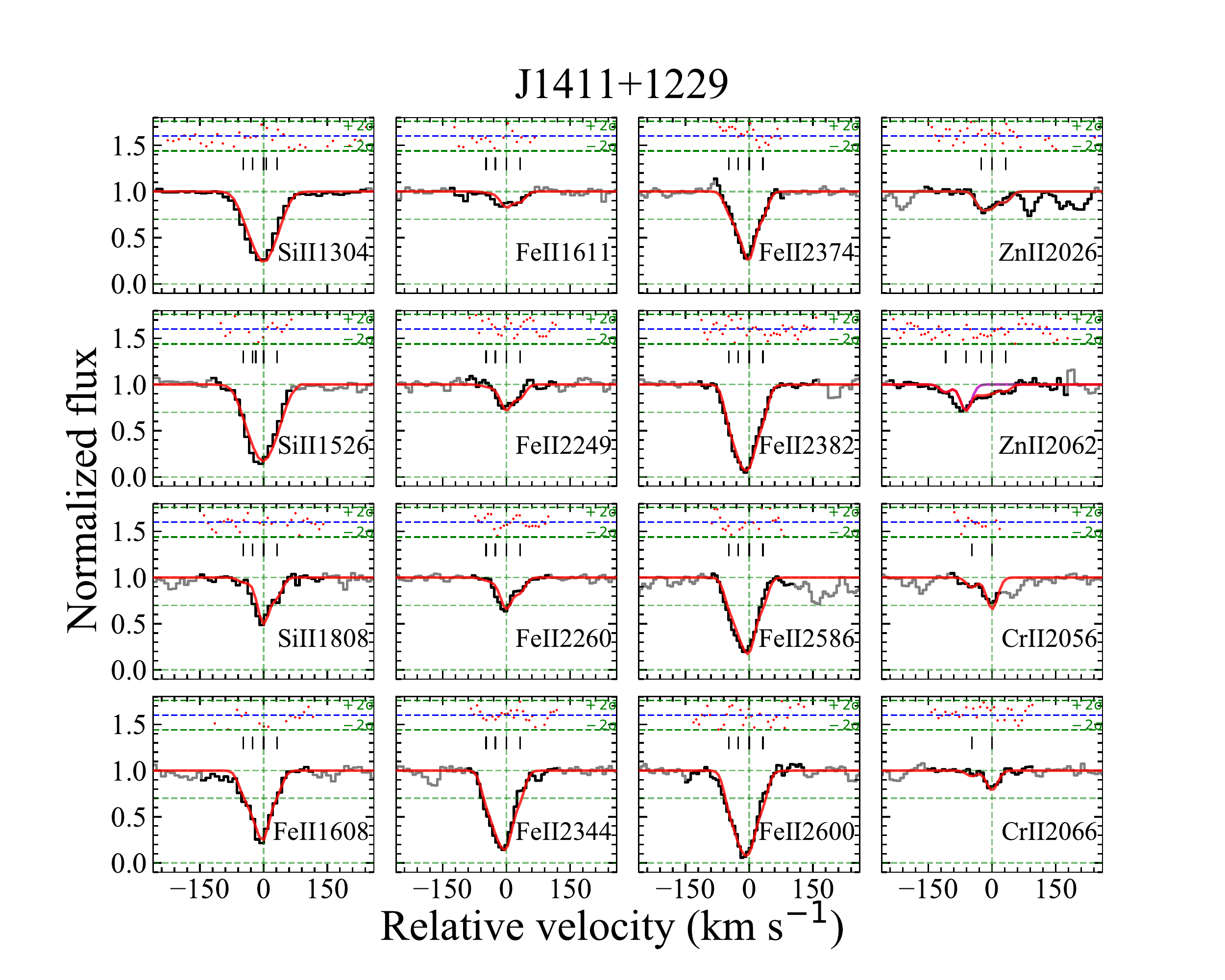}
    \caption{Same as Fig.~\ref{J0017met} for the $\zabs=2.545$ system towards J1411+1229.}
    \label{J1411met}
\end{figure}

\begin{figure}
    \centering
    \includegraphics[trim=0 0 0 0,clip,width=\hsize]{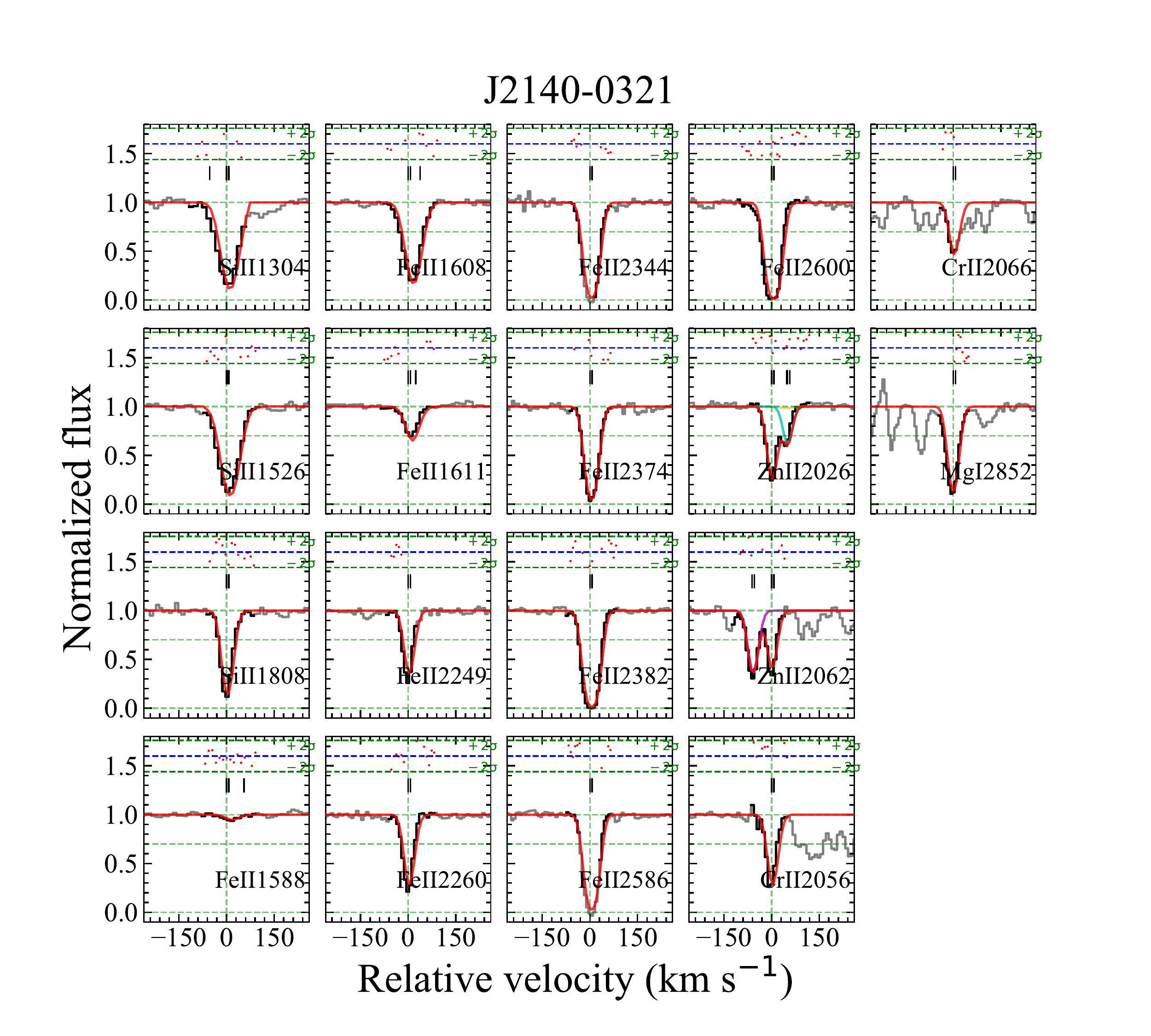}
    \caption{Same as Fig.~\ref{J0017met} for the $\zabs=2.339$ system towards J2140$-$0321.}
    \label{J2140met}
\end{figure}

\begin{figure}
    \centering
    \includegraphics[trim=0 0 0 0,clip,width=\hsize]{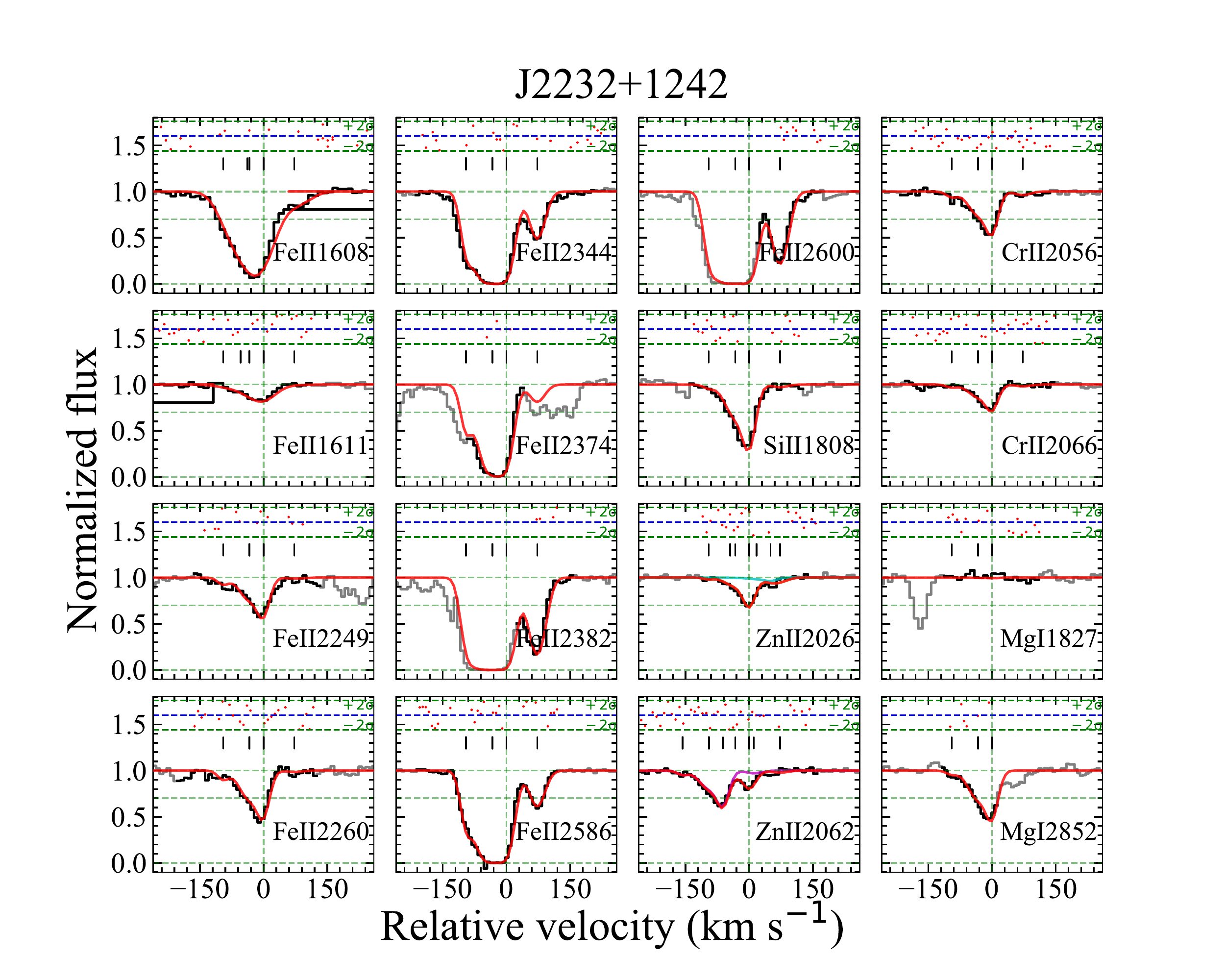}
    \caption{Same as Fig.~\ref{J0017met} for the $\zabs=2.230$ system towards J2232+1242.}
    \label{J2232met}
\end{figure}

\begin{figure}
    \centering
    \includegraphics[trim=0 0 0 0,clip,width=\hsize]{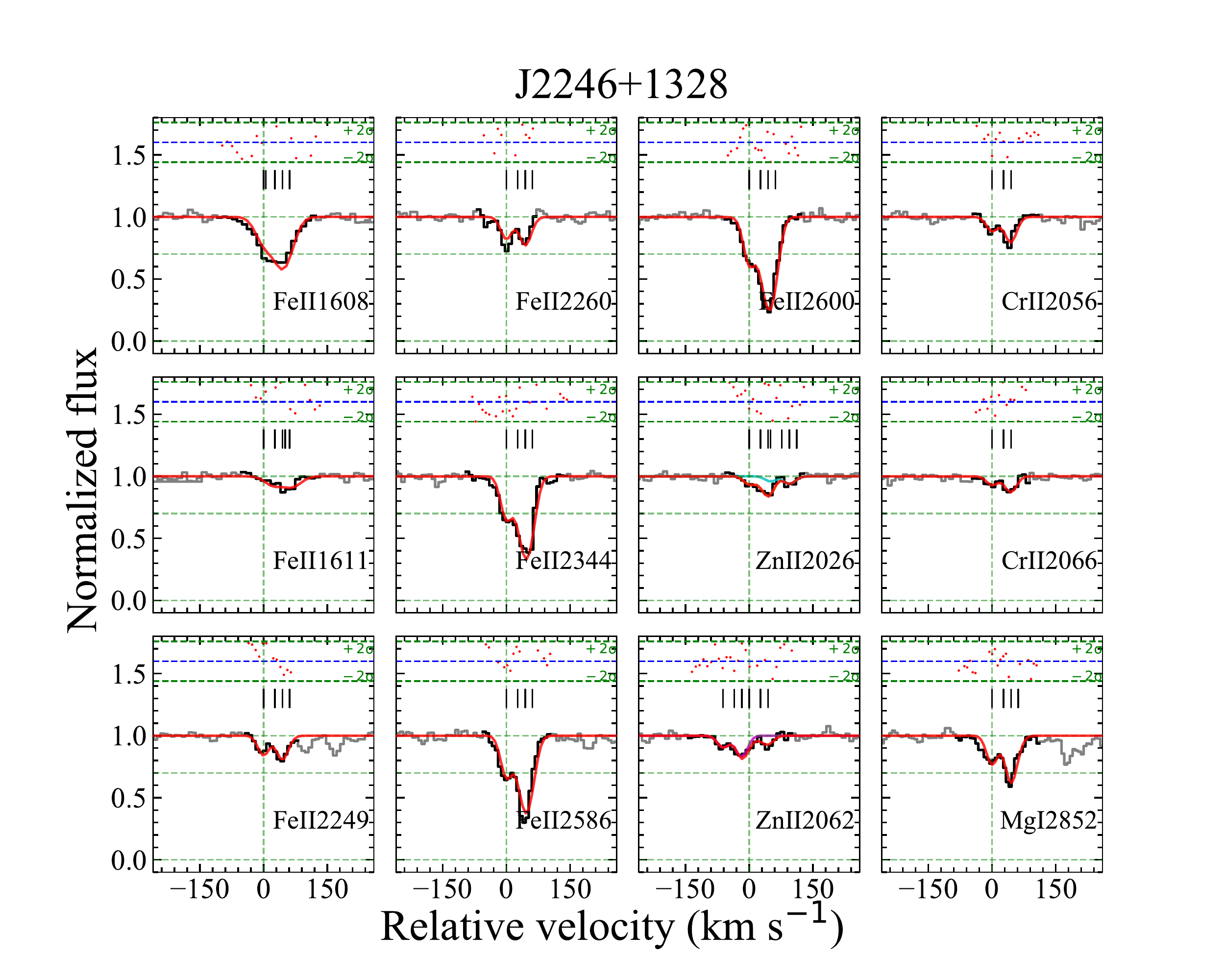}
    \caption{Same as Fig.~\ref{J0017met} for the $\zabs=2.215$ system towards J2246+1328.}
    \label{J2246met}
\end{figure}

\begin{figure}
    \centering
    \includegraphics[trim=0 0 0 0,clip,width=\hsize]{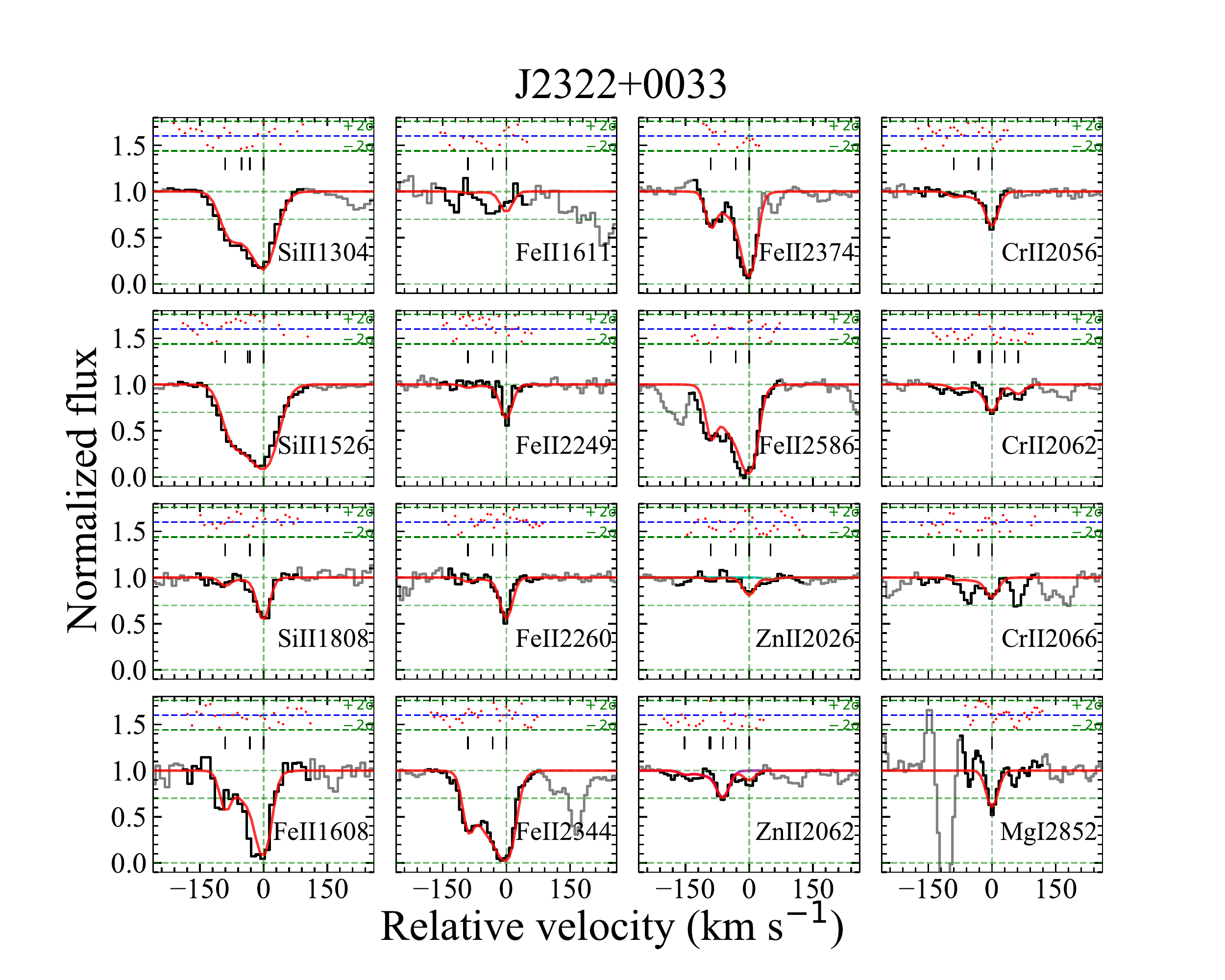}
    \caption{Same as Fig.~\ref{J0017met} for the $\zabs=2.477$ system towards J2322+0033.}
    \label{J2322met}
\end{figure}

\begin{figure}
  \centering
    \includegraphics[width=0.8\hsize]{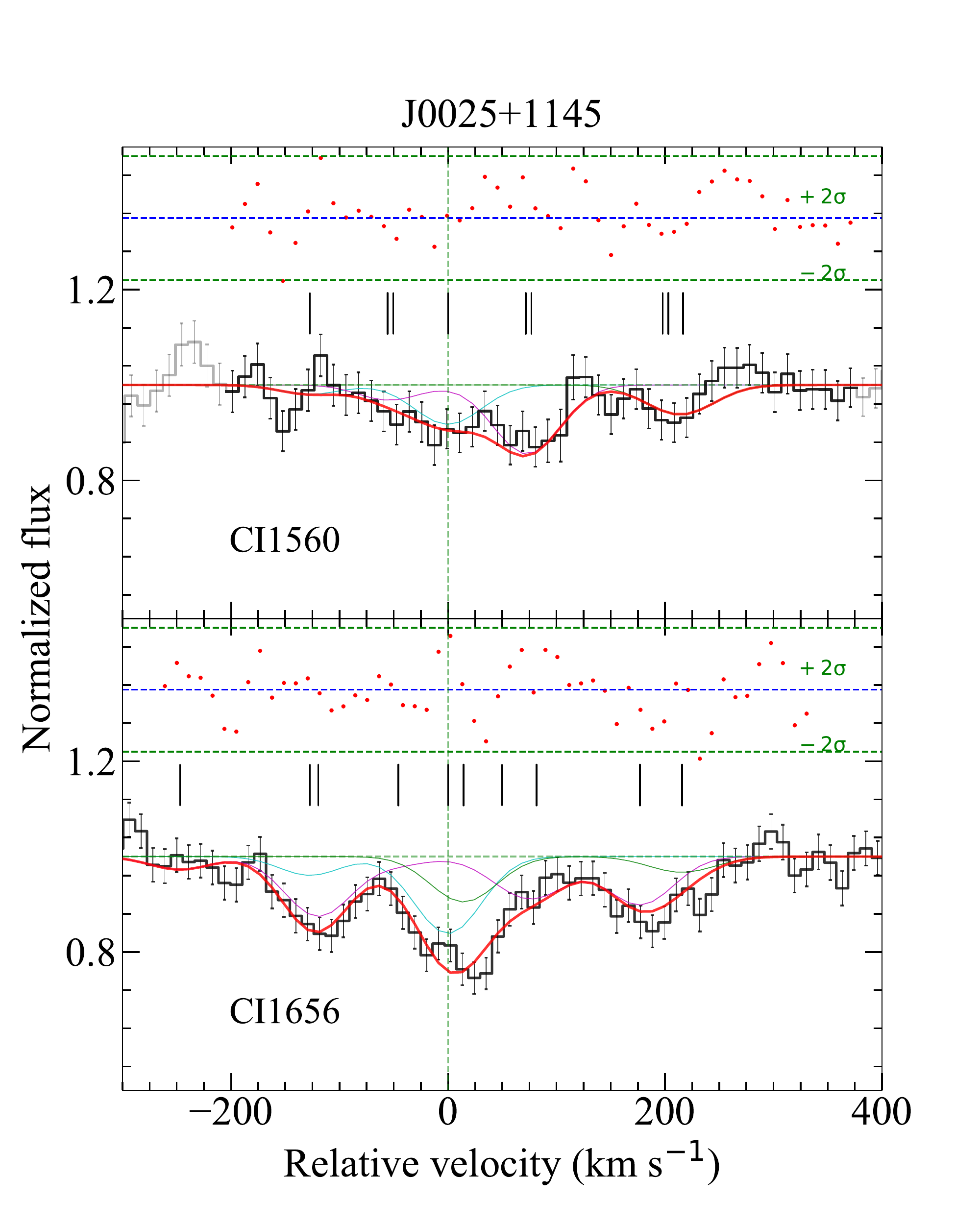}
  \caption{\ion{C}{I} lines associated with the ESDLA at $\zabs=2.304$ towards J\,0025$+$1145. The data is shown in gray and the total fit in red, with contribution from the 
  fine-structure levels $J=0$ (\ion{C}{i}), $J=1$ (\ion{C}{i}*) and $J=2$ (\ion{C}{i}**) displayed with cyan, magenta, and green, respectively.
  \label{J0025CI}}
\end{figure}

\begin{figure}
  \centering
    \includegraphics[width=0.8\hsize]{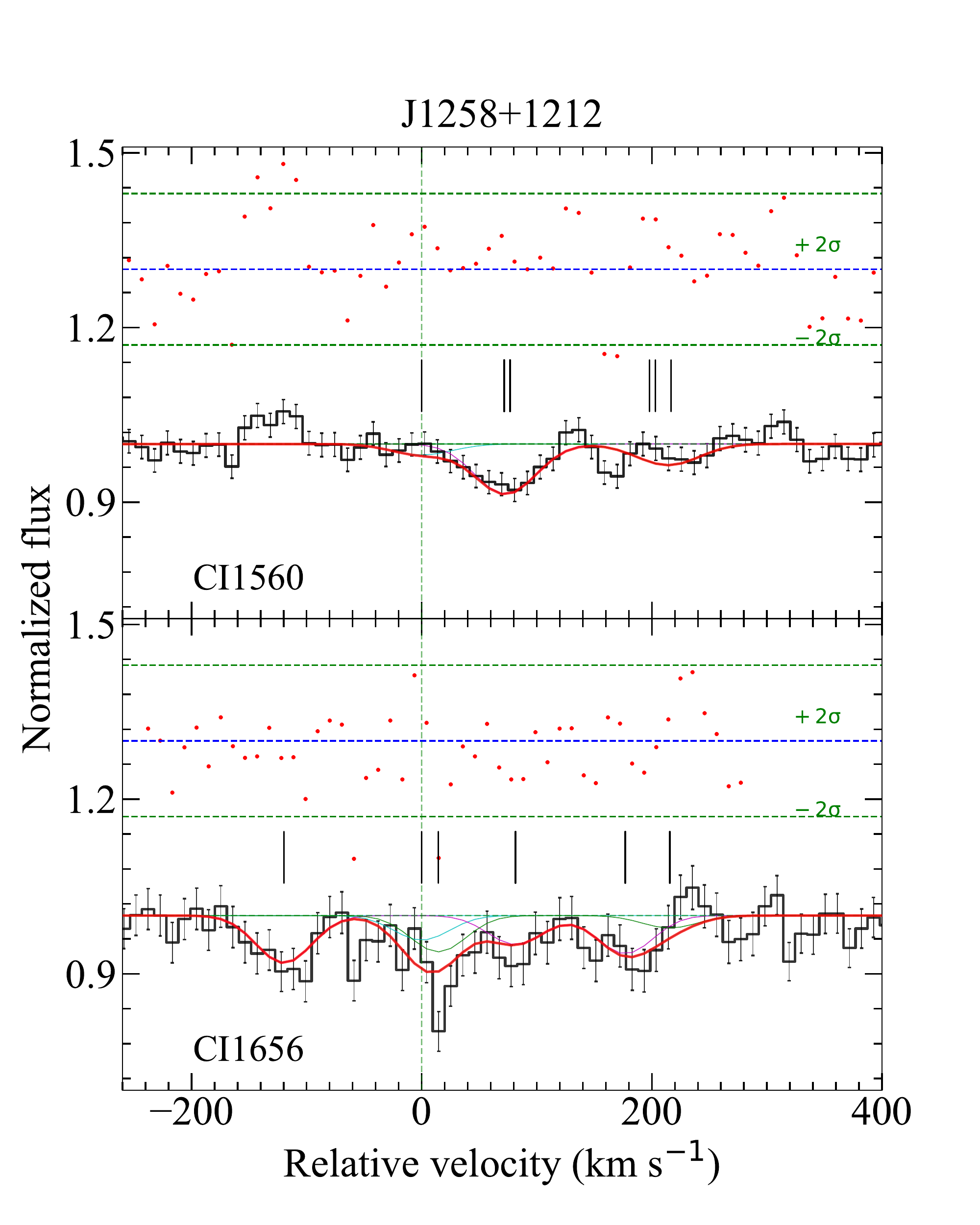}
  \caption{Same as Fig.~\ref{J0025CI} for the $\zabs=2.444$ towards J\,1258$+$1212}
  \label{J1258CI}
\end{figure}

\begin{figure}
  \centering
    \includegraphics[trim=0 50 0 50,clip,width=0.8\hsize]{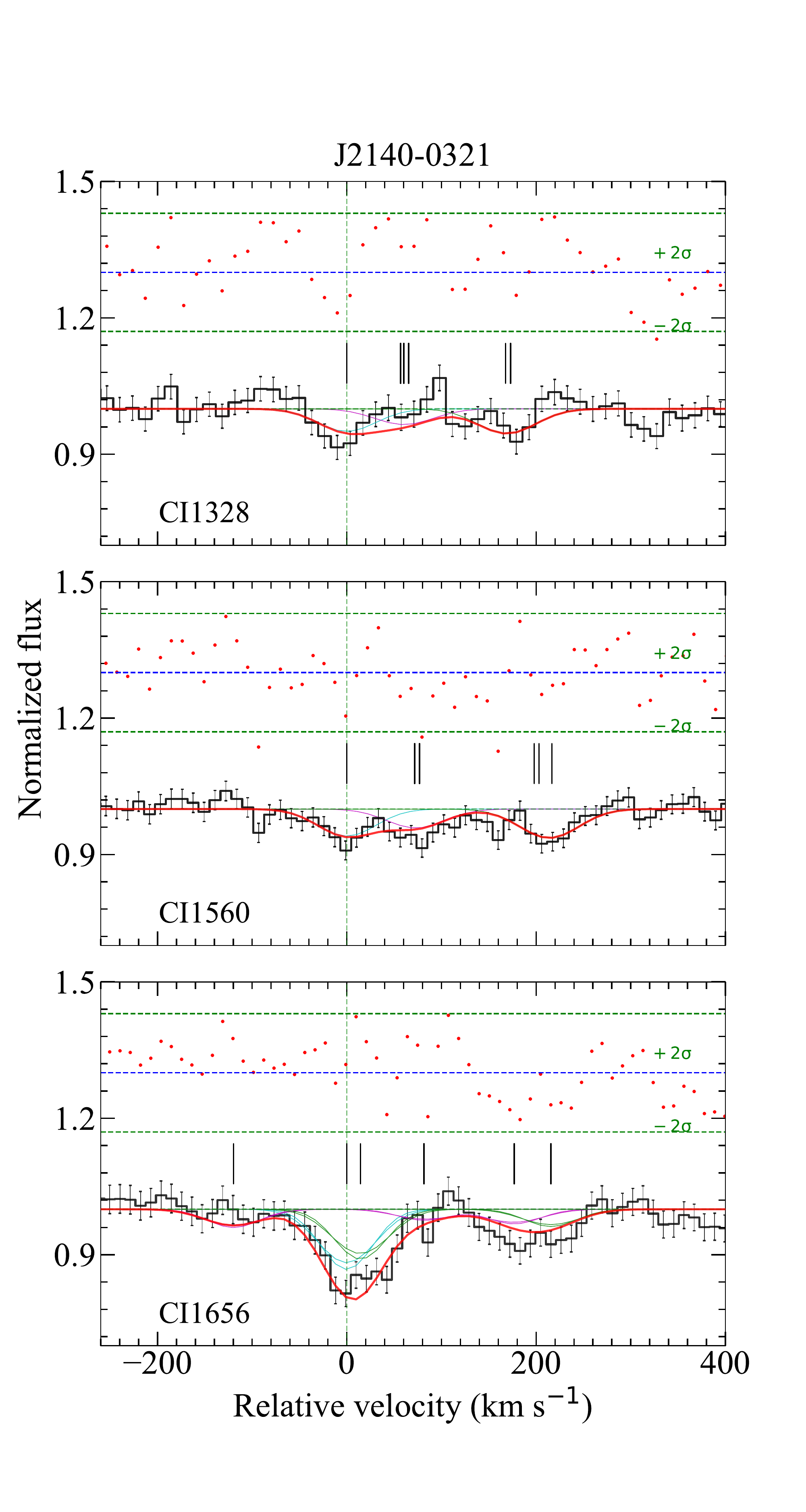}
  \caption{Same as Fig.~\ref{J0025CI} for the $\zabs=2.339$ towards J\,2140$-$0321.}
  \label{J2140CI}
\end{figure}

\begin{figure}
\centering
   \includegraphics[width = \hsize]{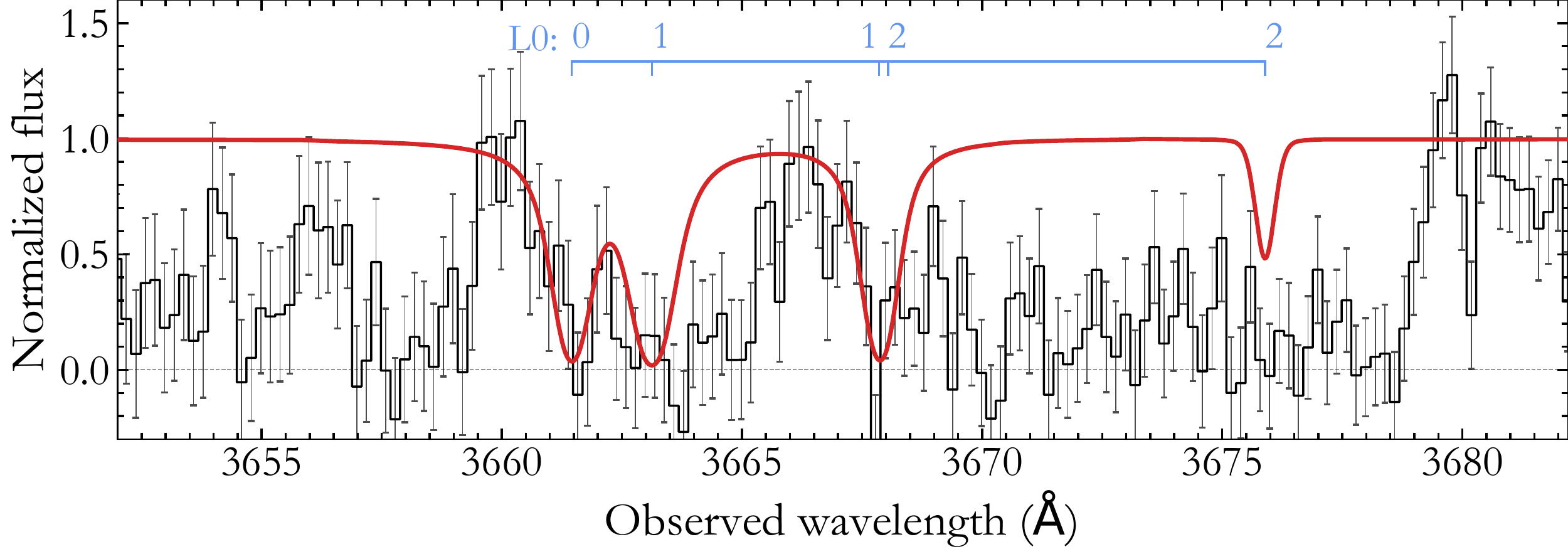}
      \caption{A portion of the X-shooter spectrum of J\,0025$+$1145 covering the L0-band absorption lines of H$_2$ from the DLA at $z=2.304(2)$. The rotational levels J are indicated above each blue tick mark. The normalised spectrum is shown in black and the synthetic profile with total H$_2$ column density $\log N (\rm H_2) = 20.1$ is overplotted in red.  \label{J0025+1145_H2_L0}}
\end{figure}

\begin{figure*}
\centering
   \includegraphics[trim=50 10 50 10,clip,width = \hsize]{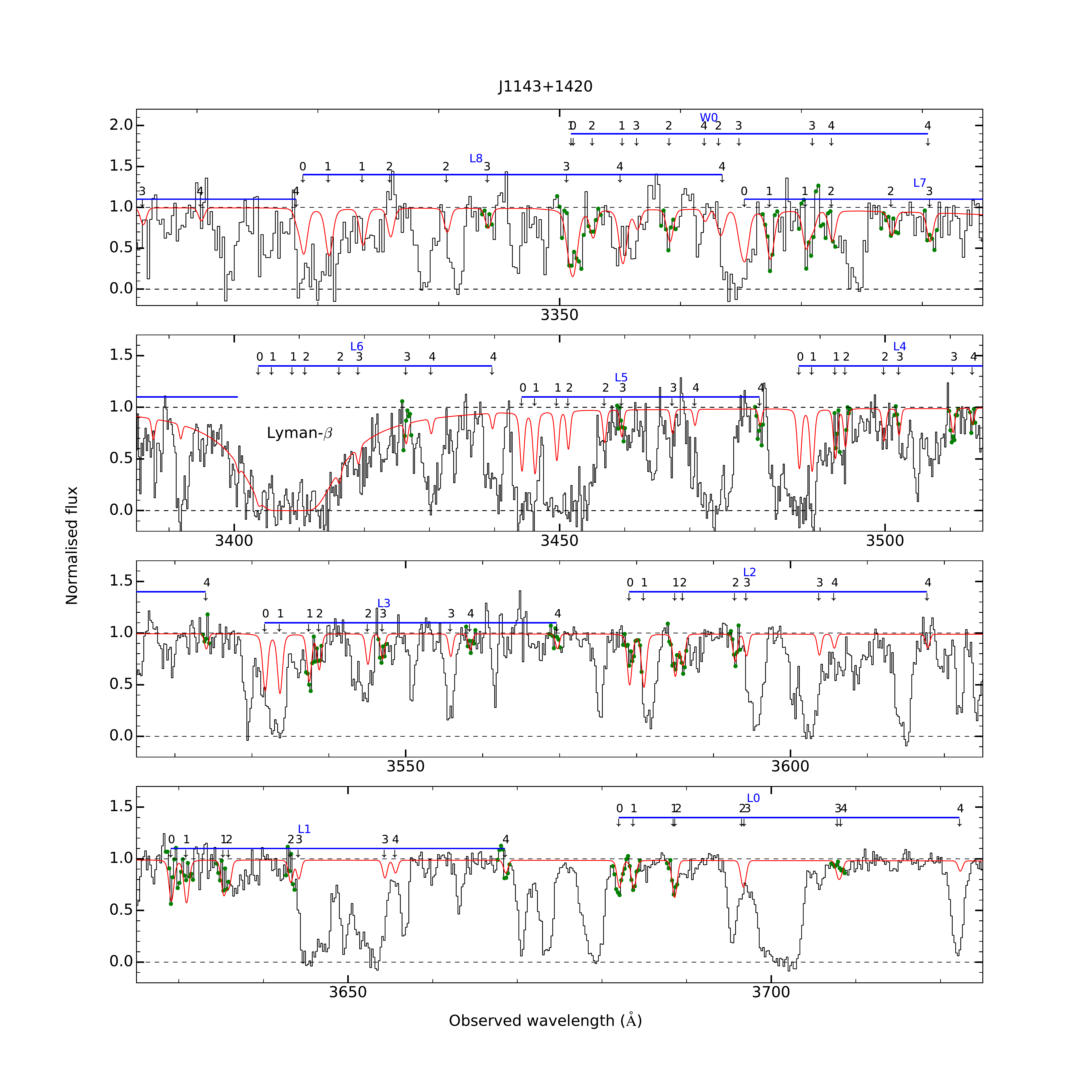}
      \caption{
      Portion of the X-shooter spectrum of J\,1143$+$1420 covering the absorption lines of H$\rm _2$ at $z=2.3225$. The normalised data are shown in black and the synthetic profile is overplotted in red. Horizontal blue segments connect rotational levels (short downwards arrows) from a given Lyman (L) or Werner (W) band, as labelled above them. The data selected for fitting are highlighted as green points. Note that the strong absorption at $\sim$3410~{\AA} is actually due to \HI\ Ly\,$\beta$. \label{J1143+1420_H2}}
\end{figure*}

\begin{figure*}
\centering
   \includegraphics[trim=0 0 0 0,clip,width=\hsize]{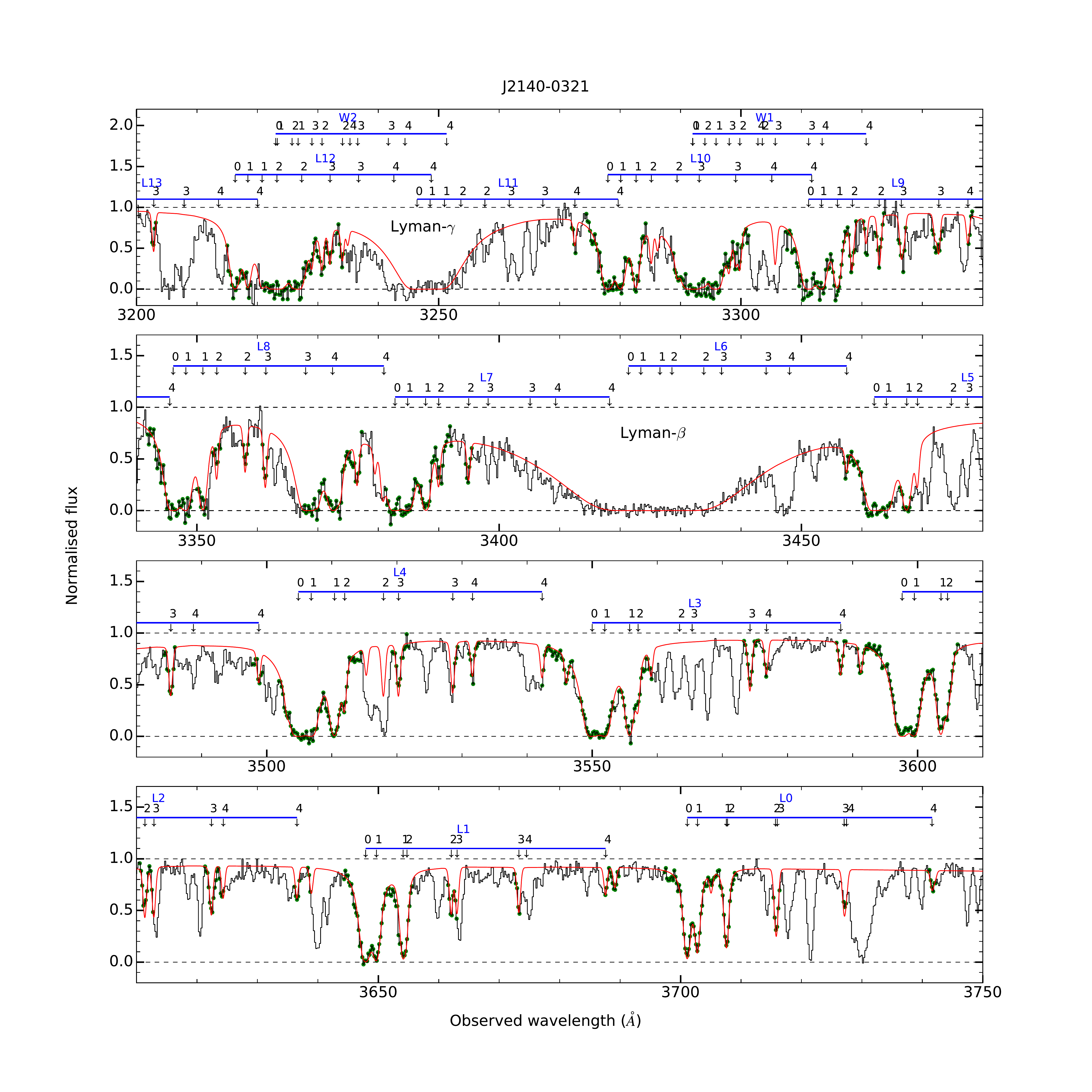}
\caption{Same as Fig.~\ref{J1143+1420_H2} for the $\zabs=2.3399$ \HH-bearing ESDLA towards J2140$-$0321. Note that the strong absorption at $\sim$3250~{\AA} and $\sim$3430~{\AA} are due to \HI\ Ly\,$\gamma$ and Ly\,$\beta$ respectively. \label{J2140-0321_H2}}
\end{figure*}

\begin{figure*}
\centering
   \includegraphics[trim=0 0 0 0,clip,width=\hsize]{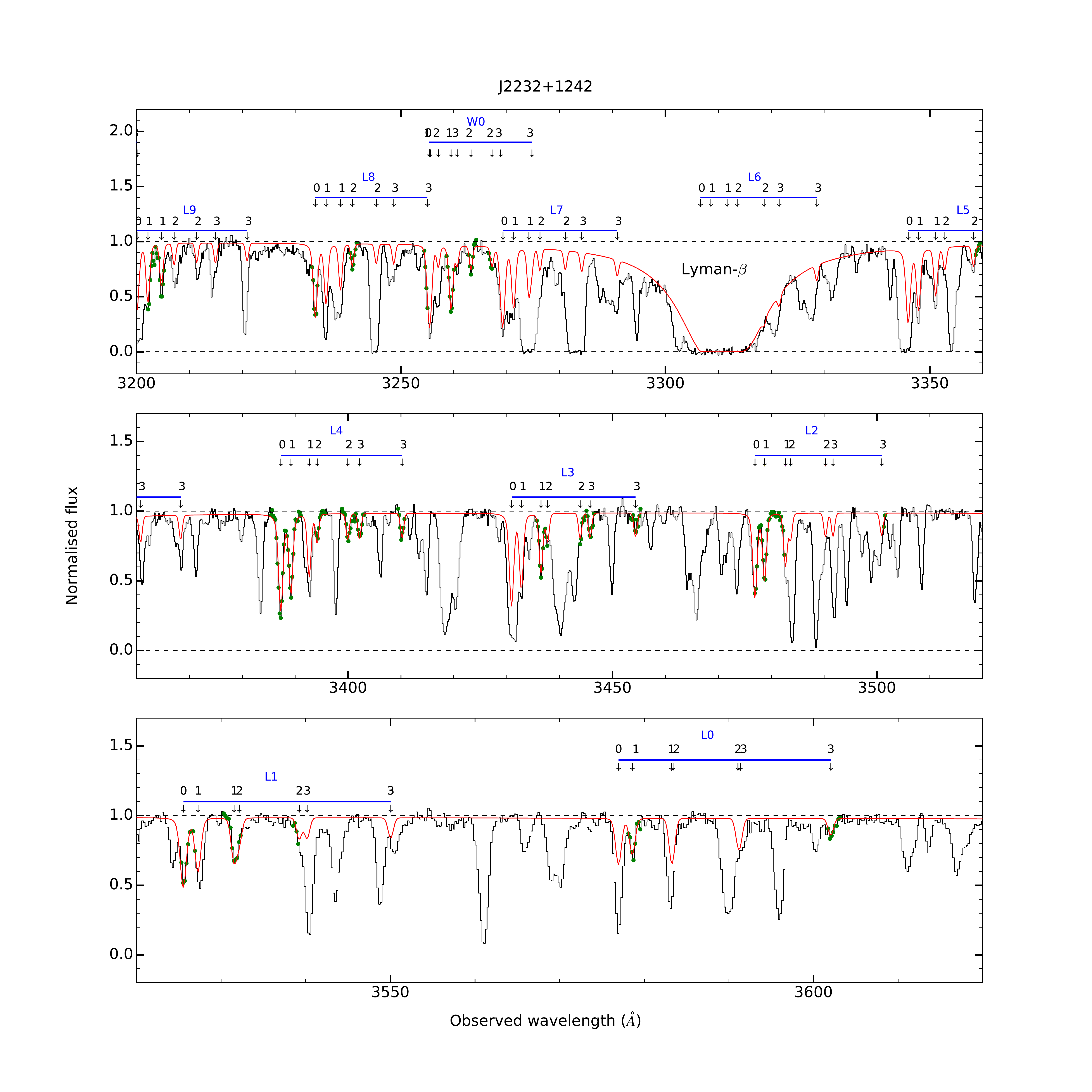}
\caption{Same as Fig.~\ref{J1143+1420_H2} for the $\zabs=2.2279$ \HH-bearing ESDLA towards J2232+1242. Note that the strong absorption at $\sim$3310~{\AA} is actually due to \HI\ Ly\,$\beta$. \label{J2232+1242_H2}}
\end{figure*}

\begin{figure}
    \centering
    \includegraphics[angle=90, width = 1.0\hsize]{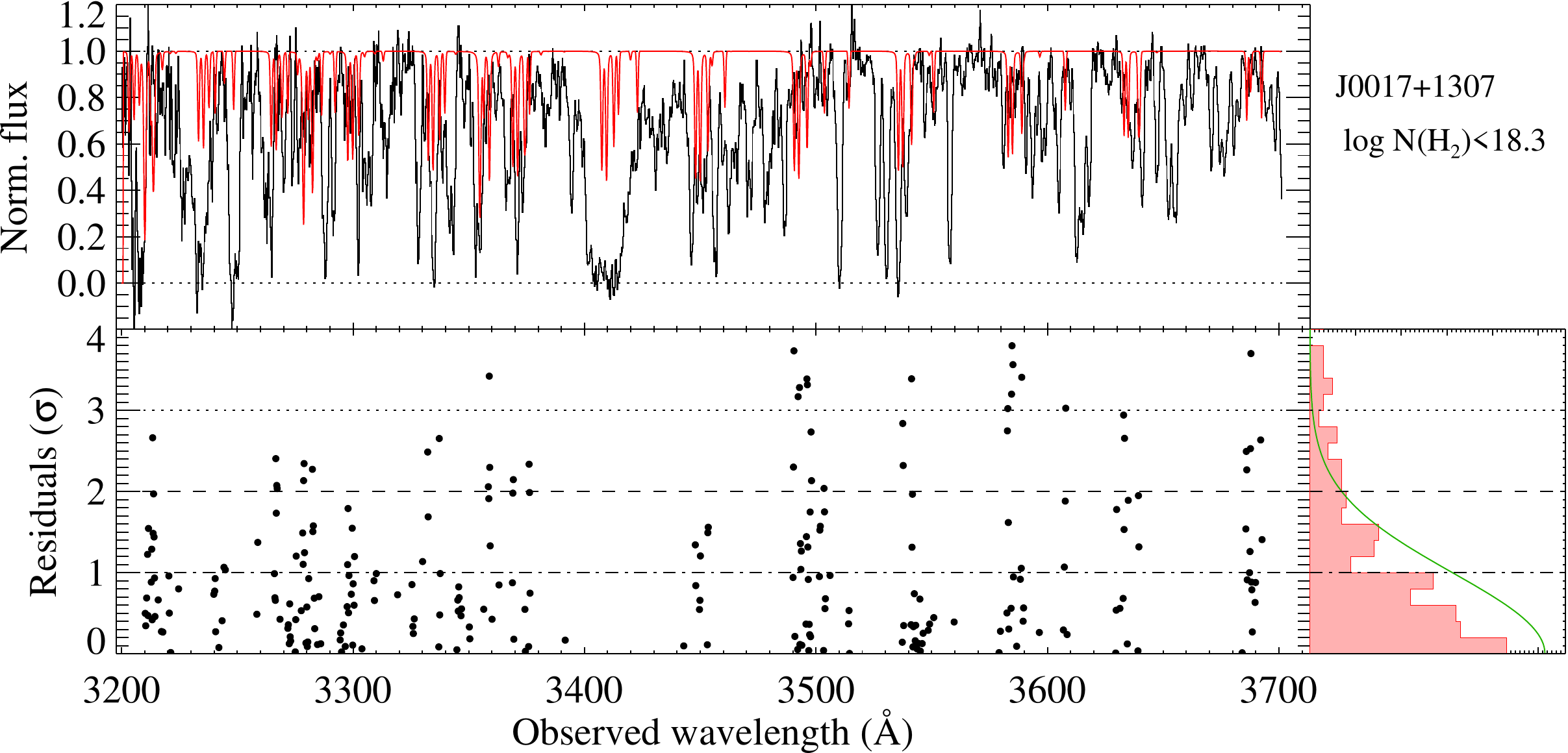}
    \caption{Top panel: Normalised spectrum of J0017+1307 over the absorber's H$_2$ Lyman-Werner band region (black) together with 
    the synthetic profile corresponding to the maximum $N(\HH)$ still consistent with the data. Bottom panel: Positive residuals  expressed in units of standard deviation as derived from the error spectrum, with the corresponding distribution 
    shown in the right panel (red histogram). The green line corresponds to the expected Gaussian distribution of the residuals.
    \label{limit:J0017}}
\end{figure}

\begin{figure}
    \centering
    \includegraphics[angle=90, width = 1.0\hsize]{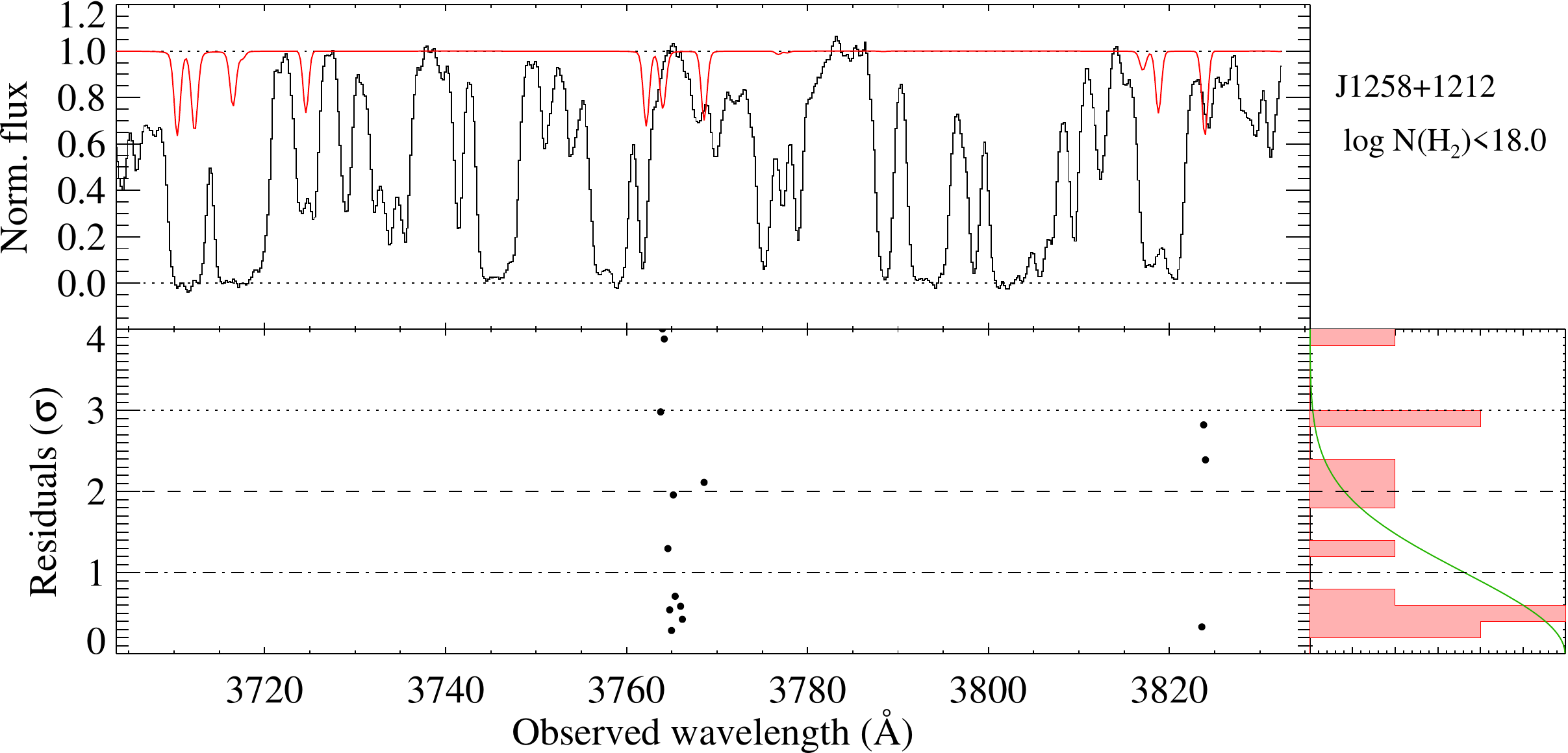}
    \caption{Same as \ref{limit:J0017} for the system towards J1258+1212. \label{limit:J1258}}
\end{figure}

\begin{figure}
    \centering
    \includegraphics[angle=90, width = 1.0\hsize]{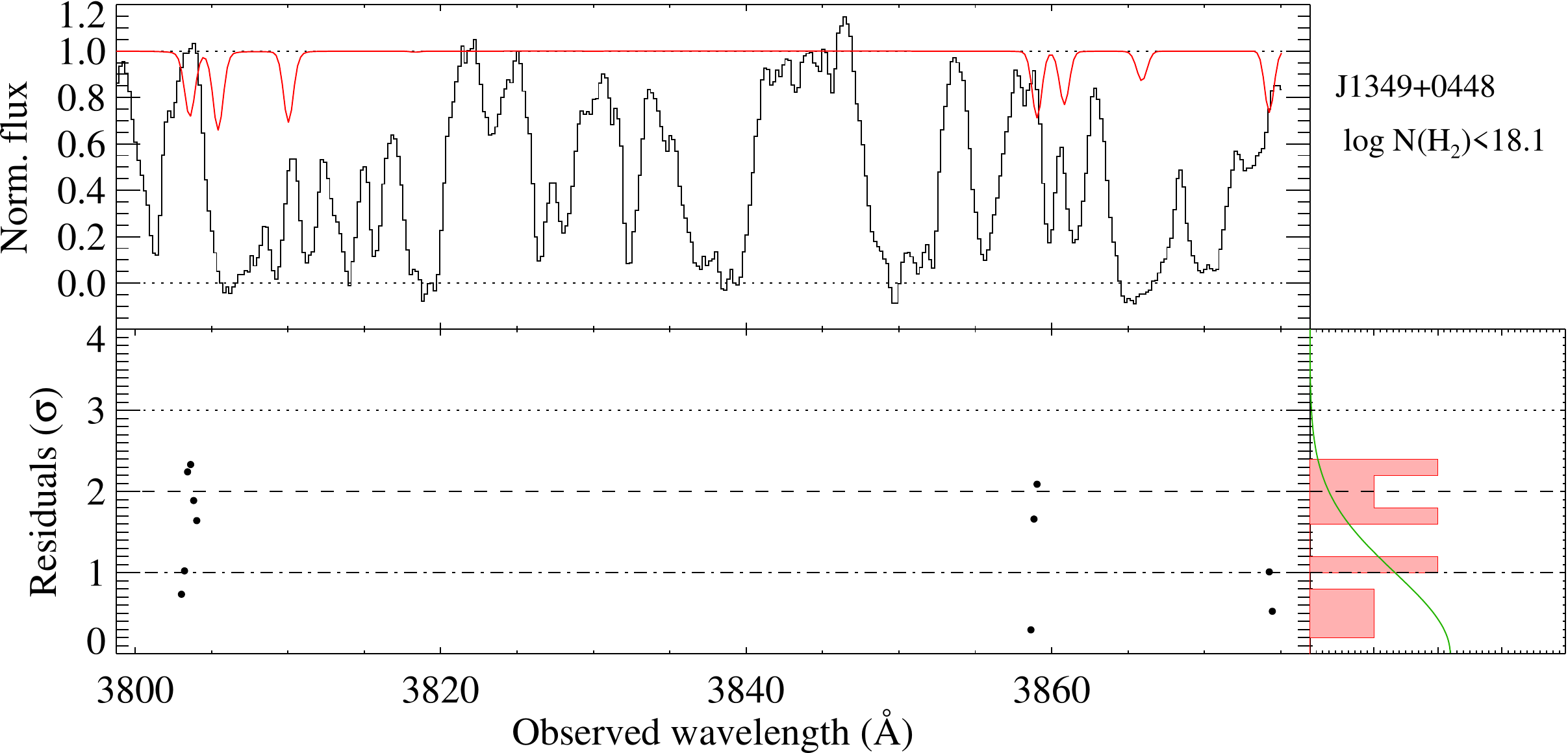}
        \caption{Same as \ref{limit:J0017} for the system towards J1349+0448. \label{limit:J1349}}
\end{figure}

\begin{figure}
    \centering
    \includegraphics[angle=90, width = 1.0\hsize]{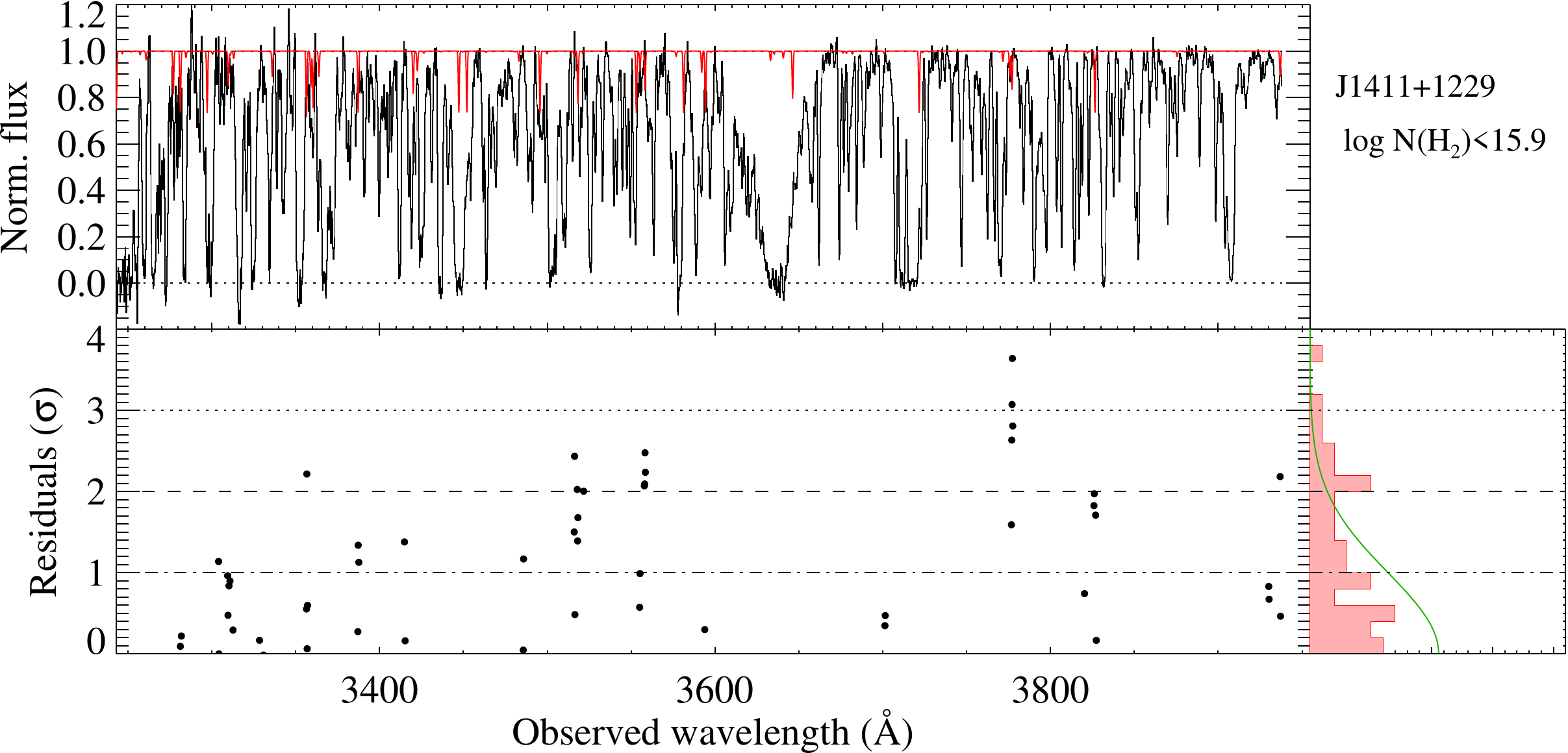}
    \caption{Same as \ref{limit:J0017} for the system towards J1411+1229. \label{limit:J1411}}
\end{figure}

\begin{figure}
    \centering
    \includegraphics[angle=90, width = 1.0\hsize]{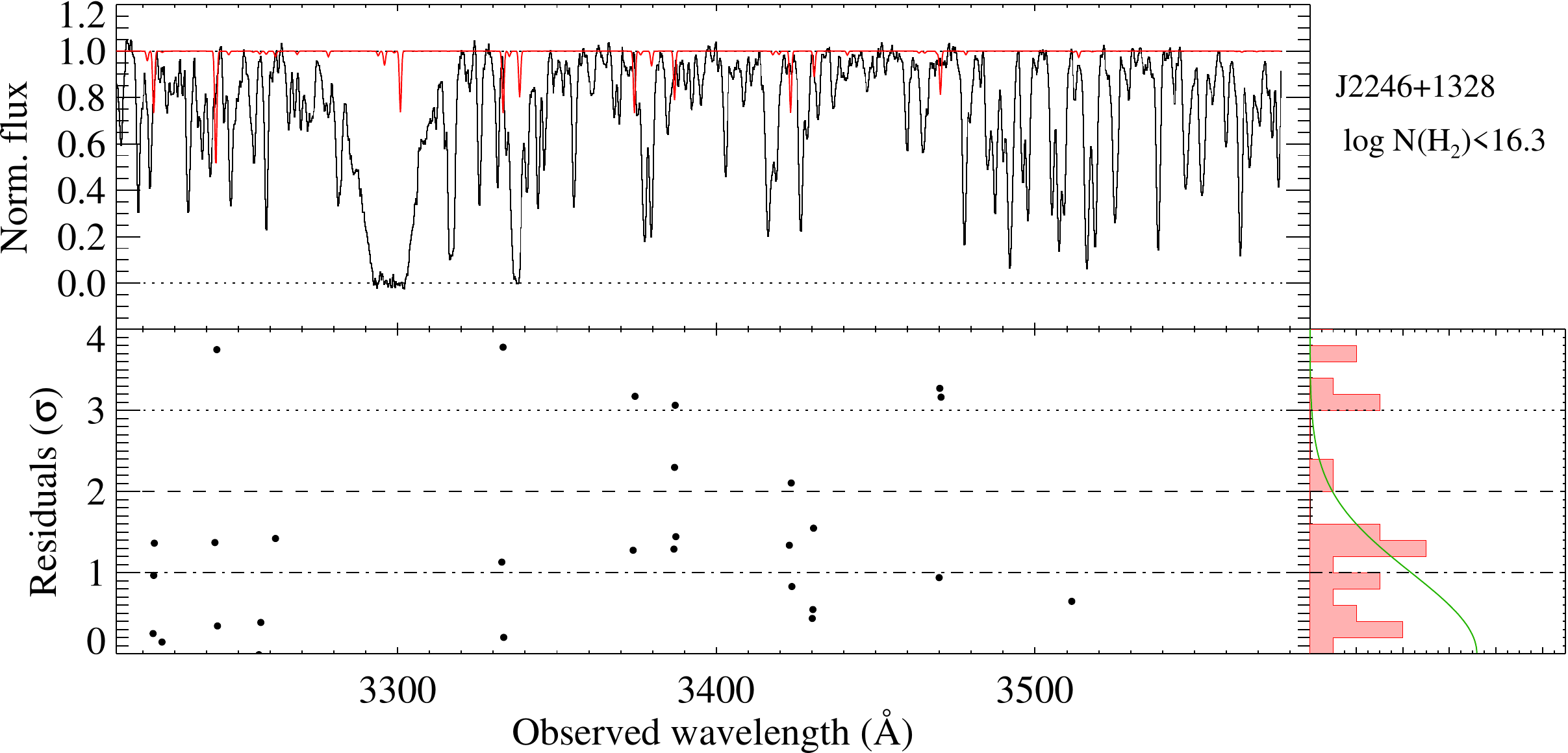}
    \caption{Same as \ref{limit:J0017} for the system towards J2246+1328. \label{limit:J2246}}
\end{figure}

\begin{figure}
    \centering
    \includegraphics[angle=90, width = 1.0\hsize]{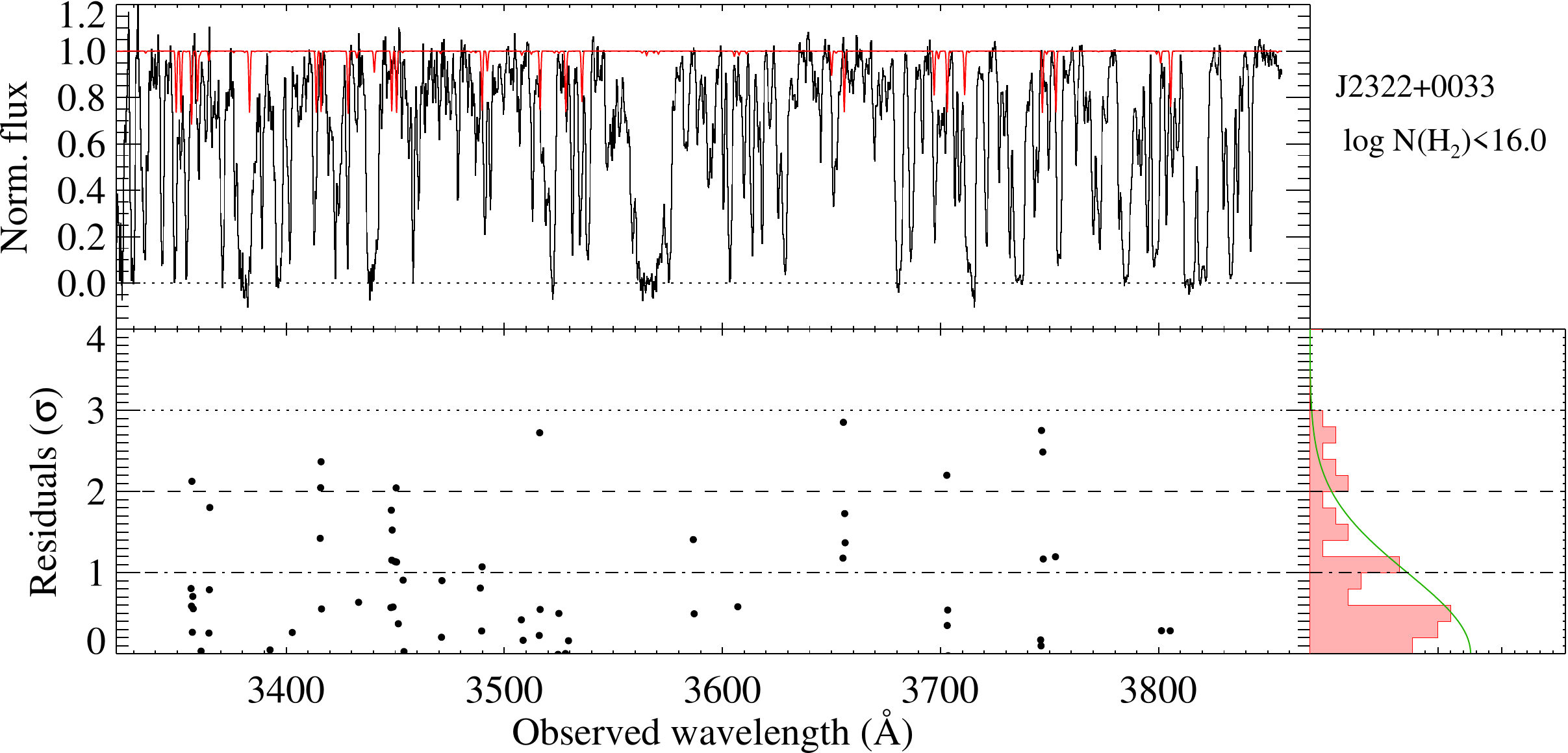}
    \caption{Same as \ref{limit:J0017} for the system towards J2322+0033. \label{limit:J2322}}
\end{figure}

We provide here details about the metals and molecular column densities
only for the 10 systems from our X-shooter observational programme analysed in
this work. The system towards J1513$+$0352 was presented in \citet{Ranjan+2018}.
Figs.~\ref{J0017met} to \ref{J2322met} show the Voigt-profile fitting to the main metal
lines using VPFIT and Table~\ref{metal_table} provides the corresponding total column densities.\\
Figs.~\ref{J0025CI}, \ref{J1258CI}, and \ref{J2140CI} present the detection of neutral carbon towards respectively
J0025$+$1145, J1143$+$1420, and J1258$+$1212. Finally, measurements of the \HH\ column densities
are presented in Figs.~\ref{J0025+1145_H2_L0} to \ref{J2232+1242_H2}, and the derivation of upper limits to $N(\HH)$ for the remaining systems is shown in Figs.~\ref{limit:J0017} to \ref{limit:J2322}.

\begin{table*}[]
\centering
\caption{Total column densities of low-ionization metal species in extremely-strong DLAs from this work} 
\label{metal_table}
\begin{tabular}{c c c c c c c}
\hline\hline
Quasar       & \zabs & N(\ion{Fe}{ii})        & N(\ion{Si}{ii})        & N(\ion{Zn}{ii})        & N(\ion{Cr}{ii})        & N(\ion{C}{i})          \\
\hline
J0017$+$1307 & 2.326 & 15.4$\pm$0.04  & 16.01$\pm$0.13 & 12.75$\pm$0.08 & 13.79$\pm$0.04 & $<$12.87       \\
J0025$+$1145 & 2.304 & 15.98$\pm$0.1  & 16.69$\pm$0.09 & 14.03$\pm$0.06 & 14.32$\pm$0.09 & 13.19$\pm$0.09 \\
J1143$+$1420 & 2.323 & 15.78$\pm$0.02 & 16.3$\pm$0.03  & 13.47$\pm$0.02 & 14.09$\pm$0.02 & $<$12.71       \\
J1258$+$1212 & 2.444 & 15.45$\pm$0.02 & 16.1$\pm$0.04  & 13.1$\pm$0.03  & 13.72$\pm$0.02 & $\sim$13.48    \\
J1349$+$0448 & 2.482 & 15.59$\pm$0.05 & 15.57$\pm$0.29 & 13.08$\pm$0.06 & 13.96$\pm$0.04 & $<$12.9        \\
J1411$+$1229 & 2.545 & 15.38$\pm$0.09 & 16.07$\pm$0.18 & 12.87$\pm$0.07 & 13.6$\pm$0.05  & $<$12.78       \\
J2140$-$0321 & 2.339 & 15.67$\pm$0.07 & 16.33$\pm$0.14 & 13.52$\pm$0.07 & 14.1$\pm$0.07  & 13.61$\pm$0.04 \\
J2232$+$1242 & 2.230 & 15.54$\pm$0.02 & 15.98$\pm$0.05 & 12.9$\pm$0.04  & 13.88$\pm$0.02 & $<$12.53       \\
J2246$+$1328 & 2.215 & 15.22$\pm$0.13 & 15.63$\pm$0.21 & 12.54$\pm$0.63 & 13.47$\pm$0.52 & $<$13.06       \\
J2322$+$0033 & 2.477 & 15.25$\pm$0.04 & 15.47$\pm$0.06 & 12.5$\pm$0.13  & 13.62$\pm$0.05 & $<$12.96   \\
\hline
\end{tabular}
\end{table*}

\begin{table*}[]
\centering
\caption{Metal lines used for the measurements of $\Delta v_{90}$} 
\label{delta_v90_table}
\begin{tabular}{c c c c c c c}
\hline\hline
Quasar & $\Delta v_{90}$ (km\,s$^{-1}$) & Transitions used  \\                 \hline
J0017$+$1307  & 120             & \NiII $\lambda$1709, \NiII$\lambda$1741, \SiII$\lambda$1808, \FeII$\lambda$2249, \MnII$\lambda$2576 \\
J0025$+$1145  & 240             & \NiII$\lambda$1709, \CrII$\lambda$2066                               \\
J1143$+$1420  & 130             & \NiII$\lambda$1454, \NiII$\lambda$1709, \NiII$\lambda$1751, \FeII$\lambda$2249           \\
J1258$+$1212  & 100             & \NiII$\lambda$1709, \NiII$\lambda$1741, \NiII$\lambda$1751, \FeII$\lambda$2249, \MnII$\lambda$2576   \\
J1349$+$0448  & 60              & \NiII$\lambda$1751, \FeII$\lambda$2249                               \\
J1411$+$1229  & 50              & \NiII$\lambda$1454, \NiII$\lambda$1709, \NiII$\lambda$1741, \FeII$\lambda$2249, \FeII$\lambda$2260 \\
J1513$+$0352  & 90              & \NiII$\lambda$1370, \NiII$\lambda$1741, \CrII$\lambda$2056, \FeII$\lambda$2249, \FeII$\lambda$2260 \\
J2140$-$0321  & 70              & \NiII$\lambda$1370, \NiII$\lambda$1454                               \\
J2232$+$1242  & 75              & \NiII$\lambda$1370, \NiII$\lambda$1709, \NiII$\lambda$1751, \CrII$\lambda$2066,          \\
J2246$+$1328  & 65              & \FeII$\lambda$1608, \SiII$\lambda$1808, \CrII$\lambda$2056, \FeII$\lambda$2260, \MnII$\lambda$2576 \\
J2322$+$0033  & 40              & \NiII$\lambda$1741, \NiII$\lambda$1751, \SiII$\lambda$1808, \CrII$\lambda$2056, \FeII$\lambda$2249 \\
\hline
\end{tabular}
\end{table*}

{}

{}

{}

{}

{}

{}

{}

{}

{}

{}

\end{appendix}

\end{document}